 \documentclass[sigconf]{acmart}
\usepackage{multirow}

\newcommand{\up}[1]{\textcolor{black}{#1}}

\copyrightyear{2026}
\acmYear{2026}
\setcopyright{cc}
\setcctype{by}
\acmConference[CHI '26]{Proceedings of the 2026 CHI Conference on Human Factors in Computing Systems}{April 13--17, 2026}{Barcelona, Spain}
\acmBooktitle{Proceedings of the 2026 CHI Conference on Human Factors in Computing Systems (CHI '26), April 13--17, 2026, Barcelona, Spain}
\acmPrice{}
\acmDOI{10.1145/3772318.3790466}
\acmISBN{979-8-4007-2278-3/2026/04}
\begin{document}

%% The "title" command has an optional parameter,
%% allowing the author to define a "short title" to be used in page headers.
\title[Writers' Professional Relationship With GenAI]{Investigating Writing Professionals' Relationships with Generative AI: How Combined Perceptions of Rivalry and Collaboration Shape Work Practices and Outcomes}

\author[R.Varanasi]{Rama Adithya Varanasi}
\affiliation{%
  \institution{Tandon School of Engineering \\ New York University}
  \city{NYC}
  \state{NY}
  \country{USA}
}
\author[O.Nov]{Oded Nov}
\affiliation{%
  \institution{Tandon School of Engineering \\ New York University}
  \city{NYC}
  \state{NY}
  \country{USA}
}
\author[B.Wiesenfeld]{Batia Mishan Wiesenfeld}
\affiliation{
 \institution{Stern School of Business \\ New York University}
 \city{NYC}
 \state{NY}
 \country{USA}
}



\begin{abstract}
This study investigates how professional writers' complex relationship with GenAI shapes their work practices and outcomes. Through a cross-sectional survey with writing professionals (n=403) in diverse roles, we show that  collaboration and rivalry orientation are associated with differences in work practices and outcomes. Rivalry is primarily associated with relational crafting and skill maintenance. Collaboration is primarily associated  with task crafting, productivity, and satisfaction, at the cost of long-term skill deterioration. Combination of the orientations (high rivalry and high collaboration) reconciles these differences, while boosting the association with the outcomes. Our findings argue for a balanced approach where high levels of rivalry and collaboration are essential to shape work practices and generate outcomes aimed at the long-term success of the job. We present key design implications on how to increase friction (rivalry) and reduce over-reliance (collaboration) to achieve a more balanced relationship with GenAI.   
\end{abstract}

\maketitle

\section{Introduction}
Generative AI (GenAI) technologies are becoming an increasingly prominent topic of debate in writing professions. Central to this debate is how GenAI is reshaping professional roles, including their behaviors, work practices, and outcomes. A stream of scholarship adopts an automation perspective, arguing that GenAI’s capabilities can displace worker responsibilities, thereby threatening tasks and processes negatively \cite{Acemoglu2024, Karunakaran2025}. In contrast, an augmentation perspective argues that GenAI can create new opportunities for workers, expanding their scope and improving outcomes \cite{Brynjolfsson2025}. However, these perspectives adopt a top-down framing that centers GenAI’s technical capabilities rather than workers’ experiences. In practice, integrations are often bottom-up, where professionals actively interpret these technologies and reshape their roles \cite{Dolata2024}.

Nascent HCI research adopts this bottom-up lens, showing how workers orientation towards GenAI and their relationship with it substantially shapes GenAI's impact on work. Some workers view GenAI as a competitor that threatens their expertise, autonomy, and professional identity, fostering a \textit{rivalrous} orientation \cite{boucher2024, Price2024resisting}. Individuals with this orientation resist GenAI when they perceive it as undermining their skills and devaluing their contributions \cite{Golgeci2025}. Conversely, workers develop \textit{collaborative} relationships with GenAI~\cite{Puerta2025}, understanding it as a supportive partner that helps them reflect on their work, refine their roles, and enhance outcomes \cite{Guo25}. A key gap in this scholarship, however, is that collaboration and rivalry have been treated as isolated phenomena. In practice, professionals often experience both simultaneously in different intensities, navigating parallel relationships with GenAI across different aspects of their work \cite{Varanasi2025, boucher2024}. Building on this work, our study investigates how writing professionals’ intertwined dual relationship with GenAI relates to their work. Specifically, we propose the following research questions: 
\medskip

\textbf{RQ1.} \textit{How do rivalry and collaboration orientations \textit{independently} associate with work practices and outcomes}? 

\smallskip

\textbf{RQ2.} \textit{How do rivalry and collaboration orientations, in \textit{combination}, associate with work practices and outcomes}?

\medskip
Work practices were measured using job crafting and skill maintenance. Job crafting refers to bottom-up behavioral changes through which workers better align their work with personal meaning, values, and identity \cite{Wrzesniewski_2001}. Skill maintenance complements crafting as an important practice that workers undertake once they have acquired and mastered a skill to reduce skill decay \cite{Arthur1998}. For work outcomes, we considered two important measures, perceptions of productivity and job satisfaction.

To answer the research questions, we conducted a cross-sectional survey with 403 professional writers. For RQ-1, we proposed several hypothesis (H1-H6) and conducted confirmatory analysis using linear regression models. For RQ-2, we engaged in an exploratory analysis guided by the RQ-1 results. In particular, we applied Response Surface Analysis (RSA), which is based on polynomial regression, to examine how the joint effects of rivalry and collaboration orientations were associated with the outcome variables. These joint effects were corroborated using cluster analysis to produce a $2X2$ combination of rivalry (low/high) and collaboration (low/high), and compared how professionals in each profile shaped their work practices and outcomes.

Overall, our results showed that both rivalry and collaboration orientations were independently associated with job crafting. However, only rivalry was associated with skill maintenance, whereas collaboration was primarily associated with productivity and satisfaction. \up{When rivalry dominated, professionals who engaged in self-directed job crafting also reported higher activity focused on maintaining their human skills, showing patterns consistent with lower skill decay, although they also reported lower satisfaction and greater strain. When collaboration dominated, professionals who engaged in work crafting reported having higher productivity and satisfaction, while also showing skill maintenance patterns associated with greater long-term skill deterioration. These findings highlight the imbalances associated with heavy reliance on a single orientation (rivalry or collaboration) with respect to calibrating work practices and outcomes}.

In contrast, the joint modeling of rivalry and collaboration showed that their combination \up{was associated with equal or greater work practices and outcomes compared to the independent models. Those respondents indicating a combination of high rivalry and collaboration with GenAI reported high levels of job crafting, skill maintenance, and productivity. Only satisfaction was lower when rivalry was high alongside collaboration than when rivalry was low and collaboration was high}. \up{Taken together, these findings suggest that a co-occurring dual relationship, i.e., contexts where rivalry aligns with the preservation of human efforts and collaboration aligns with more meaningful and satisfying work, especially at high levels, is associated with more favorable patterns in work practices and outcomes}. We conclude by offering design suggestions for workflows that cultivate healthy friction (rivalry) while reducing over-reliance (collaboration). In doing so, our study contributes to the research on GenAI in writing and future of work in the following ways:
\smallskip
\begin{itemize} 
\item Independent orientations toward collaboration or rivalry with GenAI create imbalances, \up{associated with} either work practices at the expense of outcomes, or outcomes at the expense of practices.  
\item Combination of rivalry and collaboration orientations mitigates these imbalances \up{and is associated with more holistic pathways that relate to both work practices and outcomes}.  
\item We present design recommendations for balancing the orientations of rivalry and collaboration, supporting role adaptation to changing conditions while preventing detrimental effects.  
\end{itemize}
\smallskip

\section{Prior Literature and Hypotheses}
In this section, we present related work to motivate our research questions and develop our hypotheses. First, we introduce the orientations of collaboration and rivalry towards GenAI. Next, we present prior work around both the orientations and their relationship with work practices (job crafting and skill maintenance) and work outcomes (productivity and satisfaction). 

\subsection{GenAI Orientations: Collaboration and Rivalry}
The growing ubiquity of GenAI is reshaping diverse domains of work, including creative~\cite{Varanasi2025,Kyi2025} and knowledge-based professions~\cite{lee2025impact, nasun2025}, changing work structures and practices. These sociotechnical transformations can emerge through top-down mechanisms, where organizations or clients introduce GenAI tools and prescribe associated practices for employees or contractors~\cite{Brynjolfsson2025}. Recently, these top-down mechanisms have been implemented in roles such as writing~\cite{Doshi2024} and creative arts~\cite{Heigl2025}. A significant portion of HCI research has studied the top-down deployment of GenAI tools in these work environments and resultant impact on working professionals~\cite{Cha2025}.  

At the same time, GenAI's simplistic design and high-level capabilities, make them extremely accessible for workers. They enable bottom-up practices in which individuals exercise agency by adapting GenAI to reshape workflows to achieve desirable outcomes~\cite{Parker_2025, Dolata2024}. However, these practices are shaped by how workers perceive GenAI in their work and how such perceptions guide their orientation and relationship with it. \up{Historically, HCI literature shows that individuals adopt diverse orientations toward technology. One set of orientations assumes technology as a tool~\cite{Anthony2023}. Under this framing, users focus on the instrumental and material aspects of technology, appropriating and reconfiguring its affordances within situated practice~\cite{dourish2001action}. A second set of orientations treats technology as a medium~\cite{Anthony2023}, in which users attend to its expressive and communicative properties, viewing it as enabling or constraining common ground, coordination, and interaction~\cite{baym2015personal}. In the context of AI, orientations increasingly position the system as a counterpart or social actor~\cite{Nass1996, Anthony2023}. Here, users respond to properties such as inscrutability, adaptivity, and hyper-personalization, and orient themselves towards these interactions and power dynamics by either aligning/integrating the system or challenging/avoiding it.}  As GenAI offers increasingly diverse capabilities, examining such bottom-up orientations and practices is critical because they challenge established norms and create new expectations within work roles. 

On one hand, workers perceive benefits in integrating GenAI into their work. \up{They may demonstrate a \textit{collaborative} stance by actively adopting GenAI, enhancing workflows, and using these interactions to improve outcomes~\cite{Takaffoli2024, Al2024}.} Within HCI, such studies build on rich traditions of human–AI teaming~\cite{Seeber2020, ONeill2022} (e.g., collective hybrid intelligence) and human–agent collaboration~\cite{Parasuraman2000}. This body of work emphasizes the complementarity of human and GenAI capabilities and how optimizing both can yield efficient outcomes~\cite{Puerta2025}. Trust plays a central role in shaping these collaborations, often manifesting in three forms~\cite{Seeber2020, Berretta2023}: (a) AI augmenting human limitations, (b) mutual dependency between humans and AI, or (c) human supervision of AI. Recent HCI research in this space has explored the opportunities of developing GenAI-powered tools and studying how professionals engage with them~\cite{Guo25,yun25}.

On the other hand, workers may \textit{resist} GenAI by refusing to alter existing practices or by explicitly shaping their practices through non-use, reduced use, or dysfunctional use~\cite{Samhan_2018, Price2024resisting}. Resistance literature within HCI has examined phenomena such as algorithmic aversion~\cite{Dietvorst2015, Cheng2023}, under-reliance~\cite{Schemmer2023}, non-use~\cite{Cha2025, Baumer2015}, and algorithmic contestation~\cite{Sloane2022, Karusala2024} as forms of critical responses to AI systems. \up{These studies show how individuals resist AI primarily motivated by \textit{negative} perceptions of threat, anxiety, and fear of replacement~\cite{boucher2024}. Such negative perceptions can promote AI  \textit{disengagement} among professional who reject adoption, limit reliance, subvert intended functions, or contest its legitimacy. Emerging HCI work in this space has focused on points of disengagement and friction that contribute to the resistance practices~\cite{Kobiella2024}.}  

\up{Resistance can also take the form of proactive \textit{engagement} when workers perceive a competitive threat to their occupation. One such form of AI resistance is \textit{rivalry}. It is defined as a psychologically charged  subjective competitive relationship in which individuals perceive AI a salient opponent, leading to heightened stakes, regardless of objective situation ~\cite{Kilduff2010}. Rivalry differs from fear-based resistance because it is activated not by withdrawal tendencies but by challenge-oriented responses that seek to outperform, differentiate, or symbolically counter the rival~\cite{Lee2025}. In this sense, workers resist its encroachment by leveraging competitive attitudes, reinforcing their distinct human capabilities, and protecting the boundaries of their craft. }

Perceptions of both collaboration and rivalry shape how professionals engage with GenAI in their work. \up{In the writing profession, early HCI research emphasized the collaborative orientation of writers and developed AI tools to support and enhance creative processes and outputs~\cite{Reza2024, Hoque2024, chung22}. However, more recent work has taken a critical turn by examining writing communities’ concerns about GenAI.~\citet{mirowski2023} demonstrated that screenwriters experience tensions between their creative instincts and GAI’s generative behaviors, highlighting the need for deeper understanding of how these systems are integrated into writing practice. Emerging research further documents ethical issues raised by GenAI’s use in writing communities.} 

\up{Studies of fanfiction communities show that the large-scale scraping of writers’ crafted labor for LLM training decontextualizes their work, undermines authorial agency, and detaches content from its original intent~\cite{NYT2024}. LLM outputs also blur boundaries between original, derivative, and plagiaristic writing, complicating norms of authorship, attribution, and the labor required to verify provenance~\cite{draxler24}. For example, studies of Wikipedia contributors similarly reveal increasing invisible labor among copyeditors who must delineate and manage the boundary between human- and AI-generated text~\cite{McDowell_2024}}.

Nascent research investing these positive and negative orientations around GenAI, suggests that these perceptions are not mutually exclusive; rather, individuals can experience collaboration and rivalry as distinct, co-occurring orientations toward GenAI~\cite{Varanasi2025}. Our study contributes a human-centered perspective by examining how workers adopt stances of collaboration and rivalry in their relationships with GenAI and shape their overall work.

\subsection{Work Practices: Job Crafting}
An individual's collaborative or rivalrous orientation toward a technology such as GenAI can shape how they engage with their work, influencing their practices, role interpretations, and interactions with others. For instance, a study found that perceiving AI as a collaborator was correlated with trust in AI ~\cite{Wen2025}.  Other work theorized that individuals' AI resistance is rooted in fear and perceived inefficacy ~\cite{Golgeci2025}. Human-centered work design frameworks offer a critical foundation for understanding how such differences in perceptions can translate into enacting different workplace behaviors~\cite{Wrzesniewski_2001}. Job Crafting theory is perhaps the most influential framework modeling how individuals proactively and agentically shape their work.

Job crafting’s bottom-up approach offers an alternative to earlier top-down models of work design~\cite{Parker2025}. It emphasizes the physical and cognitive changes people initiate to align their work with personal meaning, values, and identity. Unlike traditional models of work, which assume that job structures are defined by organizational mandates, job crafting highlights how workers proactively redefine their role and relational boundaries from the bottom up ~\cite{Tims_2012}. Wrzesniewski and Dutton introduced the idea of job crafting by showing how professionals invest significant energy in reconfiguring and shaping their jobs~\cite{Wrzesniewski_2001}. 

Job crafting is characterized according to two main factors. The first is based on the worker’s goal. Workers can engage in \textit{approach} crafting, which involves people actively creating opportunities to expand their work involvements in ways that align with their professional preferences, strengths, and goals, or \up{\textit{avoidance} crafting, which involves the practices workers employ to reduce negative aspects of work~\cite{Bruning_Campion_2018, zhang_2019}. Typical examples of avoidance crafting include reducing contact with demanding clients, declining optional high pressure assignments, or restructuring one’s workflow to limit emotionally taxing tasks.} The second factor relates to the work domains people recruit in their job crafting efforts. These can involve \textit{task} crafting focused on shaping one's work tasks or processes; \textit{relational} crafting focused on shaping one’s work relations; or \textit{cognitive} crafting focused on shaping one's role meaning and identity.

The extent to which workers engage in job crafting is shaped by multiple factors, including worker rank~\cite{Berg2010}, degree of autonomy~\cite{Berg2010}, and the resources available to them~\cite{Petrou2017}.  Much of this early research focused on full-time employees in traditional organizational settings, examining issues such as establishing goals and work–life balance~\cite{Bruning2019, Kossek2016, Gravador2018} and managing workplace relationships~\cite{Bizzi_2017}. As job structures have evolved in recent years, scholars have extended job crafting to non-traditional AI-mediated work contexts, such as gig work~\cite{Dominguez2016, Wong_Fieseler_2021}. Collectively, these findings underscore that job crafting is not a uniform process but one that varies across jobs, roles, and work environments~\cite{Petrou2012}. Building on this foundation, our study investigates how workers’ relationships with GenAI can shape their job crafting behaviors.

In our study context, recent research around AI resistance has indicated that a heightened sense of AI resistance increases workers need to reframe their roles, often by reconciling them with their internal values, emphasizing their unique human skills, and mentally redefining tasks in ways that differentiate workers from AI systems~\cite{Golgeci2025, Christin_2020, Varanasi2025}. In contrast, while workers who collaborate with AI also think about ways to shape their roles through GenAI, they are less likely to reframe their roles than workers who are more resistant.  Consequently, we hypothesize: 

\smallskip
\textbf{H1}: \textit{\up{Writing professionals who perceive GenAI as a rival will report higher levels of cognitive crafting compared to those who perceive GenAI as a collaborator}}.
\smallskip

Collaboration with AI can increase reliance, leading workers to invest significant time and energy in reconfiguring tasks and workflows, such as determining which tasks to delegate to AI and which to retain to perform themselves~\cite{Clarke2025, Sun2025}. These reconfigurations are representative of what the job crafting literature labels task crafting. In contrast, while AI resistance can also prompt workflow adjustments, resistant workers often respond by doubling down on practices that emphasize human autonomy and agency, maintaining existing routines rather than revising their work roles in order to protect their professional identity~\cite{Varanasi2025}. Accordingly, we hypothesize: 

\smallskip
\textbf{H2}: \textit{\up{Writing professionals who perceive GenAI as a collaborator will report higher levels of task crafting compared to those who perceive GenAI as a rival}}. 
\smallskip

With respect to interpersonal relationships, scholarship on technology and AI resistance has shown a consistent shift among workers to increase their reliance on human connections due to their common goals of competing against AI. This manifests as increased reliance on peer expertise and strengthening professional support networks to collectively safeguard their roles and reinforce their unique contributions when rejecting technological change~\cite{Khovanskaya2020}. Collaboration with AI technologies may also trigger relational changes, but these changes are often limited to establishing loose networks, such as online groups, to share best practices on utilizing AI in their workflows ~\cite{Cheng2024}. Based on these findings, we hypothesize: 

\smallskip
\textbf{H3}: \textit{\up{Writing professionals who perceive GenAI as a rival will report higher levels of relational crafting relative to those who perceive GenAI as a collaborator}}. 
\smallskip

Competitive or adversarial dynamics associated with interpersonal rivalry at work have been found to motivate individuals to engage in proactive, approach-oriented strategies as they seek to protect their standing and demonstrate their value through visible actions~\cite{Garcia2006}.  Studies of resistance to technological change further emphasize that when resisting, workers are less likely to engage in preventive or avoidance crafting, as such behaviors may be perceived as conceding ground to the competing system~\cite{mohlmann2017hands}. Conversely, collaboration with technology has been shown to elicit a broader range of adaptive strategies, where workers not only seek opportunities (approach) but also pre-emptively reduce risks (avoidance), resulting in a more balanced form of role adaptation~\cite{Leonardi_2011}. Notably, supplementing avoidance strategies with approach strategies can buffer the negative effects on work engagement resulting from pursuing avoidance crafting in isolation~\cite{Hakanen2020}. Extending these findings, we hypothesize: 

\smallskip
\textbf{H4}: \textit{\up{Writing professionals who perceive GenAI as a rival will primarily engage in approach crafting. Those who perceive GenAI as a collaborator will report both approach and avoidance crafting}}. 
\smallskip

\subsection{Work practices: Skill Maintenance}
Workers who proactively engage in collaborative and rivalry behaviors with AI can also perceive differences in various work outcomes~\cite{law2025,boucher2024}. In our study, we focused on two specific outcomes - (i) skill maintenance to counteract skill decay, and (ii) satisfaction.

\subsubsection{Skill Maintenance}
Skill decay refers to the deterioration or loss of previously acquired skills due to periods of infrequent use or complete non-use~\cite{Arthur1998}. Within occupational contexts, skill decay represents a significant concern, as it can lead to reduced competence, diminished confidence, and lower levels of autonomy~\cite{Rashid2021}, factors whose absence can accelerate  deprofessionalization~\cite{Toren1975}. Skill decay, including that associated with AI use, has been studied in the medical profession, with research examining gaps in working practice, especially technical skills, and their contribution to skill decay~\cite{Gawad2019, Ahmad2025}. Parallel findings have also emerged in other professions such as first responders~\cite{Anderson2011}, pilots~\cite{Childs1986}, and knowledge workers~\cite[Ch.1]{Arthur2013}. 

A key factor that contributes to skill decay is the degree of overlearning, necessary for skill maintenance, which is defined as the extent to which an individual practices a skill beyond initial mastery. Reduction in skill maintenance through reduced task repetitions, can contribute to skill decay~\cite{Arthur1998}. While skill decay is an important measure to capture directly, subjective perceptions are particularly difficult to measure in relation to GenAI as professionals might be unaware of the extent of their own skill decay. Moreover, when skills are AI-mediated, it can create a false impression of task repetition even though the practitioner may be unable to demonstrate the same skill independently of GenAI~\cite{Macnamara2024}. Practitioners also often downplay their level of skill decay, reflecting social desirability bias~\cite{Connolly2024}. For these reasons, skill maintenance may be a more accurate indicator of reduced skill decay~\cite{Arthur1998}.

The nature of tasks is an important factor in skill maintenance.  Cognitive skills often require significant attention and mental rehearsal compared to procedural tasks~\cite{wang2013factors}. GenAI tools are designed to emulate cognitive functions, such as recognizing patterns, reasoning through possible outcomes, and frequently guiding users toward specific actions~\cite{park2023generative}. Increasing evidence suggests that individuals who rely on GenAI  for cognitive tasks risk reducing their own repetitions of skill-maintaining tasks~\cite{lee2025impact}. In some extreme cases, professionals delegate even their broader core tasks, well beyond just cognitive ones~\cite{Varanasi2025}, risking reduced skill maintenance across a wider range of abilities. AI resistance may elicit the opposite pattern, in which professionals  focus on reinforcing their skills due to concerns about their job security~\cite{Varanasi2025}. Similar dynamics have been observed in interpersonal workplace rivalries, wherein workers focus on honing and maintaining their core competencies to gain a competitive edge~\cite{Kilduff_2014}. Based on this reasoning, we propose the following hypothesis: 

\smallskip
\textbf{H5}: \textit{\up{GenAI collaboration will be associated with lower skill maintenance, whereas GenAI rivalry will be associated with higher skill maintenance.}} 

\subsection{Work Outcomes: Productivity and Job Satisfaction}
Another factor to consider when thinking about how individuals' relationship with GenAI is associated with their work outcomes is their job productivity. 
Prior research highlights diverging relationships between collaboration and resistance toward AI and workplace productivity. Collaboration may be associated with higher productivity because collaborating professionals delegate routine or repetitive tasks to AI, enabling them to allocate greater effort to complex or creative endeavors~\cite{Clarke2025}. These collaborative dynamics reflect the complementarity between human and AI capabilities, where efficiency gains relate to optimized division of labor~\cite{Puerta2025}. In contrast, resistance behaviors may be more weakly related to productivity. Workers perceiving AI as a rival often double down on human-centric practices, emphasizing autonomy and control but foregoing opportunities to accelerate workflows~\cite{Golgeci2025}. This response aligns with broader findings in the literature on employee resistance, which links adversarial orientations to higher effort but lower productivity~\cite{mohlmann2017hands}. Synthesizing these perspectives, we suggest the following hypothesis: 

\smallskip
\textbf{H6}: \textit{\up{Workers' engagement in GenAI collaboration will be positively associated with levels of job productivity; engagement in GenAI rivalry will be less positively associated with job productivity}}. 
\smallskip

The final factor we consider is job satisfaction, which refers to how an individual perceives their overall job~\cite[Ch.1]{Spector2022}. Research around this construct has shifted from seeing job satisfaction as merely the absence of needs and towards interest in cognitive processes and attitudes~\cite[Ch.9]{Sessa2021}. In this sense, individuals’ attitudes toward job satisfaction vary along with how they actively shape their work. Professionals who engage in higher levels of job crafting often report greater job satisfaction~\cite{De2016}. This relationship is intuitive because satisfaction is tied to job characteristics, and workers who modify aspects of their jobs to better fit their needs (i.e., through job crafting) may have more positive (and fewer negative) experiences~\cite{Zito2019, Petrou2012}.

At the same time, studies on collaboration with AI show that workers often experience relief and confidence from offloading tedious tasks, which may strengthen a positive association between job satisfaction and job crafting behaviors~\cite{Sun2025}. In contrast, perceived rivalry may be associated with the opposite pattern.  Rivalry relates to heightened pressure and exhaustion, as encounters with rivals are perceived as high-stakes, which may be negatively associated with the satisfaction derived from job crafting~\cite{Kilduff2010}. Research on AI resistance reveals similar dynamics, with higher anxiety and fear stemming from uncertainty, which in turn is associated with non-use of AI tools, particularly when the implementation is top-down~\cite{Golgeci2025}. Based on these insights, we hypothesize: 

\smallskip
\textbf{H7}: \textit{\up{Workers' engagement in GenAI rivalry will be negatively associated with job satisfaction; workers' engagement in GenAI collaboration will be positively associated with job satisfaction}}.
\smallskip

\section{Methods}

\subsection{Procedure Overview}
To test our hypothesis, we conducted a cross-sectional survey study with professional writers who worked with GenAI in various capacities in their work. The study was approved by the IRB, pre-registered \footnote{https://doi.org/10.17605/OSF.IO/3BNEU}, and administered via Qualtrics. We used Prolific to recruit professional writers in line with prior HCI research \cite{Douglas2023}. We employed stratified sampling techniques to find participants that had positive as well as negative perceptions around GenAI with an aim to capture enough participants with strong rivalry and collaboration perceptions. We used Prolific’s built-in sampling tools for this purpose. First, we selected the survey participants to be representative of the U.S. population by age, sex, and race. Second, we shared the survey to only participants above 18 and were employed full-time in a field that was writing-focused (e.g., journalist, author). Third, we also defined additional custom criteria through a pre-screener to shortlist participants with negative as well positive views around AI by asking them a few questions (e.g., ``How do you view the overall effect of GenAI in your work?’’). 

\begin{table*}[!t]
  \centering
  \caption{Table listing the participant demographics of the cleaned sample (n=403). Values are counts with percentages in parentheses. Writing time is a single-item Likert measure (1:No time at all–7: All the time). Industry shows the ten most frequent categories; remaining industries are omitted for brevity.}
  \label{tab:demographics}
  \begin{tabular}{lll}
    \toprule
    \textbf{Dimension} & \textbf{Sub-dimension} & \textbf{Participants} \\
    \midrule
    Gender & Man & 178 (44.2\%) \\
           & Woman & 220 (54.6\%) \\
           & Non-binary / third gender & 5 (1.24\%) \\
    \midrule
    Age    & 18--24 & 62 (15.4\%) \\
           & 25--34 & 116 (28.8\%) \\
           & 35--44 & 101 (25.1\%) \\
           & 45--54 & 61 (15.1\%) \\
           & 55+ & 63 (15.6\%) \\
    \midrule
    Education & Less than high school & 1 (0.25\%) \\
              & High school graduate & 19 (4.71\%) \\
              & Some college, no degree & 40 (9.93\%) \\
              & Associate degree & 19 (4.71\%) \\
              & Bachelor's degree & 186 (46.2\%) \\
              & Some graduate school & 6 (1.49\%) \\
              & Master's degree & 107 (26.6\%) \\
              & Professional degree & 6 (1.49\%) \\
              & Doctoral degree & 19 (4.71\%) \\
    \midrule
    GenAI use for writing 
    (Likert, 1-7) & 4 (Significant amount of time) & 47 (11.7\%) \\
                          & 5 & 129 (32.0\%) \\
                          & 6 & 122 (30.3\%) \\
                          & 7 (All my time) & 105 (26.0\%) \\
    \midrule
    Industry (top 10) & Marketing & 65 (16.1\%) \\
                      & Online/Offline publishing & 57 (14.1\%) \\
                      & IT, Digital apps, \& Social media & 51 (12.7\%) \\
                      & Education & 36 (8.93\%) \\
                      & Arts & 33 (8.19\%) \\
                      & Consulting & 27 (6.7\%) \\
                      & Retail & 25 (6.2\%) \\
                      & Healthcare & 24 (5.96\%) \\
                      & Finance/accounting & 20 (4.96\%) \\
                      & Manufacturing & 18 (4.47\%) \\
    \bottomrule
  \end{tabular}
\end{table*}

To determine a sufficient sample size, we conducted a power analysis a priori using G*Power with a desired effect size of .25, power of .99, and error probability < 0.05.  Power analysis indicated that we needed at least 232 participants. We recruited a total of 450 writers (see Table \ref{tab:demographics}) for the survey and paid them at Prolific’s suggested rate of 14.4\$/hour (our survey took an average of 15 mins. to complete).  

% \subsection{Participants demographic} {\fixme{WIP}} 
% We standardized the roles into xxx for better analysis. For example, under business writer, we included business writer and marketing consultant; under social media writer we included article writer and blogger; and under technical writer we included science writer, research writer, etc. Exhaustive details are presented in the Table \ref{tab:demographics}.

\subsection{Measures and Instrument Development}
\up{The survey instrument for the study was developed to assess writing professionals’ perceptions of their relationship with GenAI as well as their practices and attitudes relating to the technology. This process entailed the adaptation of existing validated measures that were not originally AI-specific, such as scales assessing interpersonal relationships of rivalry and collaboration and general job crafting scales, and the conversion of skill maintenance constructs into a self-report format. Scale adaptation proceeded in multiple stages. First, an initial qualitative study involving interviews with writing professionals was conducted to identify how the focal constructs manifested in writers' work roles. Subsequently, triangulation between the coded interview data and the existing validated scales (or behavioral index) guided the adaptation of items to ensure conceptual alignment and face validity. Content validity was then evaluated by five subject matter experts and items were reworded until consensus was reached.} 

\up{We also conducted two subsequent pilot studies with 40 participants each to assess the reliability of the measures (via Cronbach’s alpha). Those items that were below the acceptable threshold of 0.70 were either discarded or updated with better measures. Based on the answering patterns, we also further simplified the wording to ensure the questions were easy to understand and added three attention check questions (1 question for every 5 mins). All items within each question block were randomized to reduce bias. The complete survey is presented in the supplementary materials.}

Overall, our survey measures focused on capturing four categories of constructs - (a) workers’ perceptions about GenAI (rivalry and collaboration), (b) their practices (job crafting), (c) their work outcomes (skill maintenance, productivity, job satisfaction), and (d) their GenAI usage rates and GenAI task involvement, included to illuminate descriptive patterns of different groups of writers. All items were measured with seven-point Likert scales. Descriptive and reliability measures of all the items are presented in the Table \ref{tab:measures-desc}. \up{For the instruments that were modified and adapted to fit the study, we have provided additional details on how we translated and validated them}. All the items are listed in the supplementary material. 

\subsubsection{Collaboration}
Collaboration was measured using a three item scale excerpted from \up{two validated scales measuring cooperation \cite{Tang_1999} and collaboration intent \cite{Schuster2021}}. Initially, we created a five item scale. However, due to low Cronbach's alpha in a pilot study, we dropped two items to obtain good reliability (cronbach's alpha = 0.93). Example items included - ``\textit{To achieve positive results in work, I must collaborate with Generative AI.}’’ \up{Additionally, as the item was translated, we also measured McDonald’s omega that does not assume tau-equivalence. The value of $\omega_{\text{total}}$ was 0.93, above the required threshold.}

\subsubsection{Rivalry}
Rivalry was measured using a six item scale adapted from the validated scale of interpersonal rivalry by \citet{Kilduff_2014}. Kilduff’s scale was originally developed to measure perceptions of rivalry between professionals in different competitive environments, \up{such as sports (e.g., ``Losing to this person felt worse than losing to other competitors'')}. \up{Three scale items were applicable to our study, and we reworded them to fit our study context. We supplemented them with three additional items derived from \citet{Varanasi2025}'s theoretical insights on AI rivalry.} Example items included - ``\textit{I consider Generative AI to be a rival at work.}’’. The Cronbach alpha of the scale was 0.77. \up{Additionally, the value of $\omega_{\text{total}}$ was 0.82}.

\subsubsection{Job Crafting}
In order to capture the change in writers’ practices, we employed the extended job crafting scale developed and validated by \citet{Bindl_2019}. While there are several established job crafting scales, this extended scale was particularly applicable to our study because of its multi-dimensionality. The scale covers both the job crafting motivation dimension (i.e., approach and avoidance) and the job crafting application dimension (i.e., task, relational, and cognitive), thereby allowing us to comprehensively measure changes in work practices. \up{The two motivation and three application dimensions yielded six sub-scales, which we adapted with minor wording changes}. We instructed respondents to answer the items in relation to their work practices since the introduction of GenAI. Example items included \textit{``I have been changing my tasks so that they are more challenging.’’} and \textit{``I have been minimizing my interactions with people related to my profession that I am not getting along with.’’}. Cronbach’s alpha for the overall scale was 0.87.

\subsubsection{Skill Maintenance}
To measure writers’ perceived skill maintenance, we developed a 21-item scale encompassing four skill dimensions relevant to professional writing: cognitive skills (5 items), creative skills (6 items), tacit skills (5 items), and social skills (5 items). Because no existing scale captured all these categories comprehensively, we derived items by synthesizing constructs from prior theoretical and empirical work. Item sources for each category included: cognitive skills \cite{lee2025impact, Varanasi2025, zhao25}; creative skills \cite{Onet2025database, Kellogg_2006, Varanasi2025}; tacit skills \cite{Onet2025database, zhao25, Kellogg_2006}; and social skills \cite{Onet2025database, Varanasi2025, zhao25}.

Participants responded to the following prompt: ``\textit{Since the emergence of Generative AI, how has the effort you invest in developing the following skills in your job changed, compared to before you used GenAI?}’’ Example items included \textit{``Developing tone/voice of characters''} and \textit{``Persuading co-workers/clients''}. To capture respondents’ underlying rationale, the scale was followed by two open-ended questions, such as: ``\textit{For the questions where you indicated spending `less effort' on developing specific skills, please explain why?}’’ The overall 21 item scale demonstrated high internal reliability (Cronbach’s alpha = 0.96, \up{$\omega_{\text{total}}$ = 0.97}). Internal reliability for subcategories is presented in the Table \ref{tab:measures-desc}. \up{Given the high alpha observed for the 21-item skill-maintenance, we conducted additional psychometric checks to assess redundancy and dimensionality. Corrected item–total correlations ranged from $.59$ to $.81$, exceeding the recommended $.30$ threshold. Inter-item correlations within the cognitive ($.63-.77$), tacit ($.59-.81$), social ($.56-.80$), and creative ($.55-.81$) domains were below the redundancy cut off ($.85$). An exploratory factor analysis extracted four factors with eigenvalues >1, consistent with the theorized cognitive, creative, tacit, and social domains. Items loaded cleanly on their intended factors (primary loadings = $.42-.90$; cross-loadings $< .30$, see Table \ref{tab:efa21} in Appendix). We also conducted CFA to verify the four-factor structure, which demonstrated an acceptable fit ($\chi^2(183) = 587.26, CFI = .95, TLI = .94, RMSEA = .074, SRMR = .056$). Please refer to the table \ref{tab:CFA21} in Appendix for comparison with other factor models. Lastly, we assessed potential common method variance using Harman’s single factor test by loading all measurement items into an unrotated one factor maximum likelihood solution; the single factor accounted for only 18.2\% of the total variance. We further estimated a single factor confirmatory factor model, which fit the data poorly, $\chi^2(4464) = 25422.93,\ \text{CFI} = .25,\ \text{TLI} = .23,\ \text{RMSEA} = .11,\ \text{SRMR} = .15$, indicating that a single common factor does not account for the observed covariances.}

\begin{table*}[!t]
  \centering
  \caption{This figure presents descriptive statistics (mean, standard deviation, 95\% confidence interval) and reliability (Cronbach's $\alpha$) for predictors and outcomes.}
  \label{tab:measures-desc}
  \begin{tabular*}{0.8\textwidth}{@{\extracolsep{\fill}} l c c c c @{}}
    \toprule
    \textbf{Variable} & \textbf{M} & \textbf{SD} & \textbf{95\% CI} & $\boldsymbol{\alpha}$ \\
    \midrule
    \textbf{Predictor Variables} \\
    \quad Rivalry              & 3.81 & 1.26 & [3.69, 3.94] & .77 \\
    \quad Collaboration        & 4.90 & 1.70 & [4.73, 5.07] & .93 \\
    \addlinespace
    \textbf{Outcome Variables} \\
    \quad Job crafting (overall)      & 4.64 & 0.88 & [4.56, 4.73] & .87 \\
    \quad \quad Cognitive crafting    & 5.22 & 0.97 & [5.12, 5.31] & .79 \\
    \quad \quad Task crafting         & 4.52 & 1.14 & [4.41, 4.64] & .79 \\
    \quad \quad Relational crafting   & 4.19 & 1.07 & [4.08, 4.29] & .72 \\
    \quad \quad Avoidance crafting    & 4.16   & 1.02   & [4.06 , 4.26] & .77 \\
    \quad \quad Approach crafting     & 5   & 1.12   & [4.89 , 5.11] & .91 \\
    \addlinespace
    \quad Skill Maintenance (overall) & 4.59 & 1.23 & [4.47, 4.71] & .96 \\
    \quad \quad Cognitive skills      & 4.69 & 1.42 & [4.56, 4.83] & .92 \\
    \quad \quad Writing skills        & 4.65 & 1.44 & [4.51, 4.79] & .93 \\
    \quad \quad Social skills         & 4.31 & 1.38 & [4.18, 4.45] & .91 \\
    \quad \quad Tacit skills          & 4.68 & 1.44 & [4.54, 4.82] & .92 \\
    \addlinespace
    \quad Satisfaction                & 5.92 & 1.01 & [5.82, 6.02] & \, .90\\
    \quad Productivity                 & 5.9 & 0.89  & [5.82, 5.99] & .89 \\
    \bottomrule
  \end{tabular*}
\end{table*}

% \subsection{Ownership} 
% Ownership was measured with a three item scale derived from a validated measure of psychological ownership \cite{Van_2004} that we adapted to our study context. Internal reliability of the items was good (Cronbach’s alpha = 0.89). Example items included \textit{``I feel a very high degree of personal ownership for the work I produce.’’}

\subsubsection{Productivity and Job Satisfaction}
In order to measure productivity, we adapted five items from the established Individual Work Performance Questionnaire \cite{Koopmans2014}. It is a commonly-used scale measuring employees' self-reports of perceived workplace productivity. Example items included\textit{``I was able to carry out my work efficiently''} and \textit{``I managed my time well.''} Cronbach’s alpha was 0.89. We measured job satisfaction with a three-item scale that captured the writer's overall satisfaction with their job. The measure was adapted from an established satisfaction sub-scale of a technostress scale developed and validated by \cite{Ragu2008}. Example items included \textit``{I like doing the things I do at work}.’’ Cronbach’s alpha was 0.90. 

\subsubsection{GenAI Usage and Task Involvement}
Finally, we also measured several variables to understand workers' GenAI usage and the nature of their professional writing work. The first set of variables in this category were around their work that capture their role, nature of work,  their participation in the union, and the GenAI use policy in their work. A second set of variables captured their background in GenAI through Likert questions capturing how capable they felt GenAI was at performing their work, their level of experience with GenAI, and their GenAI use in work. The last set of questions was an index of GenAI task involvement, which instructed respondents to indicate the level of GenAI use in different types of tasks they performed. Examples included ``\textit{Tasks in which I am an expert}'' and ``\textit{Tasks that I find boring}''.    

\subsection{Analysis} \label{methods-analysis}
All the respondents were briefed on the study and signed a consent form. In total, we received 451 responses on the Prolific platform. Out of these responses, we excluded 48 responses based on diverse exclusion criteria, including (a) failing at more than one attention check, (b) not in a writing profession, (c) low-quality free text responses, (d) duplicate responses, (e) responses with straightlining behavior, (f) outliers in response times,  and (g) participation in the pilot survey. All the subsequent confirmatory and exploratory analysis was done using R.  

\subsubsection{Confirmatory Analysis} 
To answer RQ-1, we independently analyzed the relationship between independent (i.e., perceived GenAI rivalry and collaboration) and dependent variables (i.e., job crafting, skill maintenance, productivity and job satisfaction) using multiple linear regression with ordinary least squares (OLS). In all the models, we added age, sex, and education as covariates. We report both unstandardized coefficients (in the results) and standardized coefficients (in the Tables \ref{tab:crafting-regressions} and \ref{tab:skill-maintenance}) to enable comparison of relative effect sizes. To assess whether rivalry and collaboration differed in predictive strength, we conducted a planned contrast comparing their regression coefficients $(\beta_{\text{rivalry}} - \beta_{\text{collaboration}} = 0)$ using a general linear hypothesis test (GLHT). We considered statistical significance at $p < 0.05$ and report effect sizes where applicable to indicate the strength of the relationship between the variables.

\subsubsection{Exploratory Analysis} \label{methods-exploration}
To answer RQ-2, we conducted exploratory analyses to examine the joint relations of rivalry and collaboration with respect to our dependent variables. We first tested whether such joint effects were present using Response Surface Analysis (RSA) \cite{Humberg2019}. Additionally, RSA was used to understand if joint effects were more robust than independent effects. Although RSA is well-suited for modeling joint relationships between predictors, it can be difficult to interpret when effects emerge in highly localized regions (e.g., low rivalry but high collaboration). To complement RSA and better understand such localized patterns, we employed Partition Around Medoids (PAM) clustering. PAM enabled us to identify sample-informed clusters corresponding to meaningful $2x2$ combinations of rivalry and collaboration (e.g., LowR/HighC), thereby offering an interpretable way to examine how distinct profiles related to outcome variables. Lastly, to understand why certain behavior patterns emerged among these combinations, we conducted qualitative analysis on the two open-ended questions. 

\paragraph{Response Surface Analysis} RSA is a polynomial regression–based approach designed to test non-linear associations between two predictors and an outcome, such as congruence (rivalry = collaboration), incongruence (rivlary $\neq$ collaboration), and curvature effects. Following established practice, we estimated second-order polynomial regression models of the form:

\[Z = b_{0} + b_{1}R + b_{2}C + b_{3}R^{2} + b_{4}RC + b_{5}C^{2} + \epsilon,\]

where $Z$ is the dependent variable (e.g., job crafting, skill maintenance), $R$ is rivalry, and $C$ is collaboration. Predictors were \textit{mean-centered}  prior to creating squared and interaction terms to facilitate interpretation \cite[Ch.11]{Judge2007}.  From these models, we derived the canonical RSA surface parameters ($a_{1}$–$a_{4}$), which describe slopes and curvatures along the \textit{line of congruence (LOC; $R = C$)} and the  \textit{line of incongruence (LOIC; $R = -C$)} \cite{Humberg2019}. Specifically, $a_{1}$ and $a_{2}$ capture the slope and curvature along the LOC, while $a_{3}$ and $a_{4}$ capture the slope and curvature along the LOIC. These parameters provide statistical tests of whether dependent variables are higher when rivalry and collaboration increase together, and whether mismatches between them are detrimental.

\paragraph{Partition Around Medoids (PAM) clustering} \label{pam}
When the RSA results indicated a significant association between the combined orientation of collaboration and rivalry and aggregate measures (e.g., job crafting), we further examined the relationship by identifying key clusters ($2x2$ combination of rivalry and collaboration, called profiles hereafter) and their relationship with the dependent variables. To achieve this, we ran Partition Around Medoids (PAM) clustering algorithm on our sample. We considered both probability based (e.g., Latent  Profile Analysis) as well as non-parametric (e.g., Partition Around Medoids (PAM)) clustering patterns. Probability-based clustering algorithms require that data follow multivariate normality. Collaboration values were left-skewed and failed the normality test (Shapiro–Wilk test: $W=0.92,p<.001$).  Non-parametric methods, particularly PAM, was more appropriate for clustering as it does not assume normality and is less sensitive to outliers since each cluster is represented by an actual data point (the medoid) rather than a theoretical mean \cite{Schubert2021}.

Before running the clustering algorithm, we applied box-cox transformation on collaboration values to reduce the skewness. To determine the number of profiles (k), we relied on the underlying theoretical requirement of a $2×2$ framework (low/high rivalry $×$ low/high collaboration). We then compared the solutions starting from $k=2$ to $k=6$.  The mean silhouette value peaked at $k = 3$ (0.39), with $k = 4$ (0.37) as the next best. To match with our theoretical assumptions, we retained $k=4$, as the mean silhouette value was above the acceptable range \cite{kaufman2009finding, rousseeuw1987silhouettes}. The quality of clustering was assessed using standard internal validation methods. Dunn index was 0.04 (> 0.02 for behavioral data) and Davies-Bouldin value was 1.62 (<2 for behavioral data), indicating acceptable cohesion and separation across clusters \cite{Halkidi_2001, paraschou2025}. To ensure interpretability, we examined the centroids (medoids) of each cluster and confirmed that the four clusters corresponded closely to the theoretically anticipated profiles (LowR/LowC, LowR/HighC, HighR/LowC, HighR/HighC, see Appendix Table \ref{tab:profile-means}). Each participant was assigned to one of the four PAM profiles, yielding four different groups. It is recommended that each profile constitute at least 5-8\% of the total sample size (8\% of 403 = 33) \cite{Nylund2018}. All the sizes of our profiles were well above this threshold.

To examine differences in dependent variable ratings across the PAM profiles, we fit a series of multiple linear regression models (OLS) with profile membership (\textit{LowR/\allowbreak LowC, HighR/\allowbreak LowC, LowR/\allowbreak HighC, HighR/\allowbreak HighC}) as the categorical predictor. All models controlled for age, sex, and education. To identify which profiles significantly differed from one another, we estimated the profile-specific marginal means and conducted post-hoc pairwise comparisons using Tukey’s Honest Significant Difference (HSD) adjustment \cite{field2012discovering}. We report unstandardized regression coefficients in the results and the standardized values in the summary table. We also present estimated marginal means (EMMs) with 95\% confidence intervals and Tukey-adjusted group comparisons in the figures (see Figure \ref{fig:exploration-crafting} and Figure \ref{fig:explore-maintenance}). Statistical significance was evaluated at $p < .05$, and effect sizes are reported where applicable.

\paragraph{Qualitative Open-coding}
Lastly, guided by our research questions, we also conducted thematic analysis of the two open-ended questions that captured why, or why not, respondents engaged in skill maintenance since they started using GenAI. We started the initial coding on 100 responses to understand the breadth of the responses. Subsequently, we used an abductive approach \cite{Timmermans_2012} to iteratively develop codebooks. The resultant codebook consisted of 53 codes.  Through collaborative discussions, we addressed duplicates. Using theoretical insights from our analysis and prior literature \cite{lee2025impact, Varanasi2025, boucher2024, Golgeci2025, Takaffoli2024}, we developed a thematic organization of our codes to report. Any disagreements were negotiated and resolved at each stage using peer-debriefing and prolonged engagement techniques \cite{Creswell_Miller_2000}.

\subsection{Positionality}
In this study, our objective was to understand how professionals’ orientations toward GenAI shape their work practices and subsequent outcomes. This question is particularly salient given ongoing concerns surrounding the development of large language models, such as exploitative data work \cite{varanasi2023} and training on copyrighted content \cite{Veltman2025Anthropic}, as well as their consumption, including issues of fairness \cite{Varanasi2023b}, privacy \cite{Gupta2023}, and misinformation \cite{Shin_Koerber_Lim_2025}, among others. To develop a comprehensive understanding of the research problem and produce equitable findings, we focused on capturing bottom-up and diverse worker-led perceptions and practices, whith diversity of perspectives, including workers both in favor of and opposed to incorporating GenAI into their work. To maintain neutrality, we ensured that our communication remained unbiased, neither favoring nor opposing GAI, and that participants fully understood our research while proactively addressing any concerns they had. Our analysis and interpretation were informed by our interdisciplinary research backgrounds in Human-Computer Interaction, Organizational and Management Studies, and Computer-Supported Cooperative Work. Among the authors, two identify as men and one as a woman. Two authors hold senior faculty positions at a well-established university in the Global North, while the third is an early-career scholar at the same institution.

\section{Results}
Overall, our results show that while professionals’ perceived relationship with GenAI (collaboration and rivalry) each independently \up{correlated} with their reported job crafting practices, only rivalry was \up{associated with} skill maintenance. Furthermore, the joint combination of rivalry and collaboration was more strongly associated with job crafting than to skill maintenance, where collaboration showed limited associations. Analyses of 2x2 PAM profile combinations indicated that only one combination (HighR/HighC) was \up{consistently associated} with higher levels of both work practices. In contrast, only collaboration independently \up{correlated} with productivity and job satisfaction. The joint combination of rivalry and collaboration also demonstrated a \up{stronger} association with productivity than with satisfaction. Profile analysis further indicated that LowR/HighC was \up{consistently associated} with higher productivity and satisfaction. In the following sections, we present the findings addressing RQ1 and RQ2.

  \begin{table*}
  \caption{This table presents outcomes of linear regression along with the effect sizes for (H1-H4, H6). For each predictor, the figure presents standardized coefficients ($\beta$) and standard error within the parentheses, along with the significance levels for the outcomes skill maintenance and job satisfaction. Additionally, it also provides the linear contrast between rivalry and collaboration for each outcome, along with the model fit statistics. }
  \centering
  \small
  \begin{tabular}{lcccccc}
    \toprule
    \multirow{2}{*}{Variables} & \multicolumn{6}{c}{Crafting} \\
    \cmidrule(lr){2-7}
     & Cognitive & Task & Relational & Approach & Avoidance & Productivity \\
    \midrule
    Rivalry        & 0.227$^{***}$ & 0.223$^{***}$ & 0.303$^{***}$ & 0.273$^{***}$ & 0.210$^{***}$ & 0.0455\\
                   & (0.034)       & (0.038)       & (0.038)       & (0.038)       & (0.036)      &  (0.034) \\
    Collaboration  & 0.382$^{***}$ & 0.480$^{***}$ & 0.231$^{***}$ & 0.350$^{***}$ & 0.378$^{***}$ & 0.314 $^{***}$\\
                   & (0.026)       & (0.029)       & (0.029)       & (0.029)       & (0.028)       & (0.026)\\
    Age            & -0.074        & -0.194 $^{***}$     & -0.192$^{***}$& -0.171$^{***}$& -0.131$^{**}$ & -0.036\\
                   & (0.003)       & (0.003)& (0.004)     & (0.004)$^{***}$ & (0.003)     & (0.003)\\
    Sex            & 0.011         & -0.020        & -0.080        & -0.018        & -0.049        & -0.037\\
                   & (0.083)       & (0.090)       & (0.093)       & (0.093)       & (0.088)       & (0.081)\\
    Education (Mid)& 0.027         & 0.027         & 0.086         & 0.106         & -0.040        & 0.060\\
                   & (0.118)       & (0.128)       & (0.131)       & (0.131)       & (0.125)       & (0.114)\\
    Education (High)& 0.111        & 0.085         & 0.131$^{*}$   & 0.213$^{***}$ & -0.048        & 0.062\\
                   & (0.125)       & (0.136)       & (0.140)       & (0.140)       & (0.132)       & (0.121) \\
    \midrule
    \multicolumn{7}{l}{\emph{Linear Contrast}} \\
    Rivalry $-$ Collaboration $= 0$
                   & -0.044        & -0.121$^{**}$ & 0.112$^{*}$   & 0.012         & -0.057        & \\
                   & (0.0414)      & (0.045)       & (0.046)       & (0.046)       & (0.043)       & - \\
    \midrule
    \multicolumn{7}{l}{\emph{Model Fit Statistics}} \\
    $R^2$          & 0.211         & 0.328         & 0.200         & 0.262         & 0.199         & 0.112\\
    Adjusted $R^2$ & 0.200         & 0.318         & 0.188         & 0.251         & 0.186         & 0.099\\
    F-statistic (df)
                   & 17.69 (6, 396)& 32.24 (6, 396)& 16.59 (6, 396)& 23.53 (6, 396)& 16.35 (6, 396) & (6, 396)\\
    \bottomrule
  \end{tabular}
\vspace{2pt}
  \\ \footnotesize\emph{Notes.} Entries are OLS coefficients with standard errors in parentheses. 
  $^{*}p{<}.05$, $^{**}p{<}.01$, $^{***}p{<}.001$.
  \label{tab:crafting-regressions}
\end{table*}

\subsection{RQ-1: Independent Effects of GenAI Rivalry/Collaboration} \label{RQ-1}
To address RQ1, we tested our hypotheses using multiple linear regression (OLS), examining the associations between perceived GenAI rivalry and collaboration and the outcomes of job crafting (task, relational, cognitive, approach, and avoidance), skill maintenance (cognitive, writing, social, tacit), productivity, and job satisfaction. The results indicated that both collaboration and rivalry \up{independently correlated with} all five forms of job crafting behaviors among writing professionals. However, only rivalry was \up{independently associated} with all forms of skill maintenance, whereas collaboration was \up{associated only} with cognitive skill maintenance. The pattern reversed for productivity and job satisfaction, which \up{were associated} solely with collaboration. To address RQ1, we tested our hypotheses (H1–H6) using a linear regression model (see Table \ref{tab:crafting-regressions} for an overview). Age, sex, and education were included as covariates to account for demographic variation. Although including role and industry increased model complexity, these variables did not meaningfully alter the results, so they were excluded from the final model.

\begin{table*}
  \caption{This table presents outcomes of linear regression along with the effect sizes for (H5,H7). For each predictor, the figure presents standardized coefficients ($\beta$) and standard error within the parentheses, along with the significance levels for the outcomes skill maintenance and job satisfaction. Additionally, it also provides the linear contrast between rivalry and collaboration for each outcome, along with the model fit statistics. Only rivalry predicts all types of skill maintenance, while collaboration predicts only cognitive skill maintenance and satisfaction.}
  \label{tab:skill-maintenance}
  \centering
  {\small
  \begin{tabular}{lcccccc}
    \toprule
     & \multicolumn{5}{c}{Skill Maintenance} & \\
    \cmidrule(lr){2-6}
    Variables & Overall & Cognitive & Writing & Social & Tacit & Satisfaction \\
    \midrule
    Rivalry
      & 0.234$^{***}$ & 0.141$^{**}$ & 0.205$^{***}$ & 0.235$^{***}$ & 0.229$^{***}$ & 0.014 \\
      & (0.048)       & (0.056)      & (0.056)       & (0.053)       & (0.056)       & (0.039) \\
    Collaboration
      & 0.077         & 0.133$^{**}$ & 0.000         & 0.068         & 0.079         & 0.194$^{***}$ \\
      & (0.036)       & (0.010)      & (0.042)       & (0.040)       & (0.040)       & (0.030) \\
    Age
      & 0.014        & 0.072        & 0.005         & -0.078        & 0.046         & 0.093 \\
      & (0.004)       & (0.005)      & (0.004)       & (0.005)       & (0.005)       & (0.004) \\
    Sex
      & -0.041        & 0.008        & -0.040        & -0.091        & -0.021        & 0.050 \\
      & (0.115)       & (0.134)      & (0.136)       & (0.128)       & (0.136)       & (0.095) \\
    Education (Mid)
      & 0.003         & 0.040        & -0.036        & 0.055         & -0.036        & -0.022 \\
      & (0.163)       & (0.199)      & (0.192)       & (0.180)       & (0.192)       & (0.135) \\
    Education (High)
      & 0.047         & 0.022        & 0.020         & 0.149$^{*}$   & -0.021        & 0.010 \\
      & (0.173)       & (0.201)      & (0.204)       & (0.191)       & (0.203)       & (0.143) \\
    \midrule
    \multicolumn{7}{l}{\emph{Linear Contrast}} \\
    Rivalry $-$ Collaboration $=0$
      & --            & 0.046        & --            & --            & --            & -- \\
      &               & (0.067)      &               &               &               & \\
    \midrule
    \multicolumn{7}{l}{\emph{Model Fit Statistics}} \\
    $R^2$
      & 0.065         & 0.039        & 0.048         & 0.090         & 0.060         & 0.046 \\
    Adjusted $R^2$
      & 0.050         & 0.020        & 0.033         & 0.076         & 0.045        & 0.031 \\
    F-statistic (df)
      & 4.569 (6, 396)& 2.70 (6, 396)& 3.34 (6, 396) & 6.54 (6, 396) & 4.162 (6, 396)& 3.175 (6, 396) \\
    \bottomrule
  \end{tabular}}
  \vspace{2pt}
  \\ {\footnotesize\emph{Notes.} Entries are OLS coefficients with standard errors in parentheses.
  $^{*}p{<}.05$, $^{**}p{<}.01$, $^{***}p{<}.001$.}
\end{table*}

H1 was partially supported. Both perceived rivalry ($b = 0.175, p < .001$) and perceived collaboration ($b = 0.219, p < .001$) with GenAI were significantly \up{associated} with cognitive crafting, indicating that higher levels of rivalry and collaboration each corresponded to higher levels of cognitive framing of roles and work to find meaning and purpose (Table \ref{tab:crafting-regressions}). To assess whether collaboration showed a significantly \up{stronger association} than rivalry, we conducted a pairwise linear contrast between the slopes for rivalry and collaboration using the Multcomp package in R. The difference was not statistically significant ($b = -0.044, p = .289$).

H2 was fully supported. First, both rivalry ($b = 0.202, p < .001$) and collaboration ($b = 0.323, p < .001$) with GenAI were significantly \up{associated with task crafting}, indicating that professionals reporting higher rivalry or collaboration also reported higher levels of task crafting behaviors and perceived reshaping of their everyday activities to improve their work, compared to those low in rivalry or collaboration. Second, the linear contrast revealed that collaboration showed a significantly \up{stronger association} with task crafting than rivalry ($b = 0.121, p < .01$).

H3 was also fully supported. Rivalry ($b = 0.258, p < .001$) and collaboration ($b = 0.146, p < .001$) with GenAI were \up{significantly associated with relational crafting}, indicating that higher rivalry and higher collaboration were each positively correlated with writers’ reports of modifying their work relationships. Second, linear contrast showed that rivalry was significantly more strongly associated with relational crafting than collaboration ($b = 0.112, p < .05$)

Lastly, H4 was partially supported. Rivalry was positively associated with approach ($b = 0.243, p < .001$) and avoidance ($b = 0.170, p < .001$) crafting. Similarly, collaboration was also positively associated with approach ($b = 0.231, p < .001$) and avoidance ($b = 0.228, p < .001$) crafting, suggesting that both perceptions were associated with professionals reporting efforts to expand their roles by taking on new tasks and to reduce aspects of their work that they found undesirable. Linear contrasts for approach ($b = 0.012, p = .803$) and avoidance ($b = -0.057, p = .193$) between rivalry and collaboration were not significant.

H5 was partially supported. Rivalry was significantly associated with overall skill maintenance ($b = 0.228, p < .001$), particularly cognitive skills ($b = 0.157, p < .01$), writing skills ($b = 0.235, p < .001$), social skills ($b = 0.257, p < .001$), and tacit skills ($b = 0.262, p < .001$) (see Table \ref{tab:skill-maintenance}). However, collaboration was only significantly correlated with cognitive skill maintenance ($b = 0.111, p < .01$). A linear contrast between rivalry and collaboration with respect to cognitive skills was not significant ($b = 0.046, p = 0.487$).

H6 was supported as well (see Table \ref{tab:crafting-regressions}).  GenAI collaboration was significantly associated with higher levels of productivity in their jobs ($b = 0.164, p < 0.001$), while GenAI rivalry showed no significant association with productivity ($b = 0.032, p = 0.34$). \up{We observed similar strong support for H7 (see Table \ref{tab:skill-maintenance}). GenAI collaboration was also significantly related to higher job satisfaction ($b = 0.115, p < 0.001$)}, while GenAI rivalry showed no significant association with satisfaction ($b = 0.011, p = 0.776$).

\subsection{RQ-2: Understanding the Joint Profiles of Rivalry/Collaboration} \label{demographic-analysis}
To answer RQ-2, we first tested joint effects of perceived GenAI rivalry/collaboration on job crafting, productivity, skill maintenance, and satisfaction. The analysis was followed by multiple linear regression model (OLS) with profile membership  (LowR/LowC, HighR/LowC, LowR/HighC, HighR/HighC) as the categorical predictor. The joint model of collaboration and rivalry accounted for significantly more variance in job crafting and modestly more variance in productivity, but not in skill maintenance or satisfaction, relative to the independent model in RQ1. Analysis of the $2x2$ PAM profiles of rivalry and collaboration indicated that the highest levels of crafting and skill maintenance in our sample were observed when both rivalry and collaboration were simultaneously high (HighC/HighR). However, for productivity and satisfaction, high collaboration (LowR/HighC) alone corresponded to the highest observed levels. In this section, we start by introducing each PAM profile and its demographic characteristics. In the subsequent sections, for each outcome, we present response surface analysis (RSA) to illustrate the presence of joint associations, followed by PAM profile analysis. For the skill maintenance outcome, we also present our qualitative analysis of the open ended responses.

\subsubsection{Employment}
Recall from Section \ref{pam} that our PAM clustering resulted in four relationship profiles of professionals with GenAI: (1) LowR/LowC (n=65), LowR/HighC (n=108), HighR/LowC (n=84),  HighR/HighC (n=146). Table \ref{tab:profile-means} presents the descriptive details of each profile. 

In all profiles, respondents employed in organizations as full-time employees were clearly the majority. One profile, HighR/LowC, had the highest proportion of self-employed people (see Appendix, Figure  \ref{fig:profile-worknature}). We also captured if respondents were part of a union. Interestingly, union members were predominant in profiles with higher perceptions of collaboration towards GenAI (i.e., HighR/HighC and LowR/HighC). This was surprising in light of prior literature that associates union membership with resistance to AI technologies~\cite{Cini_2023}.

\begin{figure*}
    \centering
    \includegraphics[width=1\linewidth]{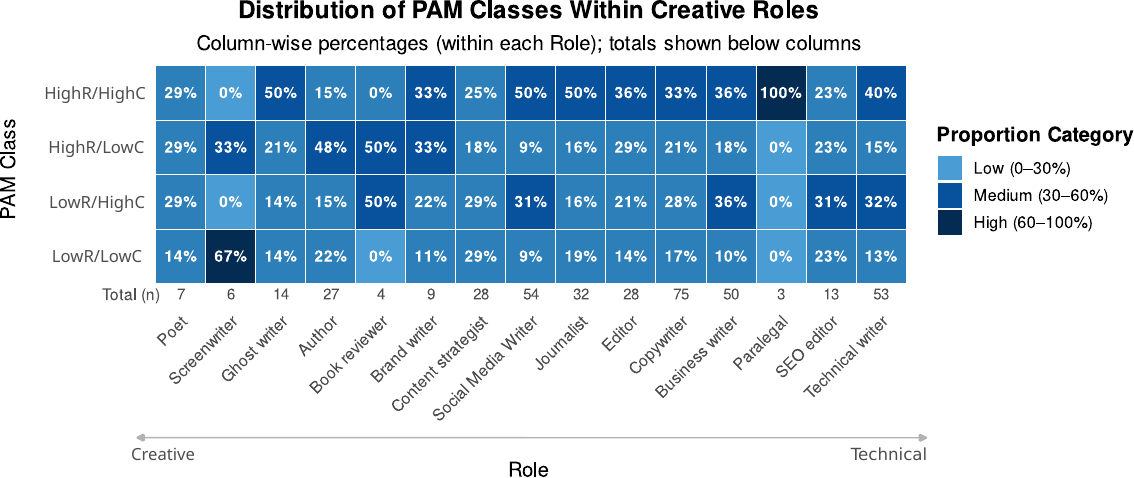}
    \caption{The figure presents the distribution of the four profiles across professionals’ roles. The y-axis shows the four profiles, while the x-axis places roles along a spectrum ranging from those with a greater emphasis on creative aspects (left) to those with a greater emphasis on technical aspects (right). Proportions of the profiles in each role are reported at three levels: low, medium, and high.}
    \label{fig:profile-role}
\end{figure*}

\begin{figure*}
    \centering
    \includegraphics[width=1\linewidth]{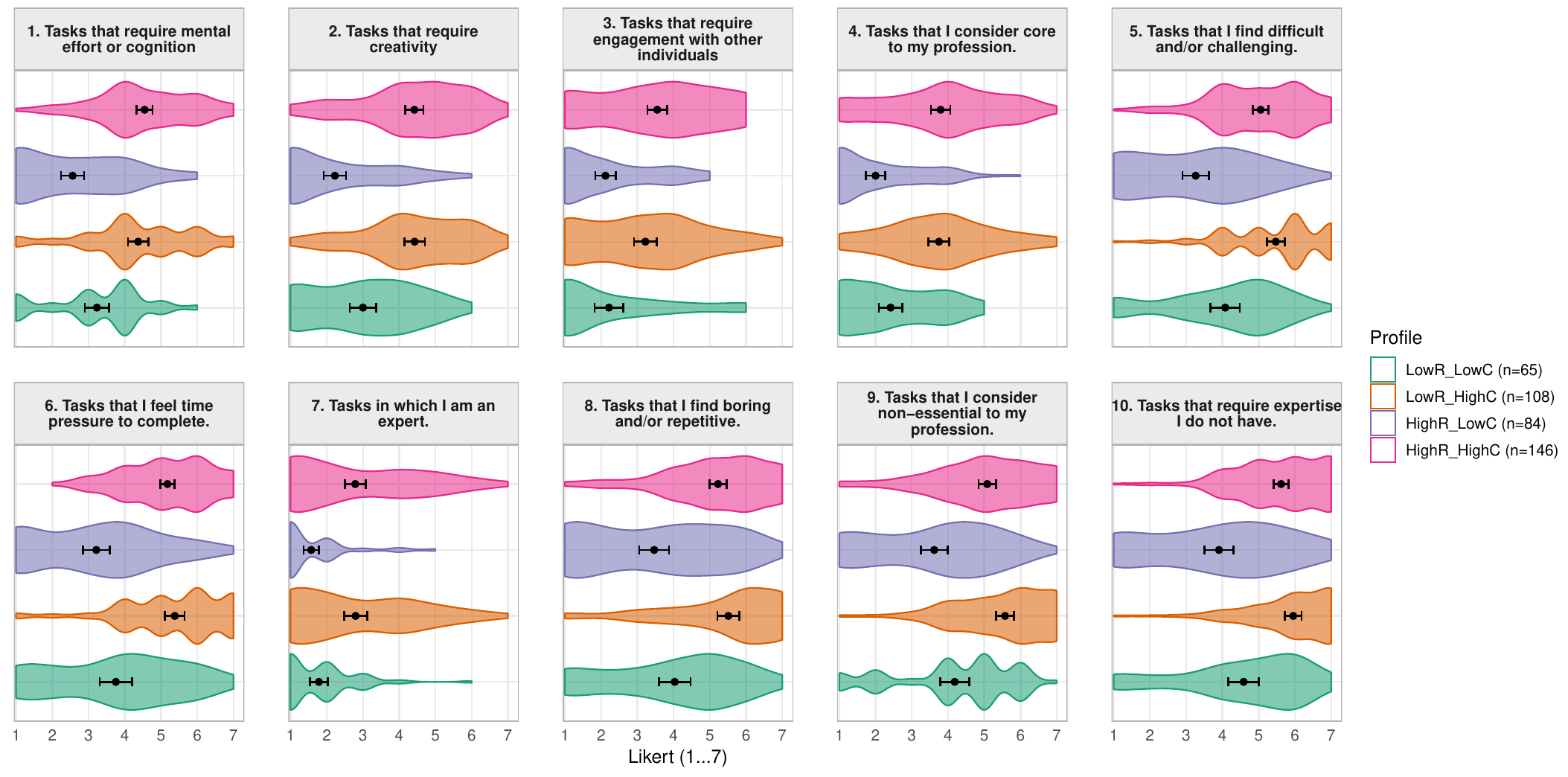}
    \caption{This figure presents the distributions of professionals' perceived use of GenAI in their work, categorized by the four profiles.  Each panel presents the mean, confidence intervals, along with the response distributions for ten different question types where they use GenAI: 1) cognition; 2) creativity; 3) Relationships; 4) Core tasks; 5) Difficult tasks; 6) Time-pressured tasks; 7) Expert tasks; 8) Boring Tasks; 9) Non-essential tasks; 10) Tasks outside respondents' expertise. Participants responded on a scale of 1: Do it entirely by themselves to 7: Do it entirely with GenAI}
    \label{fig:profile-genai-engage}
\end{figure*}

We further analyzed the distribution of the profiles across participants. Results should be interpreted with caution due to uneven distribution of participants across roles, but some notable patterns emerged (see Figure \ref{fig:profile-role}). Following prior work, roles were arranged along a spectrum from predominantly creative work \cite{florida2019rise} on the left to predominantly technical work \cite{Kynell_1996} on the right, with hybrid roles that combine both positioned in the middle \cite{cornelissen2020corporate, onet}. Participants in roles that rely relatively more on creativity (left end of the spectrum) were more concentrated in profiles characterized by high rivalry (HighR/LowC). Examples include poets and authors. Interestingly, some creative roles also showed substantial representation in collaborative profiles (HighR/HighC or LowR/HighC), such as ghostwriters and book reviewers. Conversely, participants in roles with stronger technical demands (right end of the spectrum), who were a larger proportion of our total sample, were predominantly associated with collaborative profiles (LowR/HighC or HighR/HighC). Example roles include technical writers, business writers, and SEO editors. Finally, across the spectrum we observed a relatively stable minority of participants (average 17.5\%) in the LowR/LowC profile, which indicates relatively little relational engagement with GenAI. This profile may be less common because the writing profession is among those at the forefront of disruption by GenAI.

\subsubsection{GenAI Engagement in Work} \label{profile-engagement}
Regarding GenAI use policies in the workplace, most participants reported no explicit workplace policy on GenAI use across all profiles (see Figure \ref{fig:profile-worknature}), with this being most pronounced in LowR/LowC (71\%) and HighR/LowC (61\%). Collaboration-oriented profiles (LowR/HighC and HighR/HighC) were more likely to report that their workplace policies encouraged use of GenAI, with HighR/HighC showing the highest proportion (59\%). In contrast, HighR/LowC exhibited the highest proportion of workplaces actively discouraging GenAI (23\%).  

Perceived capabilities of GenAI and their use in work (see Figure \ref{fig:profile-genai-use}) was associated with the profiles participants sorted into, with workers in collaboration-oriented profiles (HighC/LowR and HighC/HighR) reporting significantly more positive perceptions of GenAI capabilities, compared to the other profiles. In contrast, high rivalry combined with low collaboration (HighR/LowC) corresponded to the lowest capability perceptions and usage, while the LowR/LowC profile consistently fell in the middle range. Interestingly, prior experience with GenAI revealed a different pattern: both high rivalry and high collaboration profiles (HighR/LowC, LowR/HighC, and HighR/HighC) reported more experience with GenAI than the baseline (LowR/LowC), suggesting that even people engaged in rivalry exposed themselves to GenAI, even if they did not use it work contexts.

Lastly, we examined GenAI use in granular work tasks (see Figure \ref{fig:profile-genai-engage}). Across the profiles, professionals used GenAI most for  (1) tasks in which they were not experts, (2) tasks considered non-essential to their role, (3) tasks they found difficult, and (4) tasks that were repetitive. GenAI was used the least for the tasks they considered themselves experts in.  LowR/HighC reported the greatest reliance on GenAI in most of the tasks, closely followed by HighR/HighC.

\subsection{RQ-2: Joint Impacts of Rivalry/Collaboration on Job Crafting} \label{RQ-2-job-crafting}

\begin{figure*}
    \centering
    \includegraphics[width=1\linewidth]{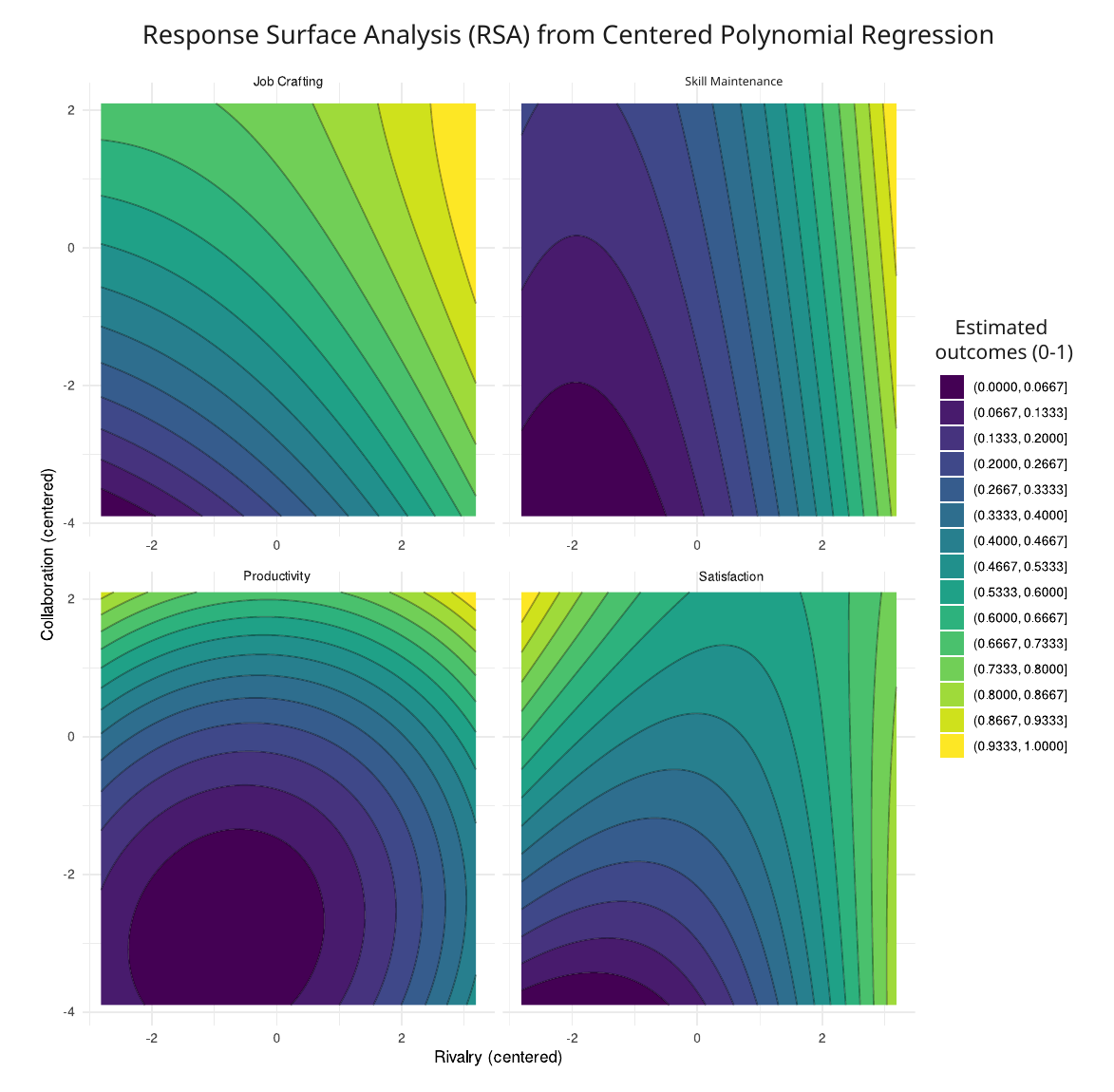}
    \caption{The figure presents visualization of the results from response surface analysis (RSA) for key aggregate measures, namely job crafting, skill maintenance, productivity, and satisfaction. The shades of blue indicate estimate ranges where the joint effects of rivalry and collaboration are the weakest, where as bright yellow indicates the strongest effects. Highest levels of collaboration and rivalry were required for job crafting, skill use, and productivity. Interestingly satisfaction and productivity also predicted by high levels of collaboration alone.}
    \label{fig:RSA}
\end{figure*}

For the overall measure of job crafting behaviors, the RSA model examining the \up{joint association} of rivalry and collaboration accounted for significant variance ($F(10,392)=21.96, p<.001$, see Appendix, Table \ref{tab:rsa_all}). Furthermore, the joint model showed improved fit relative to the independent effects model ($\Delta R^{2} = .02, F(3,392) = 4.12, p < .01$). The contour plot (Figure \ref{fig:RSA}) illustrates these joint associations, with the highest levels of job crafting observed when both rivalry and collaboration were high. Surface parameters indicated a positive slope along the line of congruence ($a_{1} = 0.42$), indicating that higher values of rivalry and collaboration corresponded to higher levels of job crafting. However, this pattern flattened at very high values ($a_{2} = -.04$), which was also reflected in the negative interaction term ($b = -0.042, p < .01$).

\begin{figure*}
    \centering
    \includegraphics[width=1\linewidth]{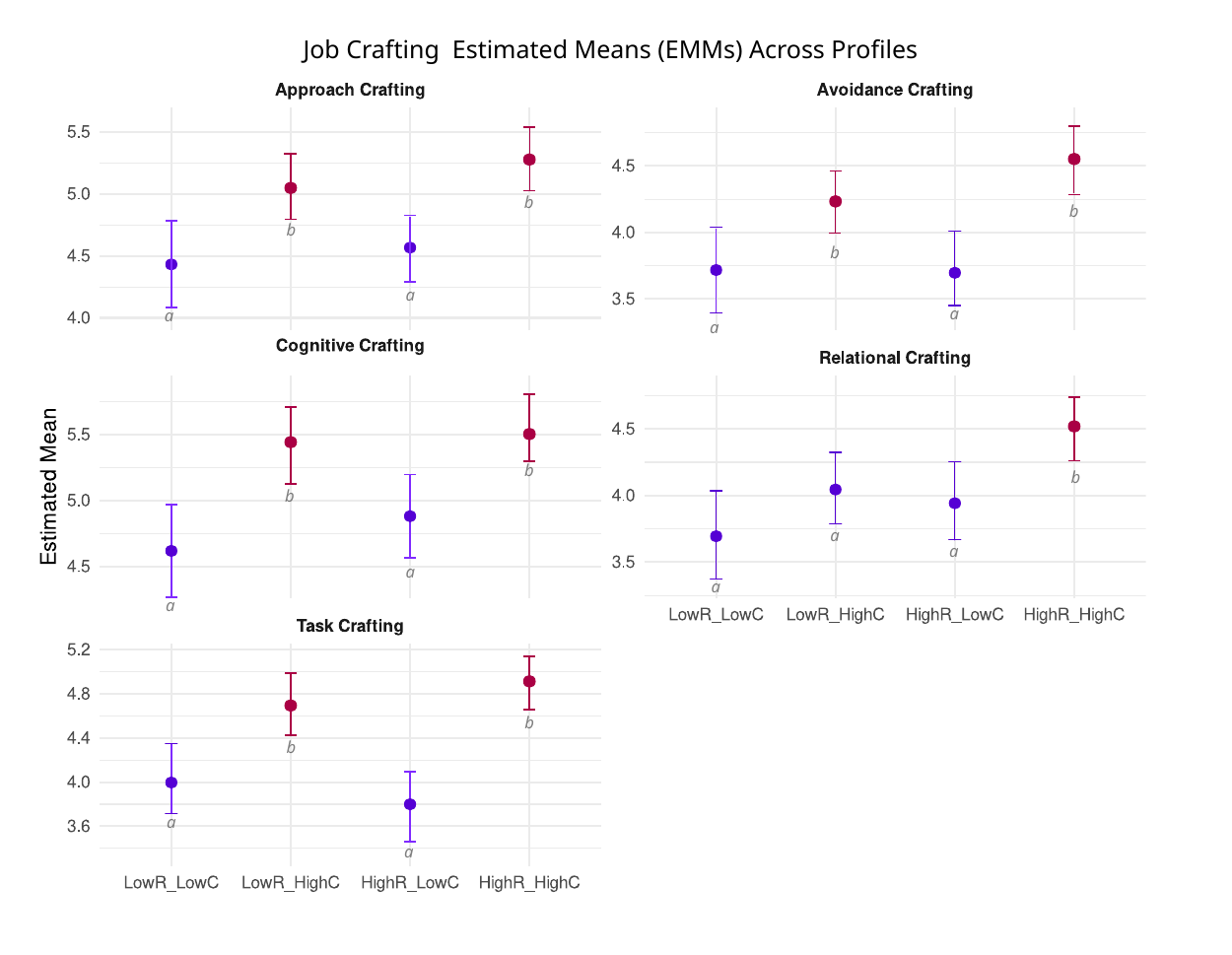}
    \caption{The figure presents job crafting estimated means (EMMs) across profiles. Additionally Tukey's HSD test is used to show groups that are significantly different from each other. These are represented by the color and the letter. For all types of crafting except relational, profiles characterized by higher collaboration were associated with higher levels of crafting behavior. For relational crafting, only the HighR/HighC profile was associated with significantly higher crafting behavior.}
    \label{fig:exploration-crafting}
\end{figure*}

\begin{figure*}
    \centering
    \includegraphics[width=1\linewidth]{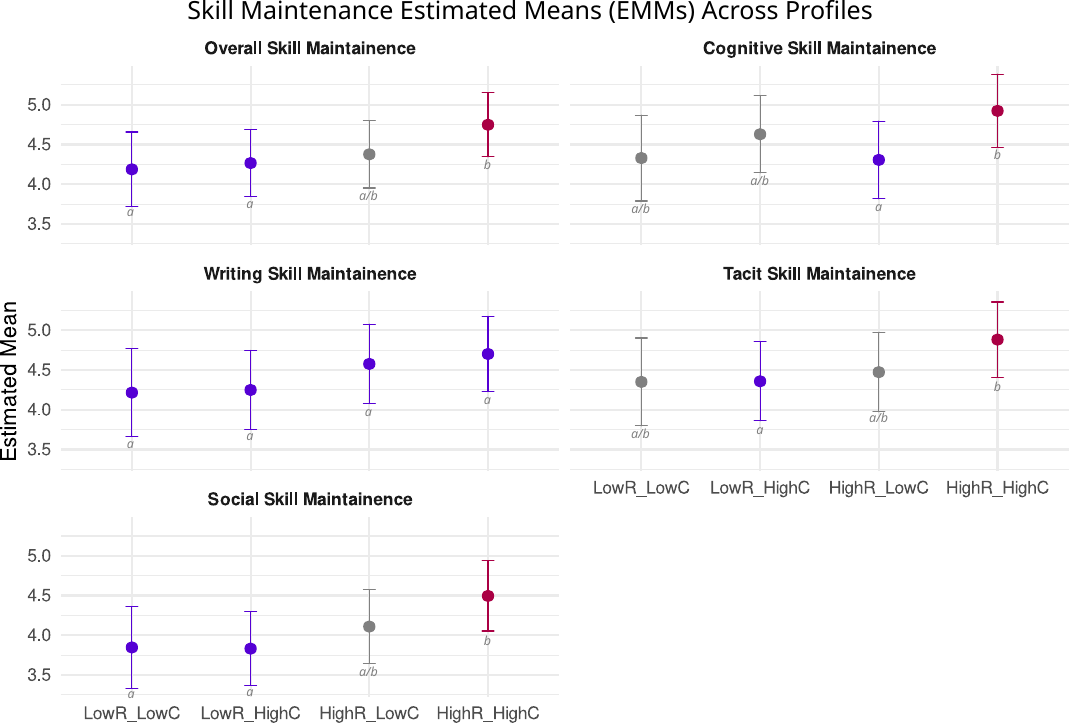}
    \caption{The figure presents skill maintenance estimated means (EMMs) across profiles. Additionally Tukey's HSD test is used to show groups that are significantly different from each other. These are represented by the color and the letter. For all types of skill maintenance except writing skill, only HighR/HighC correlated with substantial increase in skill maintenance.}
    \label{fig:explore-maintenance}
\end{figure*}

% % 
% Professional's rivalry levels were weighted towards center (median 3.8, skew - 0.05) as compared to collaboration, which were left skewed (media = 5.99, skew -.075), indicating stronger collaboration intent than rivalry. 

As the joint model accounted for significant variance in job crafting, we analyzed PAM profiles to examine how different combinations of collaboration and rivalry orientations were \up{associated with professionals’ specific job crafting behaviors}, focusing on cognitive, task, and relational crafting. Using LowR/LowC as the baseline and controlling for age, sex, and education, we conducted linear regression analyses. For all three forms of crafting, LowR/HighC (cognitive: $b = 0.741, p < .001$; task: $b = 0.648, p < .001$; relational: $b = 0.329, p < .05$) and HighR/HighC (cognitive: $b = 0.844, p < .001$; task: $b = 0.870, p < .001$; relational: $b = 0.779, p < .001$) were \up{associated with significantly} higher levels of job crafting relative to LowR/LowC (see Table \ref{tab:pam_crafting_prod}). By contrast, HighR/LowC did not differ significantly from the baseline (cognitive: $b = 0.192, p = 0.201$; task: $b = -0.210, p = 0.216$; relational: $b = 0.238, p = 0.154$).

To further examine profile differences, we conducted pairwise comparisons using Tukey’s HSD test (see Figure \ref{fig:exploration-crafting}). For cognitive crafting, profiles high in collaboration (LowR/HighC: $M = 5.46, SE = 0.16$; HighR/HighC: $M = 5.56, SE = 0.15$) scored significantly higher ($p < 0.01$) than profiles low in collaboration (LowR/LowC: $M = 4.72, SE = 0.18$ and HighR/LowC: $M = 4.91, SE = 0.16$), irrespective of rivalry. Similar results were also observed for task crafting (LowR/HighC: $M = 4.71, SE = 0.1 $; HighR/HighC: $M = 4.93, SE = 0.09$; $p < 0.01$). These results suggest that collaboration shows the strongest association with cognitive and task crafting. Professionals who perceived GenAI as a collaborator (regardless of rivalry) were more likely to cognitively reframe and rethink their work as well shape their tasks. For relational crafting, however, only the HighR/HighC profile (relational: $M = 4.51, SE = 0.08$) reported significantly higher levels of job crafting ($p < .05$) compared to the other three profiles, which did not differ from one another (see Table \ref{tab:pam_crafting_prod}). \up{Thus, the likelihood of professionals proactively redesigning their relationships was highest when both rivalry and collaboration were high}.

Approach and avoidance crafting followed a pattern similar to cognitive crafting. Relative to LowR/LowC, only LowR/HighC (approach: $b = 0.628, p < .001$; avoidance: $b = 0.499, p < .001$) and HighR/HighC (approach: $b = 0.847, p < .001$; avoidance: $b = 0.810, p < .001$) were associated with significantly higher levels of approach and avoidance crafting (see Table \ref{tab:pam_crafting_prod}). Tukey’s HSD further showed that profiles high in collaboration reported significantly higher levels of approach (LowR/HighC: $M = 5.030, SE = 0.180$; HighR/HighC: $M = 5.250, SE = 0.170$) and avoidance (LowR/HighC: $M = 4.220, SE = 0.170$; HighR/HighC: $M = 4.530, SE = 0.160$) crafting than low collaboration profiles. \up{These findings indicate that higher collaboration was consistently associated with higher levels of both approach and avoidance crafting.}

\subsection{RQ-2: Joint Impacts of Rivalry/Collaboration on Skill Maintenance} \label{Skill Maintenance}
The RSA model explained significant variance in skill maintenance, indicating that the combination of rivalry and collaboration was \up{associated with this outcome} ($F(10,392)=3.08, p<.001$, Table \ref{tab:rsa_all}). However, the RSA model did not account for additional variance beyond the independent effects model ($\Delta R^{2} = .01, F(3,392) = 1.07, p = .30$). Visual inspection of the contour map suggested that rivalry showed \up{a stronger association with skill maintenance, whereas collaboration showed comparatively limited associations} (Figure \ref{fig:RSA}). This pattern was reflected in the surface parameters, which indicated a modest positive slope along the line of congruence ($a_{1} = .27$) and along the line of incongruence ($a_{3} = .18$), suggesting that higher levels of skill use \up{corresponded more strongly to higher rivalry, with a comparatively weaker correspondence to collaboration}. Neither curvature nor interaction terms were significant, and the stationary point fell outside the observed data range, indicating limited evidence for joint associations at the aggregate level. Because potential joint patterns may be obscured in the aggregate measure if sub-dimensions (e.g., social, cognitive, or writing skills) follow distinct patterns that offset one another when averaged, we relied on PAM profiles to examine patterns across subcategories of skill maintenance.

Using the LowR/LowC group as the reference, only the HighR/\allowbreak HighC profile was associated with significantly higher overall skill maintenance ($b = 0.492, p < .01$). A similar pattern emerged for specific skills. HighR/HighC was significantly associated with higher cognitive ($b = 0.521, p < .05$), tacit ($b = 0.474, p < .05$) and social ($b = 0.590, p < .01$) skill maintenance. Tukey’s HSD test (see Figure \ref{fig:explore-maintenance}) confirmed this pattern, showing that only the HighR/HighC profile scored significantly higher than the other profiles for overall skill maintenance ($M = 4.74, SE = 0.20$), cognitive skill maintenance ($M = 4.92, SE = 0.23$), tacit skill maintenance ($M = 4.96, SE = 0.12$), and social skill maintenance ($M = 4.61, SE = 0.11$). Taken together, these results indicate that higher levels of both rivalry and collaboration were associated with higher reported levels of skill maintenance.

For writing skill maintenance, however, Tukey’s HSD confirmed that none of the profiles significantly differed, and writing skills were maintained at moderate-to-high levels across all groups, with any differences between profiles being too small to be statistically reliable. Notably, for overall, social, and tacit skill outcomes, the HighR/LowC profile occupied an intermediate position, trending higher than the LowR profiles but not differing significantly from either the LowR groups or the HighR/HighC group.

\subsubsection{Reduction in skill maintenance}
Our qualitative analysis of the open-ended responses illuminate the patterns that emerged in the quantitative findings by revealing several reasons why practitioners reduced their skill maintenance efforts. Professionals pointed to organizational shifts that diminished the perceived value of maintaining particular skills. With GenAI becoming prevalent in the workplace, they experienced increasing pressure to integrate GenAI-driven workflows. Writers noted that ``fast'' and ``casual'' content was increasingly prioritized, while ``thoughtful'', ``researched'', and ``reliable'' content was devalued. Consequently, some cognitive skills, such as idea generation, and writing skills, such as developing character tone and voice, were viewed as less worthwhile to maintain. P315, a copywriter in the healthcare industry, shared:

\begin{quote}
``\textit{Currently, I'm doing some contract work for an agency. It's very formulaic and Gen AI is required to produce material. That being the case, some of the pieces require a lot less effort because Gen AI takes over some of those central concepts. There are still opportunities for creative input, but there's a lot less room than it was five years ago. Essentially, Gen AI takes some of the ``spark’’ out of the process and, in turn, I don't invest the same type of energy that I used to.}’’
\end{quote}

Such changes in organizational directives were accompanied by increased expectations of productivity and workload, as employers and clients assumed that GenAI tools enhanced efficiency. Professionals responded to these expectations by re-categorizing the skills in which they were willing to invest time and effort to maintain. They delegated those they perceived as low-value to GenAI for the ``heavy lifting,'' while continuing to exercise and maintain those they felt were critical for their work. Skills such as proofreading, correcting writing, and summarizing were among those practitioners reduced their maintenance of and delegated to GenAI. For example, P168, a technical writer in the space research industry, shared how he let go of maintaining skills that he perceived as providing low-value work:

\begin{quote}
``\textit{Tools now handle proofreading, fact-checking, and even initial content structuring, allowing me to focus on higher-value work. For example, instead of memorizing style guides, I rely on AI for consistency checks. This shift doesn’t mean these skills are obsolete, but AI reduces the need for constant manual refinement.}’’
\end{quote}

Professionals also considered their existing skill levels when deciding which skills to maintain after the introduction of GenAI. Skills they felt ``\textit{proficient enough to not spend more time developing}'' were increasingly delegated to GenAI, reducing the effort they invested in maintaining them.

A more concerning pattern emerged among professionals who chose to stop maintaining partially developed skills they considered ``\textit{under-developed}'' or ``\textit{struggling}''. They described skill maintenance as ``time-consuming'' and noted that ``\textit{less effort is needed for those [underdeveloped] skills when [GenAI] can supplement the gaps.}''

Beyond reduced skill maintenance, some responses reflected a complete halt in upskilling efforts. Several practitioners emphasized that they lacked the time to acquire new skills and instead delegated such tasks to GenAI. For instance, P376, a business writer in digital marketing field explained: \textit{``I have spent less time learning new skills as I could easily rely on AI to perform it for me.''} Examples included social skills such as promotion and marketing, writing skills such as developing character and tone, and cognitive skills such as recollecting specific ideas or practices. In other cases, professionals perceived upskilling as futile, abandoning the effort entirely because they believed GenAI was ``significantly better at some skills'' and therefore more valuable to ``exploit.'' As P302, an SEO editor shared: \textit{``Some skills ChatGPT is good at, and I will outsource them to it so that I don't have to figure out how to do it myself.''}

\subsubsection{Increase in skill maintenance} \label{qual-increase}
Professionals who perceived GenAI as a collaborator emphasized maintaining complementary, often ``high-level and creative'' skills that they believed enhanced their synergy with the technology. Commonly cited skills were persuasive writing, storytelling, and narrative building. Investing effort into developing these skills positioned GenAI as either an ``aid'' or a ``crutch.'' For instance, P368, a copywriter in tech industry, explained:

\begin{quote}
    ``\textit{I spend more effort on refining high-level skills that AI can’t fully replicate like brand voice development, emotional storytelling, creative conception, and content strategy. I also focus on learning how to write better AI prompts, edit AI-generated copy for tone and nuance, and ensure that the final product aligns with brand goals. These human-led aspects of copywriting are now even more valuable in an AI-assisted workflow.}''
\end{quote}

These professionals also devoted significant effort to new form of skills that were not listed in the survey. They referred to as ``humanizing AI’’ skills. Examples included learning how to ``hide the evidence of [Gen]AI [writing]’’, ``keeping an eye out on [Gen]AI’’, crafting prompts that yielded outputs more relevant to their workflows, and reducing GenAI errors to guide the system more effectively. While we did not explicitly measure skill maintenance within these AI management categories, participants frequently cited them in their responses.  To advance these abilities, several professionals also invested effort in understanding the inner workings of GenAI and improving their technical proficiency. As one participant P168, an editor working in advertising and public relations, explained his humanizing efforts:

\begin{quote}
``\textit{I'm learning how to incorporate AI into our workflow. The AI program itself is pretty easy to learn, and we had a little training on how to use it with our writers. They write using AI as help, then I edit and put a personal/human flair on things. So in that sense, I'm almost putting in more effort than before to humanize the copy.}’’
\end{quote}

Interestingly, professionals who scored high on rivalry also emphasized maintaining the same high-level skills, but primarily as a way to differentiate themselves and compete with GenAI or with peers who relied on it. Common examples included cognitive skills such as critical thinking, or writing skills such as storytelling and content structuring. Social skills, such as collaborating with colleagues, negotiating, and persuading clients, were also highlighted as areas in which professionals invested effort, as they believed these gave them a critical edge over those who focused solely on GenAI-related skills. By prioritizing these abilities, professionals felt they could ``\textit{be more valuable in terms of what only humans can do at this time}''. As P135, a business writer working in consultancy services, shared:

 \begin{quote}
    ``\textit{I chose to spend less effort on developing certain skills because I wanted to focus my energy on areas that would have the greatest impact on my growth as an independent writer. My goal is to build a strong foundation in writing without relying on AI assistance. By prioritizing core skills such as critical thinking, creativity, and original expression, I can develop my own voice and style. I believe that by concentrating on these essential areas, I am better preparing myself to write authentically and effectively on my own.}''
 \end{quote}

%Interestingly, we also observed multiple instances where participants used GenAI as a guide for skill development. Those with higher perceptions of rivalry leveraged GenAI to expose their own ``weaknesses'' and ``strengths'' in their current skills to become more aware of their blindspot. They used this knowledge for skill maintenance. 

% In contrast, professionals with high collaboration perceptions used GenAI as an assistant to uncover skill gaps ``DIY'', including both the ones they needed and those they discovered serendipitously. As \fixme{PXX} shared:

% \begin{quote}
%     \textit{``\dots  Especially related to analytical reasoning, etc., that I have been more exposed to through interaction with AI on my own time, I have been putting in more effort to develop the skills I understand are being trained into Generative AI models.''}
% \end{quote}

\subsection{RQ-2: Joint Impacts of Rivalry/Collaboration on Productivity and Satisfaction} \label{RQ2-sat-prod}
The RSA model explained significant variation in productivity ($F(10,392)=6.75, p<.001$). Additionally, it improved model fit relative to the independent effects model ($\Delta R^{2} = .03, F(3,392) = 5.21, p = .002$, Table \ref{tab:rsa_all}). The contour map showed that the highest observed productivity levels were \up{associated with a combination of higher rivalry and higher collaboration} (see Figure \ref{fig:RSA}). However, relatively high levels of productivity were also observed for profiles characterized by high collaboration alone. This pattern was reflected in the positive quadratic term for collaboration ($b = .04, p < .01$), indicating that productivity levels were higher at elevated levels of collaboration. Neither the interaction term nor the quadratic term for rivalry reached significance. Surface parameters indicated a positive slope along the line of congruence ($a_{1} = .24$), suggesting that higher rivalry and collaboration \up{corresponded to higher productivity levels}, whereas the negative slope along the line of incongruence ($a_{3} = -.18$) pointed to a \up{stronger association with collaboration} than with rivalry.

\begin{figure*}
    \centering
    \includegraphics[width=1\linewidth]{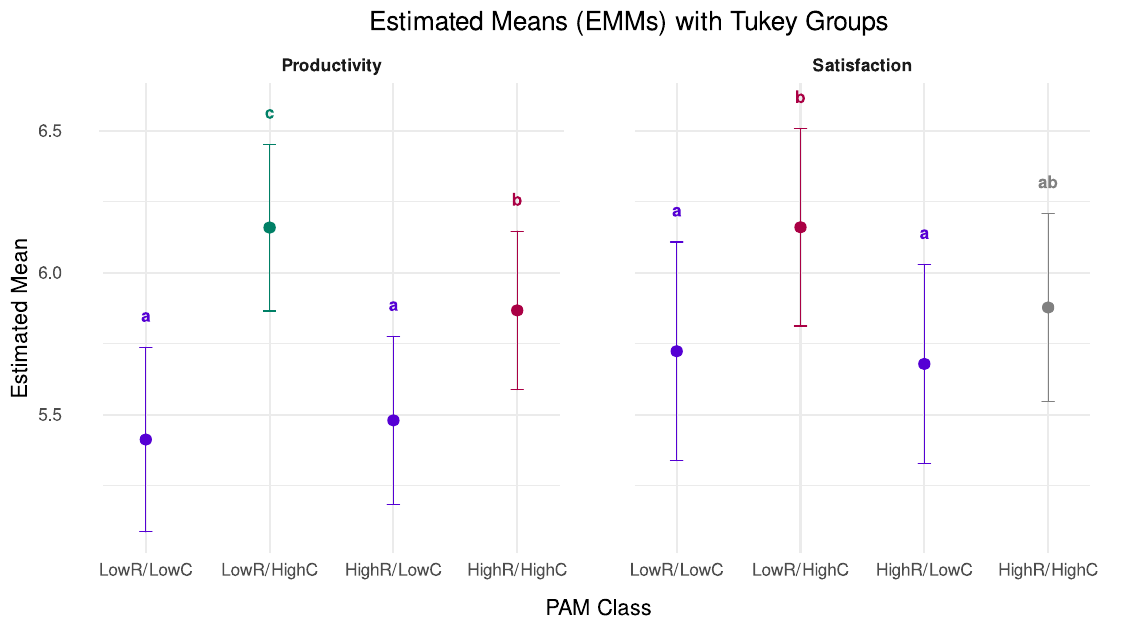}
    \caption{The figure presents productivity and skill maintenance estimated means (EMMs) across profiles. Additionally Tukey's HSD test is used to show groups that are significantly different from each other. These are represented by the color and the letter. For both productivity LowR/HighC profile projected greatest EMM scores.}
    \label{fig:tukey-prod-sat}
\end{figure*}

PAM analysis with the LowR/LowC group as the reference revealed that both the LowR/\allowbreak HighC ($b = 0.747, p < .001$) and HighR/\allowbreak HighC ($b = 0.455, p < .001$) profiles \up{were associated with significantly higher productivity}, whereas HighR/LowC was not (see Figure \ref{fig:tukey-prod-sat}). Tukey’s HSD confirmed this pattern, showing that LowR/\allowbreak HighC scored significantly higher than all other groups ($M = 6.16, SE = 0.15$), with HighR/\allowbreak HighC also scoring significantly higher than LowR/LowC ($M = 5.87, SE = 0.14$). HighR/LowC did not differ from LowR/LowC. Taken together, these results aligned with the RSA findings, indicating that the highest productivity levels were observed under high collaboration, with comparatively elevated levels also observed when both rivalry and collaboration were high.

Finally, the RSA model also explained significant variance in satisfaction ($F(10,392)=2.73, p<.01$, Table \ref{tab:rsa_all}). However, it was only marginally better than the independent-effects model ($\Delta R^{2} = .02, F(3,392) = 2.61, p = .051$). The contour map showed that the highest observed satisfaction levels occurred under low rivalry and high collaboration (see Figure \ref{fig:RSA}). The rivalry and collaboration interaction approached significance ($b = -.04, p = .065$), indicating a possible pattern in which \up{higher rivalry corresponded to lower satisfaction} at comparable levels of collaboration, though quadratic terms were not significant. Surface parameters revealed a weak positive slope along the line of congruence ($a_{1} = .12$) and a slight negative slope along the line of incongruence ($a_{3} = -.10$), indicating that satisfaction levels corresponded more \up{strongly to collaboration than to rivalry}.

PAM profile analysis showed that only the LowR/HighC profile \up{was associated with significantly higher satisfaction ($b = 0.437, p < .01$)}. HighR/LowC and HighR/HighC did not differ significantly from LowR/LowC. Tukey’s HSD test (see Figure \ref{fig:tukey-prod-sat}) corroborated this pattern, indicating that LowR/HighC scored significantly higher than both LowR/LowC ($M = 6.16, SE = 0.18$) and HighR/LowC ($M = 5.68, SE = 0.18$), while HighR/HighC ($M = 5.88, SE = 0.17$) was not distinguishable from either group. These results indicate that \up{the highest satisfaction levels were observed when collaboration was high and rivalry remained low}.
\section{Discussion}
Our results suggest there is substantial complexity in how writing professionals are orienting towards GenAI and forming relationships with it. In this section, we interpret these findings to provide (1) implications such orientations have on writing professions and shaping the profession as a result, and (2) design suggestions that can balance orientations of both rivalry and collaboration with GenAI to maximize GenAI-mediated outcomes in work. 

\subsection{Implications for Writing Profession: Balancing Orientations Towards GenAI}

At a broad level, professionals with either a rivalry or collaboration stance shaped their roles in similar ways, both by taking on new responsibilities (approach crafting, H5) and by reducing existing ones (avoidance crafting). These findings depart from prior rivalry literature \cite{Garcia2006} and our hypothesis (H4), which associated rivalry orientation only with approach crafting and not avoidance crafting. One explanation for this outcome is that a rivalrous orientation towards GenAI may prompt professionals to re-evaluate their role and reduce tasks that no longer add value in the face of GenAI competition (e.g., copyediting). Instead, they used GenAI or other technologies to handle such tasks. Section \ref{pam} presented partial evidence for this, where profiles with high rivalry reported having more GenAI experience than the baseline.

At a broad level, professionals with either a rivalry or collaboration stance reported \up{similar patterns in how they approached their roles}, both by taking on new responsibilities (approach crafting, H5) and by reducing existing ones (avoidance crafting). These findings differ from prior rivalry literature \cite{Garcia2006} and from our hypothesis (H4), which linked rivalry orientation only to approach crafting rather than avoidance crafting. One possible interpretation is that a rivalrous stance toward GenAI may be \up{associated with greater reflection} on one’s role and with decisions to reduce tasks perceived as less valuable in the context of GenAI use (e.g., copyediting), sometimes reallocating such tasks to GenAI or other technologies. Section \ref{pam} provided partial evidence consistent with this interpretation, as profiles characterized by \up{higher rivalry also reported greater GenAI experience than the baseline}.

At the same time, the two stances were \up{associated with differences in how professionals approached specific aspects of their roles}. A collaborative stance was associated with greater restructuring of workflows and optimization of processes (task crafting) compared to a rivalrous stance (H2). Exploratory evidence from Section \ref{profile-engagement} indicates that this pattern often involved collaborators describing ways they integrated GenAI into their workflows. In contrast, a rivalrous stance was associated with a stronger emphasis on relationship crafting, suggesting a greater focus on building human connections (H3). Surprisingly, and contrary to our hypothesis (H1), we found no differences between rivalry and collaboration orientations \up{in their associations with cognitive crafting}, indicating that collaboration oriented professionals reported levels of cognitive role framing similar to those of rivalry oriented professionals.

 These orientation-based differences in crafting practices also appeared in longer term patterns related to skill maintenance. Only a rivalrous stance was associated with professionals’ reports of proactively maintaining their skills (H5). One possible interpretation is that rivalry was \up{linked to greater engagement} in work practices that emphasize traditional aspects of the role (e.g., relational crafting), which corresponded to higher reported levels of skill maintenance. In contrast, collaboration was \up{associated with patterns reflecting the incorporation of GenAI into everyday work (task crafting)}, which corresponded to lower emphasis on maintaining certain traditional skills. This leaves professionals who take only a collaboration stance vulnerable, as GenAI capabilities continue to improve while their traditional skills are likely to atrophy.

\up{Evidence from the study also points to collaborators spending more time maintaining new skills required for GenAI integration, such as making GenAI outputs more human (Section \ref{qual-increase})}. These ``humanizing AI'' activities may explain why collaboration stance was associated only with cognitive skill maintenance, as these individuals still relied on their traditional cognitive skills to make sense of GenAI outputs and incorporating them in their workflows. This finding contradicts prior results suggesting collaboration stance risks disengagement from cognitive skills \cite{lee2025impact, Varanasi2025}.

The differences in crafting practices between orientations were also reflected in the outcome measures, where only a collaborative stance was primarily associated with higher levels of productivity and satisfaction in the short term (see Section \ref{RQ2-sat-prod}). This outcome highlights that while rivalrous professionals crafted their jobs with similar intensity, they did not experience the same levels of productivity or satisfaction, a combination previously linked to elevated strain and burnout risk \cite{Corbeanu_2023}. Conversely, collaborators’ higher productivity and satisfaction \up{coincided with patterns that may involve greater reliance on GenAI}. This is concerning given that many GenAI systems are inherently designed to minimize friction and streamline adoption \cite{Chen2024}. For instance, some tools offer scaffolded prompts (e.g., “Do you want me to provide an updated draft with the changes?”), making it effortless for users to interact. However, such nudges increase workers' reliance on them. Taken together, these findings point to a recurring theme of the trade-offs between \textit{short term gains} (productivity and satisfaction) and \textit{long term sustainability} (skill maintenance) for workers who lean strongly toward a particular stance in their relationship with GenAI.

In contrast, adopting a combined stance of collaboration and rivalry created a productive tension, yielding stronger associations with job crafting (Section \ref{RQ-2-job-crafting}) and productivity (Section \ref{RQ2-sat-prod}). In particular, HighR/HighC profiles showed the strongest outcomes, indicating that collaboration was strongly associated with these outcomes, with rivalry showing an additional positive association. For skill maintenance (Section \ref{Skill Maintenance}) and satisfaction (Section \ref{RQ2-sat-prod}), the joint model explained variance comparable to the independent model, producing similar results. In other words, the presence of both orientations at higher levels yielded a more balanced effect. For instance, the HighR/HighC profile showed stronger skill maintenance compared to other groups, suggesting that rivalry corresponded strongly with this outcome, with collaboration accompanying this association at elevated levels.

Designing for such balanced outcomes may require intentionally increasing friction to promote rivalry while mitigating over-reliance to keep collaboration levels in check. Such design interventions can increase intentionality and reflection, both considered essential for writing professions \cite{walker2013writing}. Emerging evidence supports this approach for writing in the classroom settings, where introducing deliberate tensions have improved long-term skill development \cite{Blasco_Charisi_2024, Wang2025}. In the next two sub-sections, we discuss a few design strategies that produce more balanced relationship with GenAI. 

\subsection{Design Implications}
As technologies approach human-level capabilities, making them entirely effortless to use may be counterproductive to the long-term outcomes that humans hope to achieve. Our findings suggest that introducing \textit{micro-frictions}, small forms of frictions that can enhance these long-term outcomes without significantly impacting the user experience. To motivate this, we lean on the rich scholarship of design friction \cite{cox2016}. Design friction is focused on intentionally introducing small elements of friction (or resistances) in technological systems to provoke reflection or challenge the user in tasks that focus on increased engagement and speed. 

\up{The current forms of GenAI technologies are generally designed to increase engagement by reducing the time between user queries and system responses \cite{Zhang2024}. For example, Large Language Models (LLMs) integrated into tools that provide autocomplete functionality (e.g., Gmail) introduce minimal cognitive friction, because no prompting is required. LLMs embedded directly into workflows (e.g., Copilot in VS Code) likewise minimize friction by running in the background and triggering automatically based on the user’s ongoing actions, reducing the need for professionals to plan or decide which tasks to delegate \cite{vaithilingam2022expectation}. In contrast, turn-taking or conversational LLMs introduce more friction, as writers must plan and articulate their interactions. Yet these tools also use explicitly supportive language that is affirmative, personal, and persuasive, making it easier for professionals to embrace and feel a sense of collaboration that promotes over reliance.}

\paragraph{Micro-frictions to promote Reflection-in-action/on-action}
One way to mitigate excessive collaboration is by designing for appropriate reliance. Professionals who reported high collaboration also demonstrated elevated task-crafting behaviors, particularly when using GenAI for creative work (see Section \ref{RQ-2-job-crafting}; Section \ref{profile-engagement}). These patterns suggest a risk of over-reliance on GenAI \cite{Passi2024}. Current design features, such as scaffolding prompts, may further exacerbate this tendency. To counterbalance it, incorporating the principles of \textit{reflection-in-action} and \textit{reflection-on-action}, put forward by \citet{Schon2017}, can serve as mechanisms for recalibrating reliance. Schön describes \textit{reflection-in-action} as practitioners' ability to make adjustments and improvement while the task is underway through the process of reflection. In contrast, \textit{reflection-on-action} involves using reflection to look back once the task is complete, such as reviewing the finished draft to assess what worked and what should change next time. Writing, as a professional craft, naturally embodies both reflection-in-action and reflection-on-action, through implicit steps that can be supported further \cite{Lowe2010}. For instance, GenAI's scafolding prompts could be reframed as reflection prompts for both in-action (e.g., “Do you want to write a copyedited draft based on the suggestions?”) and on-action (e.g., “How do you feel about the final draft you created?”). \up{A concrete way to implement such scaffolding is through Socratic-style LLMs that employ iterative questioning to encourage critical thinking and structured reflection \cite{liu2025discerning}}.

% UPsides/downside -> multiple directions. 
% autocomplete - channel peoples reponses. Inhibits creative response. 

\paragraph{Reciprocal learning}
Micro-frictions that reduce excessive collaboration can also be designed as meta-learning elements through which GenAI and humans learn from one another. Reciprocal learning is a form of learning in which humans learn from each other \cite{jorg2009thinking}. In recent years, it has been applied as a framework to facilitate learning between intelligent systems and AI \cite{zagal2021}. For example, end-user explanations in critical AI systems have enabled people to better understand how the system processes tasks and evaluates outcomes \cite{Deiw2023}. However, as GenAI becomes more pervasive in everyday work, learning processes has increasingly tilted toward the model, in which GenAI is learning from human actions, often at the expense of individuals learning from the system. When these tools are used in professional contexts, they can contribute to skill decay. To address this issue, we argue for embedding mechanisms that support bi-directional learning. One strategy is for GenAI to represent professionals’ knowledge in abstract forms that the professional can then reflect upon. Such shared representations could provide common ground for collaborative reflection and also create space for both GenAI and professionals to contest and critique outputs.

\paragraph{Bi-directional Synergy of Collaboration \& Rivalry: Communities of Practice}
Another way to balance collaboration and rivalry is by fostering bi-directional synergy between the two stances. One promising pathway is the cultivation of stronger communities of practice, where professionals with high collaboration and high rivalry can engage in information exchange and mutual support. Communities of practice, defined as groups of individuals with a shared professional interest who learn collectively by sharing lived experiences \cite{Farnsworth2016}, enable members to build shared perceptions, develop mutual understanding of each other’s mental models, and surface how GenAI both benefits and challenges their work. Such dialogue can help balance rivalry and collaboration while frequent interactions sustain both at higher levels.

Beyond mutual learning, communities of practice can contribute to the development of standardized taxonomies that capture the diverse ways professionals position themselves in relation to GenAI. These taxonomies could identify the key characteristics of rivalry and collaboration across roles, providing templates that GenAI systems could use to better recognize interaction cues. In turn, GenAI could become more receptive to whether users are engaging in collaborative or rivalrous modes and adapt its responses accordingly. Designing adversarial or complementary “layers” within GenAI systems could further ensure that the technology elicits perspectives that challenge as well as affirm, reflecting the productive tensions found in professional practice.

\section{Limitations}
Our study has a number of limitations that highlight valuable areas for future research. \up{First, we confined our respondents to writing professionals because their occupation is among the first to be severely disrupted by GenAI. Future research is required to evaluate whether our findings generalize to other professions, and to evaluate how professionals' perceptions and practices evolve over time}. \up{Second, our findings are vulnerable to common methods bias due to reliance on self-reported measures collected at a single point in time. While cross sectional surveys are valuable for capturing perceptions at scale and across diverse respondents, they reflect subjective impressions rather than objective behavior, may be influenced by common method, recall, or social desirability biases, and do not support causal inference \cite{Wang2020}. Future research using longitudinal designs, multiple raters, and behavioral data is needed to evaluate the directionality and robustness of the associations we observed.}
Our study was conducted on Prolific, and although we took measures to recruit a diverse range of writing professionals, it is possible that our sample was skewed towards professionals reporting high collaboration with GenAI due to Prolific’s tendency to attract participants who are more likely to adopt pro-GenAI stances. We also limited recruitment to participants from North America as the work practices can differ among writing professionals globally. Future studies are needed to evaluate whether our results generalize to diverse global populations. 
\up{While our measures of stances, practices and outcomes were based on constructs developed in prior research, we adapted existing scales of rivalry and collaboration stances and job crafting to make them more applicable to GenAI, and used prior behavioral skill maintenance findings that were GenAI-specific to develop our skill maintenance scale. While we found reassuring evidence of the reliability of the resulting scales, future research is needed to further validate these measures}. 

\section{Conclusion}
This study presents critical evidence on writing professionals’ complex relationship with GenAI and its impact on their work practices and outcomes. In particular, an excessive leaning toward either rivalry or collaboration with GenAI was associated with greater imbalances, where individuals prioritized long-term career development at the expense of short-term outcomes, or vice versa, but not both. By contrast, combining rivalry and collaboration orientations was associated with improved balance, enabling writers to shape their practices for long-term growth while simultaneously benefiting from short-term outcomes. Based on these results, we present key design ideas, such as introducing micro-frictions to increase rivalry while maintaining collaboration levels to improve work pathways for professional writers.

\begin{acks}
 This work was supported by NSF grants 1928614 and 2129076. We thank all the professionals who participated in the survey and the anonymous reviewers for their thoughtful feedback.
\end{acks}

\bibliographystyle{ACM-Reference-Format}
\bibliography{citations}

@article{Douglas2023, title={Data quality in online human-subjects research: Comparisons between MTurk, Prolific, CloudResearch, Qualtrics, and SONA}, volume={18}, ISSN={1932-6203}, url={https://dx.plos.org/10.1371/journal.pone.0279720}, DOI={10.1371/journal.pone.0279720}, number={3}, journal={PLOS ONE}, author={Douglas, Benjamin D. and Ewell, Patrick J. and Brauer, Markus}, editor={Hallam, Jeffrey S.}, year={2023}, month=mar, pages={e0279720}, language={en} }

@book{field2012discovering,
  title={Discovering Statistics Using R},
  author={Field, Andy and Miles, Jeremy and Field, Zoe},
  year={2012},
  publisher={Sage}
}

@article{Cini_2023, title={Resisting algorithmic control: Understanding the rise and variety of platform worker mobilisations}, volume={38}, ISSN={0268-1072, 1468-005X}, url={https://onlinelibrary.wiley.com/doi/10.1111/ntwe.12257}, DOI={10.1111/ntwe.12257}, number={1}, journal={New Technology, Work and Employment}, author={Cini, Lorenzo}, year={2023}, month=mar, pages={125–144}, language={en} }

@article{Schubert2021, title={Fast and eager k -medoids clustering: O ( k ) runtime improvement of the PAM, CLARA, and CLARANS algorithms}, volume={101}, ISSN={03064379}, url={https://linkinghub.elsevier.com/retrieve/pii/S0306437921000557}, DOI={10.1016/j.is.2021.101804}, journal={Information Systems}, author={Schubert, Erich and Rousseeuw, Peter J.}, year={2021}, month=nov, pages={101804}, language={en} }

@article{Samhan_2018, title={Revisiting Technology Resistance: Current Insights and Future Directions}, volume={22}, rights={http://creativecommons.org/licenses/by-nc/3.0/au/}, ISSN={1449-8618}, url={https://ajis.aaisnet.org/index.php/ajis/article/view/1655}, DOI={10.3127/ajis.v22i0.1655}, journal={Australasian Journal of Information Systems}, author={Samhan, Bahae}, year={2018}, month=jan }

@article{paraschou2025,
      title={Mind the XAI Gap: A Human-Centered LLM Framework for Democratizing Explainable AI}, 
      author={Eva Paraschou and Ioannis Arapakis and Sofia Yfantidou and Sebastian Macaluso and Athena Vakali},
      year={2025},
      eprint={2506.12240},
      archivePrefix={arXiv},
      primaryClass={cs.LG},
      url={https://arxiv.org/abs/2506.12240}, 
}

@article{Halkidi_2001, title={On Clustering Validation Techniques}, volume={17}, rights={https://www.springernature.com/gp/researchers/text-and-data-mining}, ISSN={0925-9902, 1573-7675}, url={https://link.springer.com/10.1023/A:1012801612483}, DOI={10.1023/A:1012801612483}, number={2–3}, journal={Journal of Intelligent Information Systems}, author={Halkidi, Maria and Batistakis, Yannis and Vazirgiannis, Michalis}, year={2001}, month=dec, pages={107–145}, language={en} }

@article{rousseeuw1987silhouettes,
  title={Silhouettes: a graphical aid to the interpretation and validation of cluster analysis},
  author={Rousseeuw, Peter J},
  journal={Journal of computational and applied mathematics},
  volume={20},
  pages={53--65},
  year={1987},
  publisher={Elsevier}
}

@article{Koopmans2014, title={Construct Validity of the Individual Work Performance Questionnaire}, volume={56}, ISSN={1076-2752}, url={https://journals.lww.com/00043764-201403000-00014}, DOI={10.1097/JOM.0000000000000113}, number={3}, journal={Journal of Occupational \& Environmental Medicine}, author={Koopmans, Linda and Bernaards, Claire M. and Hildebrandt, Vincent H. and De Vet, Henrica C. W. and Van Der Beek, Allard J.}, year={2014}, month=mar, pages={331–337}, language={en} }

@article{Karunakaran2025, title={Artificial Intelligence at Work: An Integrative Perspective on the Impact of AI on Workplace Inequality}, volume={19}, ISSN={1941-6520, 1941-6067}, url={http://journals.aom.org/doi/full/10.5465/annals.2023.0230}, DOI={10.5465/annals.2023.0230}, number={2}, journal={Academy of Management Annals}, author={Karunakaran, Arvind and Lebovitz, Sarah and Narayanan, Devesh and Rahman, Hatim A.}, year={2025}, month=july, pages={693–735}, language={en} }

@book{Arthur2013, author={Arthur Jr, Winfred and Day, Eric Anthony and Bennett Jr, Winston and Portrey, Antoinette M},
edition={0}, title={Individual and Team Skill Decay}, ISBN={9781136689406}, url={https://www.taylorfrancis.com/books/9781136689406}, DOI={10.4324/9780203576076}, publisher={Routledge}, year={2013}, month=sep, language={en} }

@inproceedings{walker2013writing,
  title={Writing and reflection},
  author={Walker, David},
  booktitle={Reflection},
  pages={52--68},
  year={2013},
  organization={Routledge}
}

@techreport{Passi2024,
  author = {Passi, Samir and Dhanorkar, Shipi and Vorvoreanu, Mihaela},
  title = {Appropriate Reliance on GenAI: Research Synthesis},
  year = {2024},
  institution = {Microsoft Research},
  url = {https://www.microsoft.com/en-us/research/wp-content/uploads/2024/03/GenAI_AppropriateReliance_Published2024-3-21.pdf}
}

@inbook{Chen2024, address={Cham}, title={Exploring a Behavioral Model of “Positive Friction” in Human-AI Interaction}, volume={14713}, ISBN={9783031613524}, url={https://link.springer.com/10.1007/978-3-031-61353-1_1}, DOI={10.1007/978-3-031-61353-1_1}, booktitle={Design, User Experience, and Usability}, publisher={Springer Nature Switzerland}, author={Chen, Zeya and Schmidt, Ruth}, editor={Marcus, Aaron and Rosenzweig, Elizabeth and Soares, Marcelo M.}, year={2024}, pages={3–22}, language={en} }

@article{Corbeanu_2023, title={The link between burnout and job performance: a meta-analysis}, volume={32}, ISSN={1359-432X, 1464-0643}, url={https://www.tandfonline.com/doi/full/10.1080/1359432X.2023.2209320}, DOI={10.1080/1359432X.2023.2209320}, number={4}, journal={European Journal of Work and Organizational Psychology}, author={Corbeanu, Andreea and Iliescu, Dragoș and Ion, Andrei and Spînu, Roxana}, year={2023}, month=jul, pages={599–616}, language={en} }

@article{Acemoglu2024, title={Learning From Ricardo and Thompson: Machinery and Labor in the Early Industrial Revolution and in the Age of Artificial Intelligence}, volume={16}, rights={http://creativecommons.org/licenses/by/4.0/}, ISSN={1941-1383, 1941-1391}, url={https://www.annualreviews.org/content/journals/10.1146/annurev-economics-091823-025129}, DOI={10.1146/annurev-economics-091823-025129}, number={1}, journal={Annual Review of Economics}, author={Acemoglu, Daron and Johnson, Simon}, year={2024}, month=aug, pages={597–621}, language={en} }

@book{kaufman2009finding,
  title={Finding groups in data: an introduction to cluster analysis},
  author={Kaufman, Leonard and Rousseeuw, Peter J},
  year={2009},
  publisher={John Wiley \& Sons}
}

@article{Parker_2025, title={Top-Down and Bottom-Up Work Design: A Multilevel Perspective on How Job Crafting and Work Characteristics Interrelate}, ISSN={0889-3268, 1573-353X}, url={https://link.springer.com/10.1007/s10869-025-10010-1}, DOI={10.1007/s10869-025-10010-1}, journal={Journal of Business and Psychology}, author={Parker, Sharon K. and Tims, Maria and Sonnentag, Sabine}, year={2025}, month=feb, language={en} }

@article{Arthur1998, title={Factors That Influence Skill Decay and Retention: A Quantitative Review and Analysis}, volume={11}, ISSN={0895-9285, 1532-7043}, url={http://www.tandfonline.com/doi/abs/10.1207/s15327043hup1101_3}, DOI={10.1207/s15327043hup1101_3}, number={1}, journal={Human Performance}, author={Arthur Jr., Winfred and Bennett Jr., Winston and Stanush, Pamela L. and McNelly, Theresa L.}, year={1998}, month=mar, pages={57–101}, language={en} }

@article{Gawad2019, title={Decay of Competence with Extended Research Absences During Residency Training: A Scoping Review}, volume={11}, ISSN={2168-8184}, DOI={10.7759/cureus.5971}, number={10}, journal={Cureus}, author={Gawad, Nada and Allen, Molly and Fowler, Amanda}, year={2019}, month=oct, pages={e5971}, language={eng} }

@article{Anderson2011, title={First aid skill retention of first responders within the workplace}, volume={19}, ISSN={1757-7241}, url={http://sjtrem.biomedcentral.com/articles/10.1186/1757-7241-19-11}, DOI={10.1186/1757-7241-19-11}, number={1}, journal={Scandinavian Journal of Trauma, Resuscitation and Emergency Medicine}, author={Anderson, Gregory S and Gaetz, Michael and Masse, Jeff}, year={2011}, pages={11}, language={en} }

@article{Childs1986, title={Flight-Skill Decay and Recurrent Training}, volume={62}, rights={https://journals.sagepub.com/page/policies/text-and-data-mining-license}, ISSN={0031-5125, 1558-688X}, url={https://journals.sagepub.com/doi/10.2466/pms.1986.62.1.235}, DOI={10.2466/pms.1986.62.1.235}, number={1}, journal={Perceptual and Motor Skills}, author={Childs, Jerry M. and Spears, William D.}, year={1986}, month=feb, pages={235–242}, language={en} }

@incollection{wang2013factors,
  title={Factors influencing knowledge and skill decay after training: A meta-analysis},
  author={Wang, Xiaoqian and Day, Eric Anthony and Kowollik, Vanessa and Schuelke, Matthew J and Hughes, Michael G},
  booktitle={Individual and team skill decay},
  pages={68--116},
  year={2013},
  publisher={Routledge}
}

@article{Brynjolfsson2025, title={Generative AI at Work}, volume={140}, rights={https://creativecommons.org/licenses/by-nc/4.0/}, ISSN={0033-5533, 1531-4650}, url={https://academic.oup.com/qje/article/140/2/889/7990658}, DOI={10.1093/qje/qjae044}, number={2}, journal={The Quarterly Journal of Economics}, author={Brynjolfsson, Erik and Li, Danielle and Raymond, Lindsey}, year={2025}, month=apr, pages={889–942}, language={en} }

@article{Doshi2024, title={Generative AI enhances individual creativity but reduces the collective diversity of novel content}, volume={10}, ISSN={2375-2548}, url={https://www.science.org/doi/10.1126/sciadv.adn5290}, DOI={10.1126/sciadv.adn5290}, number={28}, journal={Science Advances}, author={Doshi, Anil R. and Hauser, Oliver P.}, year={2024}, month=jul, pages={eadn5290}, language={en} }

@article{Heigl2025, title={Generative artificial intelligence in creative contexts: a systematic review and future research agenda}, ISSN={2198-1620, 2198-1639}, url={https://link.springer.com/10.1007/s11301-025-00494-9}, DOI={10.1007/s11301-025-00494-9}, journal={Management Review Quarterly}, author={Heigl, Rebecca}, year={2025}, month=mar, language={en} }

@inproceedings{Zhang2024, address={Luxembourg Luxembourg}, title={Explaining the Wait: How Justifying Chatbot Response Delays Impact User Trust}, ISBN={9798400705113}, url={https://dl.acm.org/doi/10.1145/3640794.3665550}, DOI={10.1145/3640794.3665550}, booktitle={ACM Conversational User Interfaces 2024}, publisher={ACM}, author={Zhang, Zhengquan and Tsiakas, Konstantinos and Schneegass, Christina}, year={2024}, month=jul, pages={1–16}, language={en} }

@article{Farnsworth2016, title={Communities of Practice as a Social Theory of Learning: a Conversation with Etienne Wenger}, volume={64}, ISSN={0007-1005, 1467-8527}, url={http://www.tandfonline.com/doi/full/10.1080/00071005.2015.1133799}, DOI={10.1080/00071005.2015.1133799}, number={2}, journal={British Journal of Educational Studies}, author={Farnsworth, Valerie and Kleanthous, Irene and Wenger-Trayner, Etienne}, year={2016}, month=apr, pages={139–160}, language={en} }

@article{Deiw2023,
author = {Dwivedi, Rudresh and Dave, Devam and Naik, Het and Singhal, Smiti and Omer, Rana and Patel, Pankesh and Qian, Bin and Wen, Zhenyu and Shah, Tejal and Morgan, Graham and Ranjan, Rajiv},
title = {Explainable AI (XAI): Core Ideas, Techniques, and Solutions},
year = {2023},
issue_date = {September 2023},
publisher = {Association for Computing Machinery},
address = {New York, NY, USA},
volume = {55},
number = {9},
issn = {0360-0300},
url = {https://doi.org/10.1145/3561048},
doi = {10.1145/3561048},
journal = {ACM Comput. Surv.},
month = jan,
articleno = {194},
numpages = {33}
}

@article{zagal2021,
author = {Zagalsky, Alexey and Te'eni, Dov and Yahav, Inbal and Schwartz, David G. and Silverman, Gahl and Cohen, Daniel and Mann, Yossi and Lewinsky, Dafna},
title = {The Design of Reciprocal Learning Between Human and Artificial Intelligence},
year = {2021},
issue_date = {October 2021},
publisher = {Association for Computing Machinery},
address = {New York, NY, USA},
volume = {5},
number = {CSCW2},
url = {https://doi.org/10.1145/3479587},
doi = {10.1145/3479587},
journal = {Proc. ACM Hum.-Comput. Interact.},
month = oct,
articleno = {443},
numpages = {36}
}

@inproceedings{cox2016,
author = {Cox, Anna L. and Gould, Sandy J.J. and Cecchinato, Marta E. and Iacovides, Ioanna and Renfree, Ian},
title = {Design Frictions for Mindful Interactions: The Case for Microboundaries},
year = {2016},
isbn = {9781450340823},
publisher = {Association for Computing Machinery},
address = {New York, NY, USA},
url = {https://doi.org/10.1145/2851581.2892410},
doi = {10.1145/2851581.2892410},
booktitle = {Proceedings of the 2016 CHI Conference Extended Abstracts on Human Factors in Computing Systems},
pages = {1389–1397},
numpages = {9},
location = {San Jose, California, USA},
series = {CHI EA '16}
}

@inproceedings{Takaffoli2024, address={Copenhagen Denmark}, title={Generative AI in User Experience Design and Research: How Do UX Practitioners, Teams, and Companies Use GenAI in Industry?}, ISBN={9798400705830}, url={https://dl.acm.org/doi/10.1145/3643834.3660720}, DOI={10.1145/3643834.3660720}, booktitle={Designing Interactive Systems Conference}, publisher={ACM}, author={Takaffoli, Macy and Li, Sijia and Mäkelä, Ville}, year={2024}, month=jul, pages={1579–1593}, language={en} }

@article{Berretta2023, title={Defining human-AI teaming the human-centered way: a scoping review and network analysis}, volume={6}, ISSN={2624-8212}, url={https://www.frontiersin.org/articles/10.3389/frai.2023.1250725/full}, DOI={10.3389/frai.2023.1250725},journal={Frontiers in Artificial Intelligence}, author={Berretta, Sophie and Tausch, Alina and Ontrup, Greta and Gilles, Björn and Peifer, Corinna and Kluge, Annette}, year={2023}, month=sep, pages={1250725} }

@article{Al2024, title={Enhancing Work Productivity through Generative Artificial Intelligence: A Comprehensive Literature Review}, volume={16}, rights={https://creativecommons.org/licenses/by/4.0/}, ISSN={2071-1050}, url={https://www.mdpi.com/2071-1050/16/3/1166}, DOI={10.3390/su16031166}, number={3}, journal={Sustainability}, author={Al Naqbi, Humaid and Bahroun, Zied and Ahmed, Vian}, year={2024}, month=jan, pages={1166}, language={en} }

@misc{Price2024resisting,
  author       = {Price, Shaun},
  title        = {Resisting AI and the Future of Work: Navigating Technological Transformation in the Modern Workplace},
  howpublished = {\url{https://metagentity.ai/articles/Resisting_AI_and_the_Future_of_Work.html}},
  note         = {Metagentity},
  year         = {2024},
  month        = jun,
  urldate      = {2025-08-26}
}

@article{ONeill2022, title={Human–Autonomy Teaming: A Review and Analysis of the Empirical Literature}, volume={64}, ISSN={0018-7208, 1547-8181}, url={https://journals.sagepub.com/doi/10.1177/0018720820960865}, DOI={10.1177/0018720820960865}, number={5}, journal={Human Factors: The Journal of the Human Factors and Ergonomics Society}, author={O’Neill, Thomas and McNeese, Nathan and Barron, Amy and Schelble, Beau}, year={2022}, month=aug, pages={904–938}, language={en} }

@article{Seeber2020, title={Machines as teammates: A research agenda on AI in team collaboration}, volume={57}, ISSN={03787206}, url={https://linkinghub.elsevier.com/retrieve/pii/S0378720619303337}, DOI={10.1016/j.im.2019.103174}, number={2}, journal={Information \& Management}, author={Seeber, Isabella and Bittner, Eva and Briggs, Robert O. and De Vreede, Triparna and De Vreede, Gert-Jan and Elkins, Aaron and Maier, Ronald and Merz, Alexander B. and Oeste-Reiß, Sarah and Randrup, Nils and Schwabe, Gerhard and Söllner, Matthias}, year={2020}, month=mar, pages={103174}, language={en} }

@article{Puerta2025, title={A Multifaceted Vision of the Human-AI Collaboration: A Comprehensive Review}, volume={13}, rights={https://creativecommons.org/licenses/by/4.0/legalcode}, ISSN={2169-3536}, url={https://ieeexplore.ieee.org/document/10857320/}, DOI={10.1109/ACCESS.2025.3536095}, journal={IEEE Access}, author={Puerta-Beldarrain, Maite and Gómez-Carmona, Oihane and Sánchez-Corcuera, Rubén and Casado-Mansilla, Diego and López-de-Ipiña, Diego and Chen, Liming}, year={2025}, pages={29375–29405} }

@article{Parasuraman2000, title={A model for types and levels of human interaction with automation}, volume={30}, rights={https://ieeexplore.ieee.org/Xplorehelp/downloads/license-information/IEEE.html}, ISSN={10834427}, url={http://ieeexplore.ieee.org/document/844354/}, DOI={10.1109/3468.844354}, number={3}, journal={IEEE Transactions on Systems, Man, and Cybernetics - Part A: Systems and Humans}, author={Parasuraman, R. and Sheridan, T.B. and Wickens, C.D.}, year={2000}, month=may, pages={286–297} }

@inproceedings{Dolata2024, address={Lisbon Portugal}, title={Development in times of hype: How freelancers explore Generative AI?}, ISBN={9798400702174}, url={https://dl.acm.org/doi/10.1145/3597503.3639111}, DOI={10.1145/3597503.3639111}, booktitle={Proceedings of the IEEE/ACM 46th International Conference on Software Engineering}, publisher={ACM}, author={Dolata, Mateusz and Lange, Norbert and Schwabe, Gerhard}, year={2024}, month=apr, pages={1–13}, language={en} }

@article{Nass1996, title={Can computers be teammates?}, volume={45}, ISSN={10715819}, url={https://linkinghub.elsevier.com/retrieve/pii/S1071581996900737}, DOI={10.1006/ijhc.1996.0073}, number={6}, journal={International Journal of Human-Computer Studies}, author={Nass, Clifford and Fogg, B.J. and Moon, Youngme}, year={1996}, month=dec, pages={669–678}, language={en} }

@book{baym2015personal,
  title={Personal connections in the digital age},
  author={Baym, Nancy K},
  year={2015},
  publisher={John Wiley \& Sons}
}

@inproceedings{Hoque2024,
author = {Hoque, Md Naimul and Mashiat, Tasfia and Ghai, Bhavya and Shelton, Cecilia D. and Chevalier, Fanny and Kraus, Kari and Elmqvist, Niklas},
title = {The HaLLMark Effect: Supporting Provenance and Transparent Use of Large Language Models in Writing with Interactive Visualization},
year = {2024},
isbn = {9798400703300},
publisher = {Association for Computing Machinery},
address = {New York, NY, USA},
url = {https://doi.org/10.1145/3613904.3641895},
doi = {10.1145/3613904.3641895},
booktitle = {Proceedings of the CHI Conference on Human Factors in Computing Systems},
articleno = {1045},
numpages = {15},
location = {Honolulu, HI, USA},
series = {CHI '24}
}

@inproceedings{Reza2024,
author = {Reza, Mohi and Laundry, Nathan M and Musabirov, Ilya and Dushniku, Peter and Yu, Zhi Yuan “Michael” and Mittal, Kashish and Grossman, Tovi and Liut, Michael and Kuzminykh, Anastasia and Williams, Joseph Jay},
title = {ABScribe: Rapid Exploration \& Organization of Multiple Writing Variations in Human-AI Co-Writing Tasks using Large Language Models},
year = {2024},
isbn = {9798400703300},
publisher = {Association for Computing Machinery},
address = {New York, NY, USA},
url = {https://doi-org.proxy.library.cornell.edu/10.1145/3613904.3641899},
doi = {10.1145/3613904.3641899},
booktitle = {Proceedings of the CHI Conference on Human Factors in Computing Systems},
articleno = {1042},
numpages = {18},
location = {Honolulu, HI, USA},
series = {CHI '24}
}

@misc{NYT2024,
  author       = {Sittenfeld, Curtis},
  title        = {Beach Read, Meet A.I.},
  howpublished = {\textit{The New York Times}},
  year         = {2024},
  month        = Aug,
  day          = {20},
  url          = {An Experiment in Lust, Regret and Kissing},
  note         = {Accessed: 2025-02-25}
}

@article{McDowell_2024, title={Wikipedia and AI: Access, representation, and advocacy in the age of large language models}, volume={30}, ISSN={1354-8565, 1748-7382}, url={https://journals.sagepub.com/doi/10.1177/13548565241238924}, DOI={10.1177/13548565241238924},  number={2}, journal={Convergence: The International Journal of Research into New Media Technologies}, author={McDowell, Zachary J}, year={2024}, month=apr, pages={751–767}, language={en} }

@inproceedings{vaithilingam2022expectation,
  title={Expectation vs. experience: Evaluating the usability of code generation tools powered by large language models},
  author={Vaithilingam, Priyan and Zhang, Tianyi and Glassman, Elena L},
  booktitle={Chi conference on human factors in computing systems extended abstracts},
  pages={1--7},
  year={2022}
}

@article{draxler24,
author = {Draxler, Fiona and Werner, Anna and Lehmann, Florian and Hoppe, Matthias and Schmidt, Albrecht and Buschek, Daniel and Welsch, Robin},
title = {The AI Ghostwriter Effect: When Users do not Perceive Ownership of AI-Generated Text but Self-Declare as Authors},
year = {2024},
issue_date = {April 2024},
publisher = {Association for Computing Machinery},
address = {New York, NY, USA},
volume = {31},
number = {2},
issn = {1073-0516},
url = {https://doi.org/10.1145/3637875},
doi = {10.1145/3637875},
journal = {ACM Trans. Comput.-Hum. Interact.},
month = feb,
articleno = {25},
numpages = {40}
}

@inproceedings{mirowski2023,
author = {Mirowski, Piotr and Mathewson, Kory W. and Pittman, Jaylen and Evans, Richard},
title = {Co-Writing Screenplays and Theatre Scripts with Language Models: Evaluation by Industry Professionals},
year = {2023},
isbn = {9781450394215},
publisher = {Association for Computing Machinery},
address = {New York, NY, USA},
url = {https://doi.org/10.1145/3544548.3581225},
doi = {10.1145/3544548.3581225},
booktitle = {Proceedings of the 2023 CHI Conference on Human Factors in Computing Systems},
articleno = {355},
numpages = {34},
location = {Hamburg, Germany},
series = {CHI '23}
}

@inproceedings{Chung22,
author = {Chung, John Joon Young and He, Shiqing and Adar, Eytan},
title = {Artist Support Networks: Implications for Future Creativity Support Tools},
year = {2022},
isbn = {9781450393584},
publisher = {Association for Computing Machinery},
address = {New York, NY, USA},
url = {https://doi.org/10.1145/3532106.3533505},
doi = {10.1145/3532106.3533505},
booktitle = {Proceedings of the 2022 ACM Designing Interactive Systems Conference},
pages = {232–246},
numpages = {15},
location = {Virtual Event, Australia},
series = {DIS '22}
}

@book{dourish2001action,
  title={Where the action is: the foundations of embodied interaction},
  author={Dourish, Paul},
  year={2001},
  publisher={MIT press}
}

@article{Anthony2023, title={“Collaborating” with AI: Taking a System View to Explore the Future of Work}, volume={34}, ISSN={1047-7039, 1526-5455}, number={5}, journal={Organization Science}, author={Anthony, Callen and Bechky, Beth A. and Fayard, Anne-Laure}, year={2023}, month=sep, pages={1672–1694}, language={en} }

@inproceedings{Kyi2025, address={Yokohama Japan}, title={Governance of Generative AI in Creative Work: Consent, Credit, Compensation, and Beyond}, ISBN={9798400713941}, url={https://dl.acm.org/doi/10.1145/3706598.3713799}, DOI={10.1145/3706598.3713799}, booktitle={Proceedings of the 2025 CHI Conference on Human Factors in Computing Systems}, publisher={ACM}, author={Kyi, Lin and Mahuli, Amruta and Silberman, M. Six and Binns, Reuben and Zhao, Jun and Biega, Asia J.}, year={2025}, month=apr, pages={1–16}, language={en} }

@inbook{law2025, address={Cham}, title={Generative AI and Changing Work: Systematic Review of Practitioner-Led Work Transformations Through the Lens of Job Crafting}, volume={15804}, ISBN={9783031928222}, url={https://link.springer.com/10.1007/978-3-031-92823-9_10}, DOI={10.1007/978-3-031-92823-9_10}, booktitle={HCI in Business, Government and Organizations}, publisher={Springer Nature Switzerland}, author={Law, Matthew and Varanasi, Rama Adithya}, editor={Siau, Keng Leng and Nah, Fiona Fui-Hoon}, year={2025}, pages={131–152}, language={en} }

@inproceedings{nasun2025,
author = {Sun, Na and Kalar, Donald},
title = {Gemini at Work: Knowledge Workers' Perceptions and Assessment of Productivity Gains},
year = {2025},
isbn = {9798400714856},
publisher = {Association for Computing Machinery},
address = {New York, NY, USA},
url = {https://doi-org.proxy.library.nyu.edu/10.1145/3715336.3735679},
doi = {10.1145/3715336.3735679},
booktitle = {Proceedings of the 2025 ACM Designing Interactive Systems Conference},
pages = {3681–3695},
numpages = {15},
location = {
},
series = {DIS '25}
}

@inproceedings{Khovanskaya2020, address={Honolulu HI USA}, title={Bottom-Up Organizing with Tools from On High: Understanding the Data Practices of Labor Organizers}, ISBN={9781450367080}, url={https://dl.acm.org/doi/10.1145/3313831.3376185}, DOI={10.1145/3313831.3376185}, booktitle={Proceedings of the 2020 CHI Conference on Human Factors in Computing Systems}, publisher={ACM}, author={Khovanskaya, Vera and Sengers, Phoebe and Dombrowski, Lynn}, year={2020}, month=apr, pages={1–13}, language={en} }

@article{Garcia2006, title={Ranks and Rivals: A Theory of Competition}, volume={32}, rights={https://journals.sagepub.com/page/policies/text-and-data-mining-license}, ISSN={0146-1672, 1552-7433}, url={https://journals.sagepub.com/doi/10.1177/0146167206287640}, DOI={10.1177/0146167206287640},number={7}, journal={Personality and Social Psychology Bulletin}, author={Garcia, Stephen M. and Tor, Avishalom and Gonzalez, Richard}, year={2006}, month=jul, pages={970–982}, language={en} }

@article{Leonardi_2011, title={When Flexible Routines Meet Flexible Technologies: Affordance, Constraint, and the Imbrication of Human and Material Agencies}, volume={35}, ISSN={02767783}, url={https://www.jstor.org/stable/10.2307/23043493}, DOI={10.2307/23043493}, number={1}, journal={MIS Quarterly}, author={Leonardi}, year={2011}, pages={147} }

@inproceedings{mohlmann2017hands,
  title={Hands on the wheel: Navigating algorithmic management and Uber drivers’},
  author={M{\"o}hlmann, Marieke and Zalmanson, Lior},
  booktitle={Autonomy’, in proceedings of the international conference on information systems (ICIS), Seoul South Korea},
  pages={10--13},
  year={2017}
}

@article{Hakanen2020, title={Interactions of Approach and Avoidance Job Crafting and Work Engagement: A Comparison between Employees Affected and Not Affected by Organizational Changes}, volume={17}, rights={https://creativecommons.org/licenses/by/4.0/}, ISSN={1660-4601}, url={https://www.mdpi.com/1660-4601/17/23/9084}, DOI={10.3390/ijerph17239084},number={23}, journal={International Journal of Environmental Research and Public Health}, author={Seppälä, Piia and Harju, Lotta and Hakanen, Jari J.}, year={2020}, month=dec, pages={9084}, language={en} }

@article{Sun2025, title={Will Employee–AI Collaboration Enhance Employees’ Proactive Behavior? A Study Based on the Conservation of Resources Theory}, volume={15}, rights={https://creativecommons.org/licenses/by/4.0/}, ISSN={2076-328X}, url={https://www.mdpi.com/2076-328X/15/5/648}, DOI={10.3390/bs15050648}, number={5}, journal={Behavioral Sciences}, author={Sun, Chenxi and Zhao, Xinan and Guo, Baorong and Chen, Ningning}, year={2025}, month=may, pages={648}, language={en} }

@article{lee2025impact,
  title={The Impact of Generative AI on Critical Thinking: Self-Reported Reductions in Cognitive Effort and Confidence Effects From a Survey of Knowledge Workers},
  author={Lee, Hao-Ping Hank and Sarkar, Advait and Tankelevitch, Lev and Drosos, Ian and Rintel, Sean and Banks, Richard and Wilson, Nicholas},
  year={2025}
}

@article{Rashid2021, title={Evaluation of Manual Skill Degradation Due to Automation in Apparel Manufacturing}, volume={11}, rights={https://creativecommons.org/licenses/by/4.0/}, ISSN={2076-3417}, url={https://www.mdpi.com/2076-3417/11/23/11098}, DOI={10.3390/app112311098},  number={23}, journal={Applied Sciences}, author={Rashid, Zahid and Rötting, Matthias}, year={2021}, month=nov, pages={11098}, language={en} }

@article{Cheng2024, title={“It would work for me too”: How Online Communities Shape Software Developers’ Trust in AI-Powered Code Generation Tools}, volume={14}, ISSN={2160-6455, 2160-6463}, url={https://dl.acm.org/doi/10.1145/3651990}, DOI={10.1145/3651990},  number={2}, journal={ACM Transactions on Interactive Intelligent Systems}, author={Cheng, Ruijia and Wang, Ruotong and Zimmermann, Thomas and Ford, Denae}, year={2024}, month=jun, pages={1–39}, language={en} }

@article{Wang2025, title={In GenAI we trust: An investigation of university students’ reliance on and resistance to generative AI in language learning}, volume={22}, ISSN={2365-9440}, url={https://educationaltechnologyjournal.springeropen.com/articles/10.1186/s41239-025-00547-9}, DOI={10.1186/s41239-025-00547-9}, number={1}, journal={International Journal of Educational Technology in Higher Education}, author={Wang, Feifei and Li, Nuoen and Cheung, Alan C. K. and Wong, Gary K. W.}, year={2025}, month=aug, pages={59}, language={en} }

@article{De2016, title={Job crafting and its impact on work engagement and job satisfaction in mining and manufacturing}, volume={19}, rights={https://creativecommons.org/licenses/by/4.0}, ISSN={2222-3436, 1015-8812}, url={https://sajems.org/index.php/sajems/article/view/1481}, DOI={10.4102/sajems.v19i3.1481}, number={3}, journal={South African Journal of Economic and Management Sciences}, author={De Beer, Leon T and Tims, Maria and Bakker, Arnold B}, year={2016}, month=sep, pages={400–412} }

@article{Varanasi2025, title={AI Rivalry as a Craft: How Resisting and Embracing Generative AI Reshape Writing Professions}, url={http://arxiv.org/abs/2503.09901}, DOI={10.1145/3706598.3714035}, note={arXiv:2503.09901}, number={arXiv:2503.09901}, publisher={arXiv}, author={Varanasi, Rama Adithya and Wiesenfeld, Batia Mishan and Nov, Oded}, year={2025}, month=mar }

@article{Toren1975, title={Deprofessionalization and its Sources: A Preliminary Analysis}, volume={2}, rights={https://journals.sagepub.com/page/policies/text-and-data-mining-license}, ISSN={0093-9285}, url={https://journals.sagepub.com/doi/10.1177/073088847500200402}, DOI={10.1177/073088847500200402}, number={4}, journal={Sociology of Work and Occupations}, author={Toren, Nina}, year={1975}, month=nov, pages={323–337}, language={en} }

@inproceedings{Karusala2024, address={Honolulu HI USA}, title={Understanding Contestability on the Margins: Implications for the Design of Algorithmic Decision-making in Public Services}, ISBN={9798400703300}, url={https://dl.acm.org/doi/10.1145/3613904.3641898}, DOI={10.1145/3613904.3641898}, booktitle={Proceedings of the CHI Conference on Human Factors in Computing Systems}, publisher={ACM}, author={Karusala, Naveena and Upadhyay, Sohini and Veeraraghavan, Rajesh and Gajos, Krzysztof Z.}, year={2024}, month=may, pages={1–16}, language={en} }

@inproceedings{Sloane2022, address={Arlington VA USA}, title={Participation Is not a Design Fix for Machine Learning}, ISBN={9781450394772}, url={https://dl.acm.org/doi/10.1145/3551624.3555285}, DOI={10.1145/3551624.3555285}, booktitle={Equity and Access in Algorithms, Mechanisms, and Optimization}, publisher={ACM}, author={Sloane, Mona and Moss, Emanuel and Awomolo, Olaitan and Forlano, Laura}, year={2022}, month=oct, pages={1–6}, language={en} }

@book{Spector2022, address={New York}, edition={1}, title={Job Satisfaction: From Assessment to Intervention}, ISBN={9781003250616}, url={https://www.taylorfrancis.com/books/9781003250616}, DOI={10.4324/9781003250616}, publisher={Routledge}, author={Spector, Paul E.}, year={2022}, month=jan, language={en} }

@article{Baumer2015, title={Why study technology non-use?}, ISSN={1396-0466}, url={https://firstmonday.org/ojs/index.php/fm/article/view/6310}, DOI={10.5210/fm.v20i11.6310}, abstractNote={This special issue provides an opportunity to rethink how we approach, study, and conceptualize human relationships with, and through, technology. The authors in this collection take a multiplicity of approaches on diverse topics to develop a rigorous theoretical understanding for non-use, setting crucial groundwork for future research.}, journal={First Monday}, author={Baumer, Eric P.S. and Ames, Morgan G. and Burrell, Jenna and Brubaker, Jed R. and Dourish, Paul}, year={2015}, month=nov }

@inproceedings{Cha2025, address={Yokohama Japan}, title={Understanding Socio-technical Factors Configuring AI Non-Use in UX Work Practices}, ISBN={9798400713941}, url={https://dl.acm.org/doi/10.1145/3706598.3713140}, DOI={10.1145/3706598.3713140}, booktitle={Proceedings of the 2025 CHI Conference on Human Factors in Computing Systems}, publisher={ACM}, author={Cha, Inha and Wong, Richmond Y.}, year={2025}, month=apr, pages={1–17}, language={en} }

@inproceedings{Schemmer2023, address={Sydney NSW Australia}, title={Appropriate Reliance on AI Advice: Conceptualization and the Effect of Explanations}, ISBN={9798400701061}, url={https://dl.acm.org/doi/10.1145/3581641.3584066}, DOI={10.1145/3581641.3584066}, booktitle={Proceedings of the 28th International Conference on Intelligent User Interfaces}, publisher={ACM}, author={Schemmer, Max and Kuehl, Niklas and Benz, Carina and Bartos, Andrea and Satzger, Gerhard}, year={2023}, month=mar, pages={410–422}, language={en} }

@inproceedings{Cheng2023, address={Hamburg Germany}, title={Overcoming Algorithm Aversion: A Comparison between Process and Outcome Control}, ISBN={9781450394215}, url={https://dl.acm.org/doi/10.1145/3544548.3581253}, DOI={10.1145/3544548.3581253}, booktitle={Proceedings of the 2023 CHI Conference on Human Factors in Computing Systems}, publisher={ACM}, author={Cheng, Lingwei and Chouldechova, Alexandra}, year={2023}, month=apr, pages={1–27}, language={en} }

@article{Dietvorst2015, title={Algorithm aversion: People erroneously avoid algorithms after seeing them err.}, volume={144}, ISSN={1939-2222, 0096-3445}, url={https://doi.apa.org/doi/10.1037/xge0000033}, DOI={10.1037/xge0000033}, number={1}, journal={Journal of Experimental Psychology: General}, author={Dietvorst, Berkeley J. and Simmons, Joseph P. and Massey, Cade}, year={2015}, pages={114–126}, language={en} }

@article{Zito2019, title={The Nature of Job Crafting: Positive and Negative Relations with Job Satisfaction and Work-Family Conflict}, volume={16}, rights={https://creativecommons.org/licenses/by/4.0/}, ISSN={1660-4601}, url={https://www.mdpi.com/1660-4601/16/7/1176}, DOI={10.3390/ijerph16071176}, number={7}, journal={International Journal of Environmental Research and Public Health}, author={Zito, Margherita and Colombo, Lara and Borgogni, Laura and Callea, Antonino and Cenciotti, Roberto and Ingusci, Emanuela and Cortese, Claudio Giovanni}, year={2019}, month=apr, pages={1176}, language={en} }

@article{Petrou2012, title={Crafting a job on a daily basis: Contextual correlates and the link to work engagement}, volume={33}, rights={http://onlinelibrary.wiley.com/termsAndConditions#vor}, ISSN={0894-3796, 1099-1379}, url={https://onlinelibrary.wiley.com/doi/10.1002/job.1783}, DOI={10.1002/job.1783}, number={8}, journal={Journal of Organizational Behavior}, author={Petrou, Paraskevas and Demerouti, Evangelia and Peeters, Maria C. W. and Schaufeli, Wilmar B. and Hetland, Jørn}, year={2012}, month=nov, pages={1120–1141}, language={en} }

@article{Ragu2008, title={The Consequences of Technostress for End Users in Organizations: Conceptual Development and Empirical Validation}, volume={19}, ISSN={1047-7047, 1526-5536}, url={https://pubsonline.informs.org/doi/10.1287/isre.1070.0165}, DOI={10.1287/isre.1070.0165}, number={4}, journal={Information Systems Research}, author={Ragu-Nathan, T. S. and Tarafdar, Monideepa and Ragu-Nathan, Bhanu S. and Tu, Qiang}, year={2008}, month=dec, pages={417–433}, language={en} }

@article{Tang_1999, title={Cooperation or Competition: A Comparison of U.S. and Chinese College Students}, volume={133}, ISSN={0022-3980, 1940-1019}, url={http://www.tandfonline.com/doi/abs/10.1080/00223989909599752}, DOI={10.1080/00223989909599752}, number={4}, journal={The Journal of Psychology}, author={Tang, Shengming}, year={1999}, month=jul, pages={413–423}, language={en} }

@book{Sessa2021, address={New York, N.Y}, title={Essentials of job attitudes and other workplace psychological constructs}, ISBN={9780367344276}, callNumber={158.7},  publisher={Routledge}, author={Sessa, Valerie I. and Bowling, Nathan A.}, year={2021}, language={eng} }

@article{Schuster2021, title={Teacher collaboration networks as a function of type of collaboration and schools’ structural environment}, volume={103}, ISSN={0742051X}, url={https://linkinghub.elsevier.com/retrieve/pii/S0742051X21000962}, DOI={10.1016/j.tate.2021.103372}, journal={Teaching and Teacher Education}, author={Schuster, Johannes and Hartmann, Ulrike and Kolleck, Nina}, year={2021}, month=jul, pages={103372}, language={en} }

@article{Bindl_2019, title={Job crafting revisited: Implications of an extended framework for active changes at work.}, volume={104}, rights={http://www.apa.org/pubs/journals/resources/open-access.aspx}, ISSN={1939-1854, 0021-9010}, url={https://doi.apa.org/doi/10.1037/apl0000362}, DOI={10.1037/apl0000362}, number={5}, journal={Journal of Applied Psychology}, author={Bindl, Uta K. and Unsworth, Kerrie L. and Gibson, Cristina B. and Stride, Christopher B.}, year={2019}, month=may, pages={605–628}, language={en} }

@misc{onet2025database,
  author       = {{National Center for O*NET Development}},
  title        = {O*NET\textsuperscript{®} Database},
  year         = {2025},
  howpublished = {O*NET Resource Center},
  note         = {Version 29.3 database [dataset]; retrieved August 6, 2025},
  url          = {https://www.onetcenter.org/database.html}
}

@inbook{Kellogg_2006, place={Cambridge}, series={Cambridge Handbooks in Psychology}, title={Professional Writing Expertise}, booktitle={The Cambridge Handbook of Expertise and Expert Performance}, publisher={Cambridge University Press}, author={Kellogg, Ronald T.}, editor={Ericsson, K. Anders and Charness, Neil and Feltovich, Paul J. and Hoffman, Robert R.Editors}, year={2006}, pages={389–402}, collection={Cambridge Handbooks in Psychology}}

@article{jorg2009thinking,
  title={Thinking in complexity about learning and education: A programmatic view},
  author={J{\"o}rg, Ton},
  journal={Complicity: An international journal of complexity and education},
  volume={6},
  number={1},
  year={2009}
}

@article{Gupta2023, title={From ChatGPT to ThreatGPT: Impact of Generative AI in Cybersecurity and Privacy}, volume={11}, rights={https://creativecommons.org/licenses/by-nc-nd/4.0/}, ISSN={2169-3536}, url={https://ieeexplore.ieee.org/document/10198233/}, DOI={10.1109/ACCESS.2023.3300381}, journal={IEEE Access}, author={Gupta, Maanak and Akiri, Charankumar and Aryal, Kshitiz and Parker, Eli and Praharaj, Lopamudra}, year={2023}, pages={80218–80245} }

@article{Shin_Koerber_Lim_2025, title={Impact of misinformation from generative AI on user information processing: How people understand misinformation from generative AI}, volume={27}, ISSN={1461-4448, 1461-7315}, url={https://journals.sagepub.com/doi/10.1177/14614448241234040}, DOI={10.1177/14614448241234040}, abstractNote={This study examines the impact of artificial intelligence (AI) on the ways in which users process and respond to misinformation in generative artificial intelligence (GenAI) contexts. Drawing on the heuristic–systematic model and the concept of diagnosticity, our approach examines a cognitive model for processing misinformation in GenAI. The study’s findings revealed that users with a high-heuristic processing mechanism, which affects positive diagnostic perception, were more likely to proactively discern misinformation than users with low-heuristic processing and low-perceived diagnosticity. When exposed to misinformation from GenAI, users’ perceived diagnosticity of misinformation can be accurately predicted by the ways in which they perform heuristic systematic evaluations. With this focus on misinformation processing, this study provides theoretical insights and relevant recommendations for firms to be more resilient in protecting users from the detrimental impacts of misinformation.}, number={7}, journal={New Media \& Society}, author={Shin, Donghee and Koerber, Amy and Lim, Joon Soo}, year={2025}, month=july, pages={4017–4047}, language={en} }

@inproceedings{Varanasi2023b,
author = {Varanasi, Rama Adithya and Goyal, Nitesh},
title = {“It is currently hodgepodge”: Examining AI/ML Practitioners’ Challenges during Co-production of Responsible AI Values},
year = {2023},
isbn = {9781450394215},
publisher = {Association for Computing Machinery},
address = {New York, NY, USA},
url = {https://doi.org/10.1145/3544548.3580903},
doi = {10.1145/3544548.3580903},
booktitle = {Proceedings of the 2023 CHI Conference on Human Factors in Computing Systems},
articleno = {251},
numpages = {17},
location = {Hamburg, Germany},
series = {CHI '23}
}

@article{Veltman2025Anthropic,
  author = {Veltman, Chloe},
  title = {Anthropic settles with authors in first-of-its-kind AI copyright infringement lawsuit},
  journal = {NPR},
  year = {2025},
  month = {September 5},
  url = {https://www.npr.org/2025/09/05/nx-s1-5529404/anthropic-settlement-authors-copyright-ai}
}

@inproceedings{varanasi2023,
author = {Varanasi, Rama Adithya and Siddarth, Divya and Seshadri, Vivek and Bali, Kalika and Vashistha, Aditya},
title = {Feeling Proud, Feeling Embarrassed: Experiences of Low-income Women with Crowd Work},
year = {2022},
isbn = {9781450391573},
publisher = {Association for Computing Machinery},
address = {New York, NY, USA},
url = {https://doi.org/10.1145/3491102.3501834},
doi = {10.1145/3491102.3501834},
booktitle = {Proceedings of the 2022 CHI Conference on Human Factors in Computing Systems},
articleno = {298},
numpages = {18},
location = {New Orleans, LA, USA},
series = {CHI '22}
}

@article{Wang2020, title={Cross-Sectional Studies}, volume={158}, ISSN={00123692}, url={https://linkinghub.elsevier.com/retrieve/pii/S0012369220304621}, DOI={10.1016/j.chest.2020.03.012}, number={1}, journal={Chest}, author={Wang, Xiaofeng and Cheng, Zhenshun}, year={2020}, month=july, pages={S65–S71}, language={en} }

@article{Nylund2018, title={Ten frequently asked questions about latent class analysis.}, volume={4}, rights={http://www.apa.org/pubs/journals/resources/open-access.aspx}, ISSN={2332-2179, 2332-2136}, url={https://doi.apa.org/doi/10.1037/tps0000176}, DOI={10.1037/tps0000176}, number={4}, journal={Translational Issues in Psychological Science}, author={Nylund-Gibson, Karen and Choi, Andrew Young}, year={2018}, month=dec, pages={440–461}, language={en} }

@article{liu2025discerning,
  title={Discerning minds or generic tutors? Evaluating instructional guidance capabilities in Socratic LLMs},
  author={Liu, Ying and Li, Can and Zhang, Ting and Wang, Mei and Zhu, Qiannan and Li, Jian and Huang, Hua},
  journal={arXiv preprint arXiv:2508.06583},
  year={2025}
}

@book{Lowe2010, address={Anderson}, series={Writing Spaces}, title={Writing Spaces 1: Readings on Writing}, ISBN={9781602351851}, publisher={Parlor Press, LLC}, author={Lowe, Charles}, year={2010}, collection={Writing Spaces}, language={eng} }

@article{Blasco_Charisi_2024, title={The Impact of Large Language Models on Students: A Randomised Study of Socratic vs. Non-Socratic AI and the Role of Step-by-Step Reasoning}, url={https://www.ssrn.com/abstract=5040921}, DOI={10.2139/ssrn.5040921}, author={Blasco, Andrea and Charisi, Vicky}, year={2024} }

@book{Schon2017, edition={0}, title={The Reflective Practitioner}, ISBN={9781351883160}, url={https://www.taylorfrancis.com/books/9781351883160}, DOI={10.4324/9781315237473}, publisher={Routledge}, author={Schön, Donald A.}, year={2017}, month=mar, language={en} }

@inproceedings{zhao25,
author = {Zhao, Zixin and Masson, Damien and Kim, Young-Ho and Penn, Gerald and Chevalier, Fanny},
title = {Making the Write Connections: Linking Writing Support Tools with Writer Needs},
year = {2025},
isbn = {9798400713941},
publisher = {Association for Computing Machinery},
address = {New York, NY, USA},
url = {https://doi.org/10.1145/3706598.3713161},
doi = {10.1145/3706598.3713161},
booktitle = {Proceedings of the 2025 CHI Conference on Human Factors in Computing Systems},
articleno = {1216},
numpages = {21},
location = {
},
series = {CHI '25}
}

@misc{onet,
  author       = {U.S. Department of Labor, Employment and Training Administration},
  title        = {O*NET OnLine},
  year         = {Year},
  url          = {https://www.onetonline.org},
  note         = {Accessed: Month Day, Year}
}

@book{cornelissen2020corporate,
  title={Corporate Communication: A Guide to Theory and Practice},
  author={Cornelissen, J.},
  isbn={9781529712674},
  url={https://books.google.com/books?id=z0i3DwAAQBAJ},
  year={2020},
  publisher={SAGE Publications}
}

@book{Kynell_1996, address={Norwood, NJ}, series={The Ablex communication, culture \& information series}, title={Writing in a milieu of utility: the move to technical communication in American engineering programs, 1850 - 1950}, ISBN={9781567502640}, publisher={Ablex Publ. Corp}, author={Kynell, Teresa C.}, year={1996}, collection={The Ablex communication, culture \& information series}, language={eng} }

@book{florida2019rise,
  title={The Rise of the Creative Class},
  author={Florida, R.},
  isbn={9781541617735},
  lccn={2019300475},
  url={https://books.google.com/books?id=UkODDwAAQBAJ},
  year={2019},
  publisher={Basic Books}
}

@inproceedings{yun25,
author = {Yun, Bhada and Feng, Dana and Chen, Ace S. and Nikzad, Afshin and Salehi, Niloufar},
title = {Generative AI in Knowledge Work: Design Implications for Data Navigation and Decision-Making},
year = {2025},
isbn = {9798400713941},
publisher = {Association for Computing Machinery},
address = {New York, NY, USA},
url = {https://doi-org.proxy.library.nyu.edu/10.1145/3706598.3713337},
doi = {10.1145/3706598.3713337},
booktitle = {Proceedings of the 2025 CHI Conference on Human Factors in Computing Systems},
articleno = {634},
numpages = {19},
location = {
},
series = {CHI '25}
}

@article{Connolly2024, title={(Inaccurate) Beliefs about Skill Decay}, url={https://www.ssrn.com/abstract=4916412}, DOI={10.2139/ssrn.4916412}, author={Connolly, Daniel and Horn, Samantha and Loewenstein, George F.}, year={2024} }

@article{Ahmad2025, title={Endoscopist deskilling: an unintended consequence of AI-assisted colonoscopy?}, ISSN={24681253}, url={https://linkinghub.elsevier.com/retrieve/pii/S2468125325001645}, DOI={10.1016/S2468-1253(25)00164-5}, journal={The Lancet Gastroenterology \& Hepatology}, author={Ahmad, Omer F}, year={2025}, month=aug, pages={S2468125325001645}, language={en} }

@article{Macnamara2024, title={Does using artificial intelligence assistance accelerate skill decay and hinder skill development without performers’ awareness?}, volume={9}, ISSN={2365-7464}, url={https://cognitiveresearchjournal.springeropen.com/articles/10.1186/s41235-024-00572-8}, DOI={10.1186/s41235-024-00572-8}, number={1}, journal={Cognitive Research: Principles and Implications}, author={Macnamara, Brooke N. and Berber, Ibrahim and Çavuşoğlu, M. Cenk and Krupinski, Elizabeth A. and Nallapareddy, Naren and Nelson, Noelle E. and Smith, Philip J. and Wilson-Delfosse, Amy L. and Ray, Soumya}, year={2024}, month=jul, pages={46}, language={en} }

@misc{park2023generative,
      title={Generative Agents: Interactive Simulacra of Human Behavior}, 
      author={Joon Sung Park and Joseph C. O'Brien and Carrie J. Cai and Meredith Ringel Morris and Percy Liang and Michael S. Bernstein},
      year={2023},
      eprint={2304.03442},
      archivePrefix={arXiv},
      primaryClass={cs.HC},
      url={https://arxiv.org/abs/2304.03442}, 
}

@inproceedings{boucher2024,
author = {Boucher, Josiah D and Smith, Gillian and Telliel, Yunus Do\u{g}an},
title = {Is Resistance Futile?: Early Career Game Developers, Generative AI, and Ethical Skepticism},
year = {2024},
isbn = {9798400703300},
publisher = {Association for Computing Machinery},
address = {New York, NY, USA},
url = {https://doi.org/10.1145/3613904.3641889},
doi = {10.1145/3613904.3641889},
booktitle = {Proceedings of the CHI Conference on Human Factors in Computing Systems},
articleno = {173},
numpages = {13},
location = {Honolulu, HI, USA},
series = {CHI '24}
}

@article{Wong_Fieseler_2021, title={From crafting what you do to building resilience for career commitment in the gig economy}, volume={31}, ISSN={0954-5395, 1748-8583}, url={https://onlinelibrary.wiley.com/doi/10.1111/1748-8583.12342}, DOI={10.1111/1748-8583.12342}, number={4}, journal={Human Resource Management Journal}, author={Wong, Sut I and Kost, Dominique and Fieseler, Christian}, year={2021}, month=nov, pages={918–935}, language={en} }

@article{Dominguez2016, title={Why Individuals Participate in Micro-task Crowdsourcing Work Environment: Revealing Crowdworkers’ Perceptions}, volume={17}, ISSN={15369323}, url={http://aisel.aisnet.org/jais/vol17/iss10/3/}, DOI={10.17705/1jais.00441}, number={10}, journal={Journal of the Association for Information Systems}, author={California State University, Dominguez Hills and Deng, Xuefei and Joshi, K.D. and Washington State University}, year={2016}, month=nov, pages={648–673} }

@article{Bizzi_2017, title={Network characteristics: When an individual’s job crafting depends on the jobs of others}, volume={70}, ISSN={0018-7267, 1741-282X}, url={http://journals.sagepub.com/doi/10.1177/0018726716658963}, DOI={10.1177/0018726716658963},  number={4}, journal={Human Relations}, author={Bizzi, Lorenzo}, year={2017}, month=apr, pages={436–460}, language={en} }

@article{Gravador2018, title={Work-life balance crafting behaviors: an empirical study}, volume={47}, rights={https://www.emerald.com/insight/site-policies}, ISSN={0048-3486}, url={https://www.emerald.com/insight/content/doi/10.1108/PR-05-2016-0112/full/html}, DOI={10.1108/PR-05-2016-0112}, number={4}, journal={Personnel Review}, author={Gravador, Luz Nario and Teng-Calleja, Mendiola}, year={2018}, month=may, pages={786–804}, language={en} }

@article{Kossek2016, title={Filling the Holes: Work Schedulers As Job Crafters of Employment Practice in Long-Term Health Care}, volume={69}, ISSN={0019-7939, 2162-271X}, url={http://journals.sagepub.com/doi/10.1177/0019793916642761}, DOI={10.1177/0019793916642761}, number={4}, journal={ILR Review}, author={Kossek, Ellen Ernst and Piszczek, Matthew M. and McAlpine, Kristie L. and Hammer, Leslie B. and Burke, Lisa}, year={2016}, month=aug, pages={961–990}, language={en} }

@article{Bruning2019, title={Exploring job crafting: Diagnosing and responding to the ways employees adjust their jobs}, volume={62}, ISSN={00076813}, url={https://linkinghub.elsevier.com/retrieve/pii/S0007681319300734}, DOI={10.1016/j.bushor.2019.05.003}, number={5}, journal={Business Horizons}, author={Bruning, Patrick F. and Campion, Michael A.}, year={2019}, month=sep, pages={625–635}, language={en} }

@misc{clarke2025,
      title={Beyond Replacement or Augmentation: How Creative Workers Reconfigure Division of Labor with Generative AI}, 
      author={Michael Clarke and Michael Joffe},
      year={2025},
      eprint={2505.18938},
      archivePrefix={arXiv},
      primaryClass={cs.CY},
      url={https://arxiv.org/abs/2505.18938}, 
}

@article{Christin_2020, title={The ethnographer and the algorithm: beyond the black box}, volume={49}, ISSN={0304-2421, 1573-7853}, url={https://link.springer.com/10.1007/s11186-020-09411-3}, DOI={10.1007/s11186-020-09411-3}, number={5–6}, journal={Theory and Society}, author={Christin, Angèle}, year={2020}, month=oct, pages={897–918}, language={en} }

@article{Golgeci2025, title={Confronting and alleviating AI resistance in the workplace: An integrative review and a process framework}, volume={35}, ISSN={10534822}, url={https://linkinghub.elsevier.com/retrieve/pii/S1053482224000652}, DOI={10.1016/j.hrmr.2024.101075}, number={2}, journal={Human Resource Management Review}, author={Golgeci, Ismail and Ritala, Paavo and Arslan, Ahmad and McKenna, Brad and Ali, Imran}, year={2025}, month=jun, pages={101075}, language={en} }

@article{Wen2025, title={Trust and AI weight: human-AI collaboration in organizational management decision-making}, volume={3}, ISSN={2813-771X}, url={https://www.frontiersin.org/articles/10.3389/forgp.2025.1419403/full}, DOI={10.3389/forgp.2025.1419403}, journal={Frontiers in Organizational Psychology}, author={Wen, Yanjun and Wang, Jiale and Chen, Xiaoxi}, year={2025}, month=jun, pages={1419403} }

@article{Berg2010, title={When Callings Are Calling: Crafting Work and Leisure in Pursuit of Unanswered Occupational Callings}, volume={21}, ISSN={1047-7039, 1526-5455}, url={https://pubsonline.informs.org/doi/10.1287/orsc.1090.0497}, DOI={10.1287/orsc.1090.0497}, number={5}, journal={Organization Science}, author={Berg, Justin M. and Grant, Adam M. and Johnson, Victoria}, year={2010}, month=oct, pages={973–994}, language={en} }

@article{Tims_2012, title={Development and validation of the job crafting scale}, volume={80}, rights={https://www.elsevier.com/tdm/userlicense/1.0/}, ISSN={00018791}, url={https://linkinghub.elsevier.com/retrieve/pii/S0001879111000789}, DOI={10.1016/j.jvb.2011.05.009}, number={1}, journal={Journal of Vocational Behavior}, author={Tims, Maria and Bakker, Arnold B. and Derks, Daantje}, year={2012}, month=feb, pages={173–186}, language={en} }

@article{Petrou2017, title={Weekly job crafting and leisure crafting: Implications for meaning‐making and work engagement}, volume={90}, rights={http://onlinelibrary.wiley.com/termsAndConditions#vor}, ISSN={0963-1798, 2044-8325}, url={https://bpspsychub.onlinelibrary.wiley.com/doi/10.1111/joop.12160}, DOI={10.1111/joop.12160}, number={2}, journal={Journal of Occupational and Organizational Psychology}, author={Petrou, Paraskevas and Bakker, Arnold B. and Van Den Heuvel, Machteld}, year={2017}, month=jun, pages={129–152}, language={en} }

@article{Parker2025, title={Top-Down and Bottom-Up Work Design: A Multilevel Perspective on How Job Crafting and Work Characteristics Interrelate}, ISSN={0889-3268, 1573-353X}, url={https://link.springer.com/10.1007/s10869-025-10010-1}, DOI={10.1007/s10869-025-10010-1}, journal={Journal of Business and Psychology}, author={Parker, Sharon K. and Tims, Maria and Sonnentag, Sabine}, year={2025}, month=feb, language={en} }

@article{Bruning_Campion_2018, title={A Role–resource Approach–avoidance Model of Job Crafting: A Multimethod Integration and Extension of Job Crafting Theory}, volume={61}, ISSN={0001-4273, 1948-0989}, url={http://journals.aom.org/doi/10.5465/amj.2015.0604}, DOI={10.5465/amj.2015.0604}, number={2}, journal={Academy of Management Journal}, author={Bruning, Patrick F. and Campion, Michael A.}, year={2018}, month=apr, pages={499–522}, language={en} }

@article{zhang_2019, title={Reorienting job crafting research: A hierarchical structure of job crafting concepts and integrative review}, volume={40}, ISSN={0894-3796, 1099-1379}, url={https://onlinelibrary.wiley.com/doi/10.1002/job.2332}, DOI={10.1002/job.2332}, number={2}, journal={Journal of Organizational Behavior}, author={Zhang, Fangfang and Parker, Sharon K.}, year={2019}, month=feb, pages={126–146}, language={en} }

@article{Wrzesniewski_2001, title={Crafting a Job: Revisioning Employees as Active Crafters of Their Work}, volume={26}, ISSN={03637425}, url={http://www.jstor.org/stable/259118?origin=crossref}, DOI={10.2307/259118}, number={2}, journal={The Academy of Management Review}, author={Wrzesniewski, Amy and Dutton, Jane E.}, year={2001}, month=apr, pages={179} }

@article{Kilduff_2014, title={Driven to Win: Rivalry, Motivation, and Performance}, volume={5}, ISSN={1948-5506, 1948-5514}, url={https://journals.sagepub.com/doi/10.1177/1948550614539770}, DOI={10.1177/1948550614539770}, number={8}, journal={Social Psychological and Personality Science}, author={Kilduff, Gavin J.}, year={2014}, month=nov, pages={944–952}, language={en} }

@inproceedings{Kobiella2024, address={New York, NY, USA}, series={CHI ’24}, title={“If the Machine Is As Good As Me, Then What Use Am I?” – How the Use of ChatGPT Changes Young Professionals’ Perception of Productivity and Accomplishment}, ISBN={9798400703300}, url={https://doi.org/10.1145/3613904.3641964}, DOI={10.1145/3613904.3641964}, booktitle={Proceedings of the CHI Conference on Human Factors in Computing Systems}, publisher={Association for Computing Machinery}, author={Kobiella, Charlotte and Flores López, Yarhy Said and Waltenberger, Franz and Draxler, Fiona and Schmidt, Albrecht}, year={2024}, month=may, pages={1–16}, collection={CHI ’24} }

@inproceedings{Guo25,
author = {Guo, Alicia and Sathyanarayanan, Shreya and Wang, Leijie and Heer, Jeffrey and Zhang, Amy X.},
title = {From Pen to Prompt: How Creative Writers Integrate AI into their Writing Practice},
year = {2025},
isbn = {9798400712890},
publisher = {Association for Computing Machinery},
address = {New York, NY, USA},
url = {https://doi-org.proxy.library.nyu.edu/10.1145/3698061.3726910},
doi = {10.1145/3698061.3726910},
booktitle = {Proceedings of the 2025 Conference on Creativity and Cognition},
pages = {527–545},
numpages = {19},
location = {
},
series = {C\&C '25}
}

@article{Lee2025, title={The Student Physician–Patient–AI Relationship}, volume={2}, ISSN={2836-9386}, url={https://ai.nejm.org/doi/10.1056/AIp2500260}, DOI={10.1056/AIp2500260}, number={8}, journal={NEJM AI}, author={Lee, Albert H. and Link, Collin D. and Hersh, David and Angoff, Nancy R.}, year={2025}, month=jul, language={en} }

@article{Kilduff2010, title={The Psychology of Rivalry: A Relationally Dependent Analysis of Competition}, volume={53}, ISSN={0001-4273, 1948-0989}, url={http://journals.aom.org/doi/10.5465/amj.2010.54533171}, DOI={10.5465/amj.2010.54533171}, number={5}, journal={Academy of Management Journal}, author={Kilduff, Gavin J. and Elfenbein, Hillary Anger and Staw, Barry M.}, year={2010}, month=oct, pages={943–969}, language={en} }

@book{Judge2007, edition={0}, title={Perspectives on Organizational Fit},  author={Ostroff, C. and Judge, T.A.}, ISBN={9781136679223}, series={SIOP Organizational Frontiers Series}, url={https://www.taylorfrancis.com/books/9781136679223}, DOI={10.4324/9780203810026}, publisher={Taylor \& Francis}, year={2007}, month=jun, language={en} }

@article{Humberg2019,
  author    = {Sarah Humberg and Jens B. Asendorpf and Mitja D. Back},
  title     = {Response surface analysis in personality and social psychology: Checklist and clarifications for the case of congruence hypotheses},
  journal   = {Social Psychological and Personality Science},
  year      = {2019},
  volume    = {10},
  number    = {3},
  pages     = {409--419},
  doi       = {10.1177/1948550618757600},
  url       = {https://doi.org/10.1177/1948550618757600}
}

@article{Timmermans_2012, title={Theory Construction in Qualitative Research: From Grounded Theory to Abductive Analysis}, volume={30}, ISSN={0735-2751, 1467-9558}, url={http://journals.sagepub.com/doi/10.1177/0735275112457914}, DOI={10.1177/0735275112457914},author={Timmermans, Stefan and Tavory, Iddo}, year={2012}, month=sep, pages={167–186}, language={en} }

@article{Creswell_Miller_2000, title={Determining Validity in Qualitative Inquiry}, volume={39}, ISSN={0040-5841, 1543-0421}, url={http://www.tandfonline.com/doi/abs/10.1207/s15430421tip3903_2}, DOI={10.1207/s15430421tip3903_2}, number={3}, journal={Theory Into Practice}, author={Creswell, John W. and Miller, Dana L.}, year={2000}, month=aug, pages={124–130}, language={en} }

\clearpage
\appendix

% \paragraph{GenAI Relationship Profiles: Descriptive Statistics}

\begin{table*} [t]
  \centering
  \caption{Means and standard deviations of rivalry and collaboration values across four GenAI relationship profiles (based on the PAM profiles calculated in Section \ref{methods-analysis}). The table presents total sample count, followed by means and standard deviation of each profile.}
  \label{tab:profile-means}
  \begin{tabular}{lccccc}
    \toprule
    \multirow{2}{*}{Profiles (n)} & 
    \multicolumn{2}{c}{Rivalry} & 
    \multicolumn{2}{c}{Collaboration} \\
    \cmidrule(lr){2-3} \cmidrule(lr){4-5}
     & Mean & SD & Mean & SD \\
    \midrule
    LowR/LowC (65)   & 2.66 & 0.61 & 3.86 & 1.12 \\
    LowR/HighC (108) & 2.66 & 0.74 & 6.33 & 0.56 \\
    HighR/LowC (84)  & 4.37 & 0.91 & 2.53 & 1.07 \\
    HighR/HighC (146)& 4.86 & 0.71 & 5.67 & 0.82 \\
    \bottomrule
  \end{tabular}
\end{table*}

% \large \textbf{GenAI Relationship Profiles: Work Details}
\begin{figure*} [t]
    \centering
    \includegraphics[width=1\linewidth]{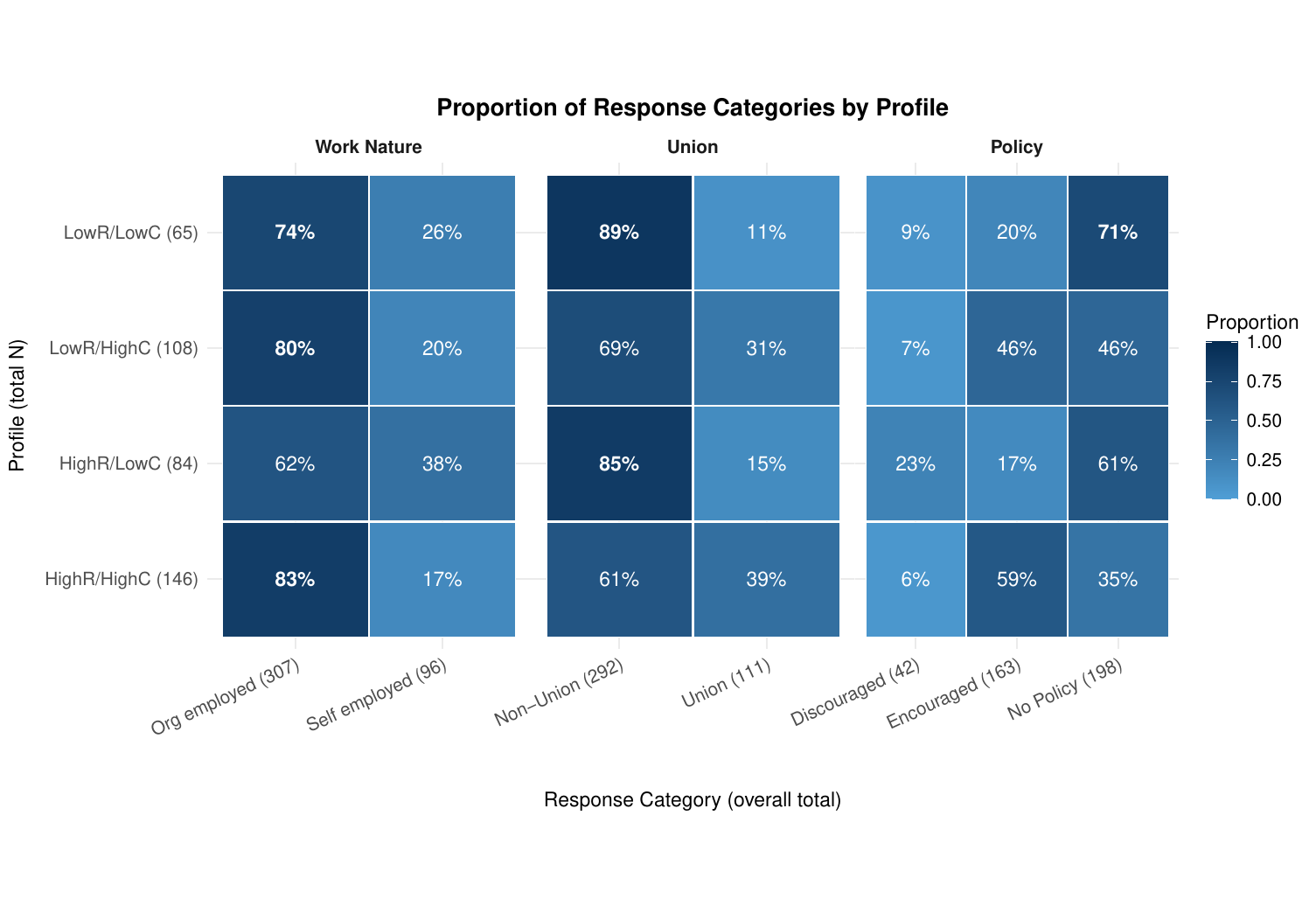}
    \caption{The figure depicts the distribution of the four profiles across three variables. The first panel indicates workplace policies regarding GenAI, whether its use was discouraged, encouraged, or not specified. The second panel presents professionals’ union membership status. The third panel illustrates the nature of employment, distinguishing between organizational employment and self-employment.}
    \label{fig:profile-worknature}
\end{figure*}

% \large \textbf{GenAI Relationship Profiles: GenAI Background} 

\begin{figure*} [t]
    \centering
    \includegraphics[width=1\linewidth]{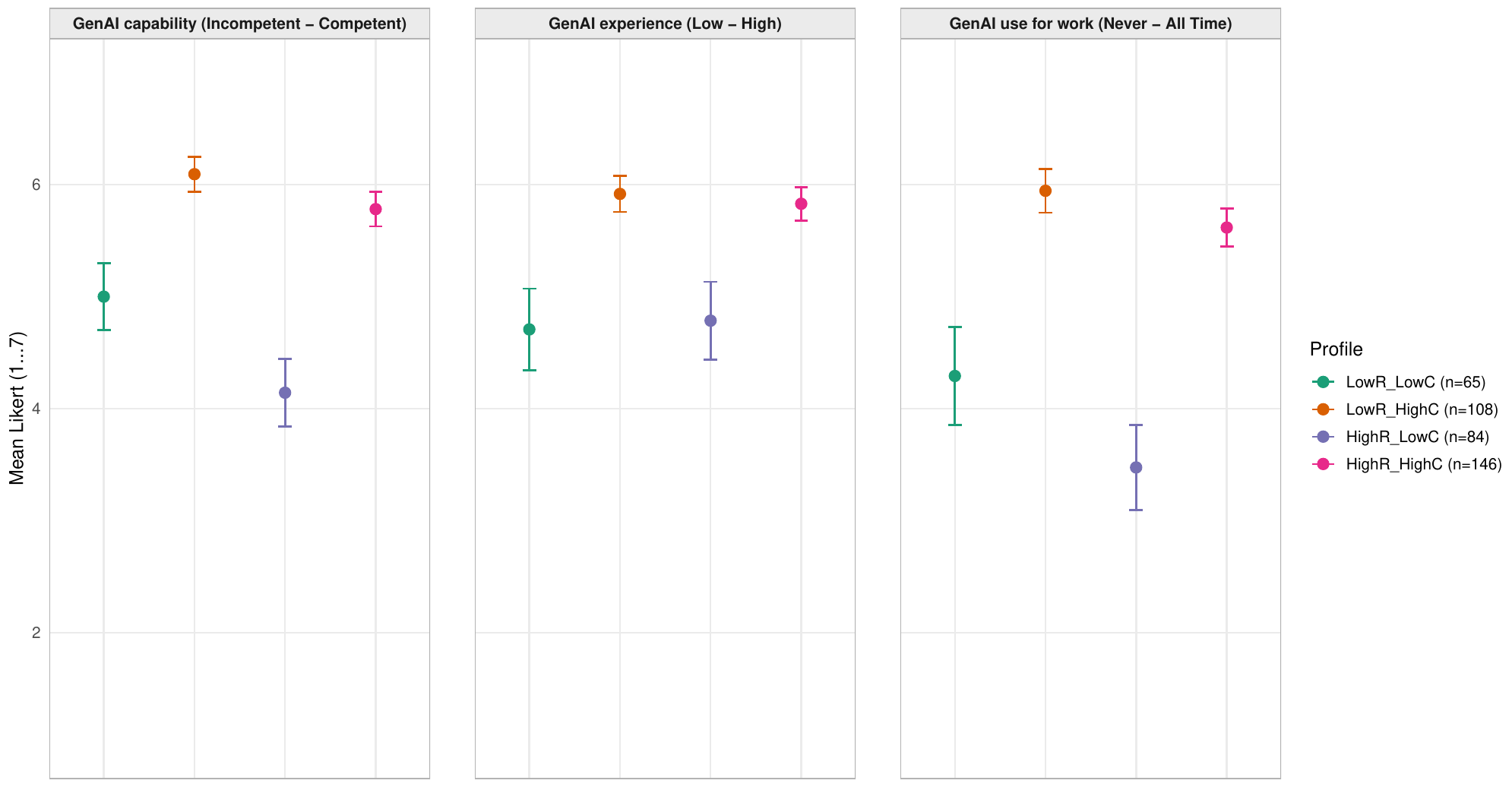}
    \caption{The figure presents the means and confidence intervals for the four profiles across three variables capturing GenAI background. The first panel reflects how capable professionals perceived GenAI to be in their work (1 = incompetent, 7 = competent). The second panel shows professionals’ overall perceived experience with GenAI in their work (1 = low, 7 = high). The third panel depicts the frequency of GenAI use for work (1 = never, 7 = all the time).}
    \label{fig:profile-genai-use}
\end{figure*}

% \large \textbf{Response Surface Analysis of Outcome Variables}
\begingroup
\setlength{\intextsep}{0pt}% space above/below [H] floats
\setlength{\textfloatsep}{0pt}% (harmless here, affects normal floats)
\begin{table*}[t]
\centering
\caption{Response Surface Analysis (RSA) regression results across different outcomes. Entries are OLS coefficients with standard errors in parentheses.}
\label{tab:rsa_all}
\small
\begin{tabular}{lcccc}
\toprule
 & \textbf{Job Crafting} & \textbf{Skill Maintenance} & \textbf{Productivity} & \textbf{Satisfaction} \\
\midrule
Rivalry (R)                     & $0.216^{***} \; (0.029)$ & $0.225^{***} \; (0.048)$ & $0.031 \; (0.033)$ & $0.014 \; (0.040)$ \\
Collaboration (C)               & $0.200^{***} \; (0.027)$ & $0.047 \; (0.045)$       & $0.213^{***} \; (0.031)$ & $0.111^{**} \; (0.037)$ \\
$R^2$                           & $0.024 \; (0.020)$       & $0.058 \; (0.034)$       & $0.038 \; (0.023)$   & $0.046 \; (0.028)$ \\
$C^2$                           & $-0.024^{*} \; (0.012)$ & $-0.003 \; (0.020)$    & $0.040^{**} \; (0.014)$  & $-0.003 \; (0.017)$ \\
$R \times C$                    & $-0.042^{**} \; (0.016)$ & $-0.002 \; (0.026)$   & $-0.012 \; (0.018)$     & $-0.040^{\dagger} \; (0.022)$ \\
Age                             & $-0.012^{***} \; (0.003)$ & $0.001 \; (0.005)$     & $-0.003 \; (0.003)$     & $0.006^{\dagger} \; (0.004)$ \\
Sex (Female)                    & $-0.072 \; (0.072)$    & $-0.079 \; (0.121)$    & $-0.037 \; (0.084)$     & $0.115 \; (0.100)$ \\
Sex (Non-binary)                & $-0.101 \; (0.326)$    & $-0.401 \; (0.547)$    & $-0.147 \; (0.379)$     & $0.026 \; (0.451)$ \\
Education (Mid)                 & $0.096 \; (0.097)$       & $0.014 \; (0.163)$       & $0.127 \; (0.113)$        & $-0.038 \; (0.134)$ \\
Education (High)                & $0.253^{*} \; (0.103)$   & $0.122 \; (0.173)$       & $0.140 \; (0.120)$        & $0.039 \; (0.143)$ \\
\midrule
$R^2$                           & $0.359$ & $0.073$ & $0.147$ & $0.065$ \\
Adj. $R^2$                      & $0.343$ & $0.049$ & $0.125$ & $0.041$ \\
$F$ (model df, resid df)        & $21.96 \; (10, 392)^{***}$ & $3.08 \; (10, 392)^{***}$ & $6.75 \; (10, 392)^{***}$ & $2.73 \; (10, 392)^{**}$ \\
\addlinespace
$\Delta R^2$ vs. linear model   & $0.020^{**}$ & $0.008 \; (n.s.)$ & $0.034^{**}$ & $0.019^{\dagger}$ \\
$\Delta F$ (3, 392)             & $4.12^{**}$  & $1.07 \; (n.s.)$  & $5.21^{**}$  & $2.61^{\dagger}$ \\
\bottomrule
\end{tabular}

\begin{flushleft}
\footnotesize
\textit{Notes.} Predictors were mean-centered before computing squares and the interaction.  Coefficients are unstandardized; standard errors in parentheses. $\Delta R^2$ rows compare the RSA model to the linear model with only $R$ and $C$ (plus the same controls). $^{\dagger}p<.10$, $^{*}p<.05$, $^{**}p<.01$, $^{***}p<.001$.
\end{flushleft}
\end{table*}
\endgroup

% \large \textbf{Regression Analysis of PAM profile}
\begingroup
\setlength{\intextsep}{0pt}% space above/below [H] floats
\setlength{\textfloatsep}{0pt}% (harmless here, affects normal floats)
\begin{table*} [t]
\centering
\small
\caption{Standardized coefficients ($\beta$) with standard errors in parentheses for predictors (baseline = LowR/LowC) on crafting outcomes and productivity. Significance: * $p<.05$, ** $p<.01$, *** $p<.001$.}
\label{tab:pam_crafting_prod}
\begin{tabular}{lccccc c}
\toprule
& \multicolumn{5}{c}{\textbf{Crafting}} & \textbf{Outcome} \\
\cmidrule(lr){2-6}\cmidrule(lr){7-7}
& Task & Relationship & Cognitive & Approach & Avoidance & Productivity \\
\midrule
\textbf{PAM item class} \\
\quad LowR/HighC        & 0.251*** (0.162) & 0.136* (0.160)   & 0.338*** (0.144) & 0.248*** (0.164) & 0.217** (0.152)  & 0.373*** (0.134) \\
\quad HighR/LowC        & -0.075 (0.169)   & 0.090 (0.167)    & 0.080 (0.150)    & 0.049 (0.171)    & -0.004 (0.158)   & 0.031 (0.140) \\
\quad HighR/HighC       & 0.367*** (0.156) & 0.349*** (0.153) & 0.418*** (0.137) & 0.364*** (0.156) & 0.382*** (0.145) & 0.247*** (0.128) \\
\addlinespace[2pt]
\textbf{Controls} \\
\quad Age                      & -0.194*** (0.004) & -0.186*** (0.004) & -0.073 (0.003) & -0.169*** (0.004) & -0.126** (0.004) & -0.043 (0.003) \\
\quad Sex: Female              & -0.019 (0.103)    & -0.084. (0.102)   & 0.002 (0.091)  & -0.020 (0.104)    & -0.054 (0.097)   & -0.026 (0.085) \\
\quad Sex: Non-binary/Third    & -0.042 (0.464)    & -0.030 (0.458)    & 0.021 (0.411)  & -0.021 (0.468)    & -0.015 (0.434)   & -0.042 (0.384) \\
\quad Education: Mid           & 0.053 (0.138)     & 0.100 (0.137)     & 0.037 (0.123)  & 0.121. (0.140)    & -0.023 (0.130)   & 0.060 (0.114) \\
\quad Education: High          & 0.100 (0.148)     & 0.140* (0.146)    & 0.114. (0.131) & 0.224*** (0.149)  & -0.042 (0.138)   & 0.056 (0.122) \\
\midrule
$R^2$                 & 0.222 & 0.142 & 0.154 & 0.176 & 0.145 & 0.118 \\
Adjusted $R^2$        & 0.206 & 0.124 & 0.137 & 0.160 & 0.128 & 0.100 \\
F-statistic (df)      & 14.04 (8,394) & 8.13 (8,394) & 8.96 (8,394) & 10.54 (8,394) & 8.35 (8,394) & 6.61 (8,394) \\
\bottomrule
\end{tabular}
\end{table*}
\endgroup

\begingroup
\setlength{\intextsep}{0pt}% space above/below [H] floats
\setlength{\textfloatsep}{0pt}% (harmless here, affects normal floats)
\begin{table*}[t]
\centering
\small
\caption{Standardized coefficients ($\beta$) with standard errors in parentheses for predictors (baseline = LowR/LowC) on skill maintenance outcomes and satisfaction. Significance: * $p<.05$, ** $p<.01$, *** $p<.001$.}
\label{tab:pam_skill_sat}
\begin{tabular}{lcccc c}
\toprule
& \multicolumn{4}{c}{\textbf{Skill Maintenance}} & \textbf{Outcome} \\
\cmidrule(lr){2-5}\cmidrule(lr){6-6}
& Writing & Social & Cognitive & Tacit & Satisfaction \\
\midrule
\textbf{PAM item class} \\
\quad LowR/HighC        & -0.001 (0.229)  & -0.009 (0.213)  & 0.081 (0.224)  & -0.000 (0.229)  & 0.192** (0.159) \\
\quad HighR/LowC        & 0.096 (0.239)   & 0.075 (0.223)   & -0.010 (0.234) & 0.029 (0.239)   & -0.018 (0.166) \\
\quad HighR/HighC       & 0.133 (0.218)   & 0.206** (0.203) & 0.179* (0.213) & 0.158* (0.218)  & 0.073 (0.151) \\
\addlinespace[2pt]
\textbf{Controls} \\
\quad Age                      & 0.009 (0.005)   & -0.071 (0.005) & 0.080 (0.005) & 0.053 (0.005) & 0.085. (0.004) \\
\quad Sex: Female              & -0.031 (0.146)  & -0.087. (0.136) & 0.012 (0.143) & -0.019 (0.146) & 0.053 (0.101) \\
\quad Sex: Non-binary/Third    & -0.051 (0.655)  & -0.053 (0.611) & -0.011 (0.641) & -0.030 (0.655) & -0.004 (0.454) \\
\quad Education: Mid           & -0.033 (0.195)  & 0.064 (0.182) & 0.043 (0.191) & -0.027 (0.195) & -0.018 (0.136) \\
\quad Education: High          & 0.024 (0.209)   & 0.155* (0.195) & 0.017 (0.204) & -0.015 (0.209) & 0.016 (0.145) \\
\midrule
$R^2$                 & 0.024 & 0.072 & 0.032 & 0.027 & 0.043 \\
Adjusted $R^2$        & 0.004 & 0.053 & 0.012 & 0.007 & 0.023 \\
F-statistic (df)      & 1.22 (8,394) & 3.80 (8,394) & 1.63 (8,394) & 1.34 (8,394) & 2.20 (8,394) \\
\bottomrule
\end{tabular}
\end{table*}
\endgroup

% \large \textbf{Item-Total for Skill Maintenance (21 items)} 
\begingroup
\setlength{\intextsep}{0pt}% space above/below [H] floats
\setlength{\textfloatsep}{0pt}% (harmless here, affects normal floats)
\begin{table*}[t]
\centering
\caption{Corrected Item–Total Correlations for Skill-Maintenance Items}
\label{item-total21}
\begin{tabular}{lcc}
\toprule
\textbf{Item} & \textbf{Corrected Item–Total} & \textbf{Mean (SD)} \\
\midrule
(Cog1) Breaking down a problem & .735 & 4.61 (1.63) \\
(Cog2) Generating  ideas/concepts/inspirations & .717 & 4.74 (1.67) \\
(Cog3) Organization/translation of ideas into narratives  & .711 & 4.63 (1.61) \\
(Cog4) Evaluating and quality checking of ideas & .689 & 4.93 (1.58) \\
(Cog5) Recalling a particular idea & .689 & 4.56 (1.60) \\
(Tacit1) Managing time effectively & .760 & 4.72 (1.71) \\
(Tacit2) Planning and managing writing tasks & .793 & 4.56 (1.73) \\
(Tacit3) Managing  resources for writing & .779 & 4.54 (1.65) \\
(Tacit4) Learning and apply new techniques & .789 & 4.84 (1.65) \\
(Tacit5) Engaging in self-reflection & .674 & 4.73 (1.53) \\
(Relation1) Negotiation with co-workers/clients & .601 & 4.24 (1.54) \\
(Relation2) Persuading co-workers/clients & .614 & 4.30 (1.58) \\
(Relation3) Collaborating with other professionals  & .683 & 4.48 (1.61) \\
(Relation4) Seeking or providing writing assistance & .723 & 4.27 (1.72) \\
(Relation5) Instructing or managing other professionals & .585 & 4.26 (1.57) \\
(Creative1) Comprehending written content & .788 & 4.62 (1.64) \\
(Creative2) Writing persuasively & .788 & 4.67 (1.65) \\
(Creative3) Writing with clarity  & .797 & 4.87 (1.68) \\
(Creative4) Developing tone/voice of characters & .720 & 4.51 (1.59) \\
(Creative5) Storytelling and narrative building & .760 & 4.57 (1.67) \\
(Creative6) Proofreading and correcting writing & .706 & 4.66 (1.80) \\
\bottomrule
\end{tabular}
\end{table*}
\endgroup

% \large \textbf{Exploratory Factor Analysis for Skill Maintenance (21 items)}
\begingroup
\setlength{\intextsep}{0pt}% space above/below [H] floats
\setlength{\textfloatsep}{0pt}% (harmless here, affects normal floats)
\begin{table*}[t]
\centering
\caption{Exploratory Factor Analysis (Maximum Likelihood, Oblimin Rotation) of Skill-Maintenance Items. SS loadings: ML1 = 3.622, ML3 = 3.286, ML4 = 2.959, ML2 = 2.985; Cumulative variance explained: 61.2\%}
\label{tab:efa21}
\begin{tabular}{lcccc}
\toprule
\textbf{Item} & \textbf{ML1} & \textbf{ML3} & \textbf{ML4} & \textbf{ML2} \\
\midrule
(Creative1) Comprehending written content & \textbf{.590} & .165 & .122 & .018 \\
(Creative2) Writing persuasively & \textbf{.813} & -.012 & .055 & .041 \\
(Creative3) Writing with clarity \& precision & \textbf{.900} & .020 & .018 & -.048 \\
(Creative4) Developing tone/voice of characters & \textbf{.771} & -.095 & .019 & .127 \\
(Creative5) Storytelling and narrative building & \textbf{.806} & .029 & .022 & -.003 \\
(Creative6) Proofreading and correcting writing & \textbf{.676} & .150 & .011 & -.038 \\
\midrule
(Cog1) Breaking down a problem & -.074 & \textbf{.846} & .130 & -.001 \\
(Cog2) Generating  ideas/concepts/inspirations & .068 & \textbf{.800} & -.007 & .017 \\
(Cog3) Organization/translation of ideas into narratives & -.026 & \textbf{.867} & .017 & .026 \\
(Cog4) Evaluating and quality checking of ideas & 0.177 & \textbf{.751} & -.072 & -.029 \\
(Cog5) Recalling a particular idea & .047 & \textbf{.683} & -.007 & .132 \\
\midrule
(Tacit1) Managing time effectively & .012 & .034 & \textbf{.734} & .102 \\
(Tacit2) Planning and managing writing tasks & .036 & .041 & \textbf{.873} & -.048 \\
(Tacit3) Managing  resources for writing & .016 & .003 & \textbf{.873} & -.007 \\
(Tacit4) Learning and apply new techniques & .116 & .086 & \textbf{.701} & -.004 \\
(Tacit5) Engaging in self-reflection & .150 & -.117 & \textbf{.529} & .237 \\
\midrule
(Relation1) Negotiation with co-workers/clients & .014 & .040 & -.061 & \textbf{.890} \\
(Relation2) Persuading co-workers/clients & .072 & -.004 & -.069 & \textbf{.892} \\
(Relation3) Collaborating with other professionals & -.043 & .083 & .193 & \textbf{.696} \\
(Relation4) Seeking or providing writing assistance & -.033 & -.021 & .103 & \textbf{.789} \\
(Relation5) Instructing or managing other professionals  & .098 & .234 & .168 & \textbf{.423} \\
\\
\bottomrule
\end{tabular}
\end{table*}
\endgroup

% \large \textbf{Confirmatory Factor Analysis Comparison for Skill Maintenance (21 items)}
\begingroup
\setlength{\intextsep}{0pt}% space above/below [H] floats
\setlength{\textfloatsep}{0pt}
\begin{table*}[t]
\centering
\caption{Comparison of Competing CFA Models for the Skill-Maintenance Scale}
\label{tab:CFA21}
\begin{tabular}{lccccccc}
\toprule
\textbf{Model} & \textbf{$\chi^2$} & \textbf{df} & \textbf{CFI} & \textbf{TLI} & \textbf{RMSEA} & \textbf{SRMR} & \textbf{BIC} \\
\midrule
1-factor  & 2104.40 & 189 & 0.741 & 0.713 & 0.159 & 0.089 & 27205.33 \\
2-factor  & 1798.56 & 188 & 0.783 & 0.757 & 0.146 & 0.079 & 26905.50 \\
3-factor  & 846.59  & 186 & 0.911 & 0.899 & 0.094 & 0.060 & 25965.52 \\
4-factor (hypothesized) & \textbf{587.26} & \textbf{183} & \textbf{0.945} & \textbf{0.937} & \textbf{0.074} & \textbf{0.056} & \textbf{25724.19} \\
\bottomrule
\end{tabular}
\end{table*}
\endgroup

\end{document}